\documentclass[final,3p,times,authoryear]{elsarticle}

\usepackage[english]{babel}
\usepackage[bookmarks=false,hidelinks]{hyperref}
\usepackage{booktabs}
\usepackage{bm}
\usepackage{newtxmath}
\usepackage{newtxtext}
\usepackage{amsmath}
\usepackage[usenames]{color}
\usepackage{nicefrac}

\newcommand{\kB}{k_\mathrm{B}}
\newcommand{\Ts}{T_\mathrm{s}}
\newcommand{\Tb}{T_\mathrm{b}}
\newcommand{\nne}{n_\mathrm{e}}
\newcommand{\nnii}{n_\mathrm{i}}
\newcommand{\xr}{x_\mathrm{r}}
\newcommand{\gammar}{\gamma_\mathrm{r}}
\newcommand{\rhob}{\rho_\mathrm{b}}
\newcommand{\gcc}{\mbox{g~cm$^{-3}$}}
\newcommand{\slfrac}[2]{\left.#1\middle/#2\right.}
\newcommand{\beq}{\begin{equation}}
\newcommand{\eeq}{\end{equation}}
\newcommand{\bea}{\begin{eqnarray}}
\newcommand{\eea}{\end{eqnarray}}
\newcommand{\req}[1]{Eq.\ (\ref{#1})}
\newcommand{\omc}{\omega_\mathrm{c}}
\newcommand{\mel}{m_\mathrm{e}}

\newcommand{\dd}{\mathrm{d}}
\newcommand{\vB}{\bm{B}}
\newcommand{\Sdot}{\dot{S}}

\renewcommand{\theequation}{\thesection.\arabic{equation}}

\journal{Physics Reports}

\begin{document}

\begin{frontmatter}

\title{Heat Blanketing Envelopes of Neutron Stars}

\author[a1]{M. V. Beznogov}
\ead{mikavb89@gmail.com}
\author[a2]{A. Y. Potekhin}
\ead{palex@astro.ioffe.ru}
\author[a2]{D. G. Yakovlev\corref{cor1}}
\ead{yak@astro.ioffe.ru}
\cortext[cor1]{Corresponding author}

\address[a1]{Universidad Nacional Aut\'onoma de M\'exico,
 Instituto de Astronom\'{\i}a, AP 70-264, CDMX  04510, M\'exico}
\address[a2]{Ioffe Institute, Politekhnicheskaya 26,
 194021 St Petersburg, Russia}

\date{Received: date / Accepted: date}

\begin{abstract}
Near the surface of any neutron star there is a thin heat blanketing
envelope  that produces substantial thermal insulation of warm neutron 
star interiors and that relates the internal  temperature of the star to
its effective surface temperature.  Physical processes in the blanketing
envelopes are reasonably clear but the chemical composition is not. The
latter circumstance complicates inferring physical parameters  of matter
in the stellar interiors from observations of the thermal surface
radiation of the stars and urges one to elaborate the models of
blanketing envelopes. We outline physical properties of these envelopes,
particularly, the equation of state, thermal conduction, ion
diffusion and others. Various models of heat blankets are reviewed, such
as composed of separate layers of different elements, or containing 
diffusive binary ion mixtures  in or out of diffusion equilibrium.  The
effects of strong magnetic fields in the envelopes are outlined as well
as the effects of high temperatures which induce strong  neutrino
emission in the envelopes themselves. Finally, we discuss how the
properties of the heat blankets affect thermal evolution of neutron
stars and the ability to infer important information on internal
structure of neutron stars from observations.
\end{abstract}

\begin{keyword}
Neutron stars
\sep
Diffusion
\sep
Cooling
\end{keyword}

\end{frontmatter}

\tableofcontents

\section{Introduction: Neutron stars, their superdense matter
and thermal insulation}
\label{introduc}

Neutron stars (e.g., \citealt{ST83,HPY07}) are the most compact stellar
objects, with typical masses $M \sim (1-2) \, M_\odot$,
where $M_\odot$ is the solar mass, and radii $R \sim
(10-15)$ km. Their mean mass density  $\rho $ is a few $\rho_0$, where
$\rho_0 \approx 2.8 \times 10^{14}$ g~cm$^{-3}$ is the density of
standard saturated nuclear matter. The central density of massive
neutron stars exceeds $\rho_0$ by about one order of magnitude. Neutron
stars are born in supernova explosions and demonstrate a wealth of
manifestations observed by the methods of multiwavelength astronomy
(from radio to gamma rays) and by gravitational observatories;
they are also expected to be observed by neutrino
observatories.

According to current theories (e.g., \citealt{HPY07}), a neutron star
can roughly  be divided into a relatively light and thin crust (about
1\% by mass and 10\% by radius) and a massive and bulky core. The core
is thought to be liquid; it contains strongly degenerate nucleons,
electrons, muons, and (possibly) other particles like hyperons and/or
deconfined quarks, which are distributed uniformly on the microscopic
scale. In contrast, the crust contains aggregates of nucleons of
microscopic scale (nuclei). The ``crust'' in the broad sense consists of
liquid ``ocean'', solid layers (the crust in the proper sense), and
possibly layers of non-spherical nuclei that behave like liquid crystals
(so called ``pasta phases'', which constitute a ``mantle'' of the star).
The crust is divided into the ``outer crust'', where the nuclei are
immersed in the fluid of electrons, and the ``inner crust'', which
additionally contains the fluid of free neutrons (and possibly free
protons in the deepest layers).

Since the core contains superdense matter, it is the most mysterious and
interesting part of neutron stars; the composition, equation of state
and many other properties of this matter are poorly known. Their study 
is of fundamental importance for  astrophysics and other branches of
physics including nuclear physics, physics of elementary particles, and
condensed matter physics. The basic problems are that (i) it is
difficult to explore  the superdense matter in terrestrial laboratories;
(ii) its properties cannot be calculated reliably because of the absence
of practical theory for describing strong interactions of baryons with
account of many-body effects. Some properties of superdense matter can
be studied in collider experiments on heavy ions collisions,  on neutron
skin measurements and in some other ways  (e.g.,
\citealt{2016LattimerPrakash,Mondal_etal16,2017Oertel}), but these data
are incomplete.  Very dense matter ($\rho \gtrsim 10^4 \rho_0$), which
can appear  after the deconfinement of quarks, can be analyzed by the
methods of perturbative quantum chromodynamics (e.g.,
\citealt{HPY07,MachleidtEntem11}). However, the matter of the most
interesting,  intermediate density, $\rho \sim (3 - 10)\, \rho_0$,
cannot be accurately studied in laboratory experiment and theory but it
can be investigated using observations of neutron stars.

The outer crust extends from the  stellar surface to the neutron drip
density $\rho_\mathrm{nd}\approx 4 \times 10^{11}$ \gcc; its  mass is
$\sim 10^{-5}\,M_\odot$. The inner crust extends from $\rho_\mathrm{nd}$
to the crust/core interface at
$\rho_\mathrm{cc}\approx(1-2)\times10^{14}$ \gcc. The electrons are
mainly degenerate except for the very surface layer. At densities
$\rho \ll 10^{6}$ \gcc\ they are non-relativistic; at higher $\rho$
they become relativistic, and the atomic nuclei become progressively
more neutron-rich.  The free neutrons and the nucleons within the nuclei
can be in superfluid state.

The astrophysical methods to explore the nature of superdense matter in
neutron stars are complex. Generally, they consist of modeling various
processes accessible in observations of neutron stars. Then one can
compare theoretical models with observations and select those  models
which are most suitable. In particular, one can model thermal evolution
of neutron stars with different microphysics of  matter and confront
such models with the measurements of surface temperatures and ages of
neutron stars (as reviewed by
\citealt{YP04,Page09,Tsuruta09,PPP2015,Geppert17}).
This method has been used for many decades and faces many difficulties. 
We will focus on one important obstacle associated with our poor
knowledge of chemical composition of heat blanketing envelopes of
neutron stars.

A heat blanketing (thermally insulating) envelope is situated under
the stellar atmosphere and is thin; its mass is $\lesssim 10^{-6} M_\odot
$.  The surface temperature of the star ($\Ts$), that can be measured,
is typically much smaller than the internal temperature ($\Tb$).  The
relation between $\Ts$ and $\Tb$ is a complex problem, because the 
composition of the heat blankets is often unknown.  By varying the
composition one can obtain different temperatures $\Tb$ at a fixed
temperature $\Ts$, with different conclusions on properties of
superdense matter. In addition, the composition of the heat blanket may
vary in time owing to accretion, diffusion and nuclear burning. These
effects are not easy to study, particularly, because the plasma of ions
can be strongly non-ideal. 

Therefore, to explore properties of superdense matter inside 
neutron stars one needs a reliable theory of heat blanketing
envelopes. While the main features of superdense matter are still basically unclear, 
the properties of heat blankets are based on
a much more elaborated physics of not very dense plasma. However,
the problem of heat blankets creates 
a really serious obstacle to investigate the superdense matter. Our aim is to
describe the current status of the problem. 

In Section~\ref{str-str_equations} we outline the basic
equations of neutron star structure and evolution. Section~\ref{therm-blanket} describes the main properties of the heat
blankets. In Section~\ref{therm-analyt} we consider a simple
semi-analytic model by \citet{VP} of a non-magnetic heat blanket
which explains its main features without detailed
numerical computations. Then we discuss 
(Section~\ref{therm-env}) the 
properties of non-magnetic blanketing envelopes 
constructed by \citet{PCY97}; they 
consist either of iron, or of the layers of lighter
elements (hydrogen, helium, carbon)
and possibly the layer of iron at the bottom. Section~\ref{sec:2} 
is focused on diffusion of ions in the surface layers
of neutron stars. In Section~\ref{sec:4} we describe 
diffusive blanketing envelopes of non-magnetic
neutron stars; these envelopes have been computed for binary
ionic mixtures (H-He, He-C, C-Fe).
The
next Section~\ref{therm-magn-env} is devoted to the envelopes
of magnetized stars. Unlike the non-magnetic envelopes, where
the temperature distribution is isotropic (spherically symmetric),
the temperature distribution in this case can be highly
anisotropic in response to anisotropic character of heat
transport in a magnetic field. In the end of this
section we outline also other models of heat blankets --
for hot stars and magnetars (where the neutrino emission
can be important in the blanket itself); for accreting neutron
stars, where nuclear burning in the blanket can be significant;
and for some other cases.    
In Sections~\ref{sec:5} and \ref{sec:OtherBlankets} 
we present some illustrative examples
how the blankets may affect neutron star evolution. Finally, we
conclude 
in Section~\ref{sec:Conclude}. 
 Some aspects
of the heat-blanket theory are applicable also for white dwarfs. 
In \ref{app:Lambda_Eff} and \ref{app:TsTb} 
we present, respectively, analytic fitting formulas for the diffusion
coefficient in a binary ion mixture and for the relations 
between the surface and internal temperatures for binary heat-blanketing
envelopes.
 
\section{Equations of neutron star structure and thermal evolution}
\setcounter{equation}{0}
\label{str-str_equations}

Let us present general equations of neutron star structure and thermal
evolution. To shorten this introductory task we will restrict ourselves
to spherically symmetric stars with spherically symmetric temperature
distribution inside them, neglecting the effects of magnetic fields and
rotation.  The effects of strong magnetic fields will be briefly
discussed in Section~\ref{therm-magn-env}.

\subsection{Hydrostatic equilibrium}
\label{str-hydrostatic}
 
Neutron stars are relativistic objects and should be studied using General 
Relativity. The importance of relativistic
effects is characterized by the parameter $r_\mathrm{g}/R$, where $R$
is the stellar radius, $r_\mathrm{g}=2GM/c^2$ 
is the gravitational radius, and $G$ the gravitational constant.
Typically, $r_\mathrm{g}/R \sim$ 0.2\,--\,0.4 for neutron stars,
and $r_\mathrm{g}/R \ll 1$ for all other stars.

The metric within or around 
a stationary and spherically symmetric star is (e.g., \citealt{HPY07})
\begin{equation}
    \mathrm{d}s^2=c^2 \, \mathrm{d}t^2 \, \exp(2 \Phi)
         - \exp(2\lambda) \, \mathrm{d}r^2 - r^2 \,
         (\mathrm{d} \theta^2 + \sin^2 \theta \, \mathrm{d} \phi^2),
\label{str-metrics}
\end{equation}
where $t$ is a time-like coordinate (Schwarzschild time for
a distant observer), $r$ is a radial
coordinate, $\theta$ and $\phi$ are the polar angle and azimuth,
respectively, while $\Phi=\Phi(r)$ and $\lambda=\lambda(r)$
are two metric functions of $r$. The angular geometry (with respect
to $\theta$ and $\phi$) is the same as in flat space-time
because of spherical symmetry, but space-time
is generally curved along  $r$ and $t$ ``directions.''
In flat space-time, we would have $\Phi=\lambda=0$.

It is well known that $r$ in Eq.\ (\ref{str-metrics}) is
the circumferential radius which
determines proper length of the circle, $2 \pi r$;
proper area of a spherical surface
at given $r$ is $4 \pi r^2$. The proper radial length
from the stellar center, 
$l= \int_0^r \, \exp \lambda \,\mathrm{d}r$, is generally
different from $r$. 
Hence $\lambda(r)$ determines curvature in the radial direction. 
It is related to
the gravitational mass $m(r)$
contained inside a sphere with radial coordinate $r$,
\begin{equation}
      \exp \lambda = \left( 1 - \frac{2 G m}{r c^2} \right)^{-1/2}.
\label{str-lambda}
\end{equation}
The gravitational mass
is  smaller than the baryon mass (``rest mass'') due to
gravitational mass defect. 

A proper radial length element $\mathrm{d}l$ and
a proper volume $\mathrm{d}V$ between close spherical shells
are
\begin{equation}
     \mathrm{d}l= \frac{ \mathrm{d}r }{\sqrt{1 - 2Gm/(c^2r)}}, 
\quad
     \mathrm{d}V= \frac{ 4 \pi r^2 \, \mathrm{d}r
      }{
      \sqrt{1- 2Gm/(c^2 r)}}.
\label{str-proper}
\end{equation}
A proper time
interval in a local rest-frame is 
\begin{equation}
     \mathrm{d}\tau = \mathrm{d}t \, \exp {\Phi(r)}.
\label{str-tau}
\end{equation}
Therefore, $\Phi(r)$ determines gravitational 
dilatation of time and gravitational redshift of signals.
If a local source produces a periodic signal
of frequency $\omega_0$ at $r=r_0$, a distant observer 
($r \to \infty$, $\Phi \to 0$) will detect a signal of frequency
 $\omega_\infty=\omega_0 \, \exp {\Phi(r_0)}$.

The equations of hydrostatic structure of the star
follow directly from Einstein equations, 
\begin{eqnarray}
   \frac{\mathrm{d}P }{\mathrm{d}r} & = &
     - \frac{G \rho m }{r^2 } \left( 1 + \frac{P }{\rho c^2} \right)
     \left( 1 + \frac{4 \pi P r^3 }{m c^2 } \right)
     \left( 1 - \frac{2 G m }{c^2 r} \right)^{-1},
\label{str-P} \\
   \frac{\mathrm{d}m }{\mathrm{d}r} & = & 4 \pi r^2 \rho,
\label{str-m} \\
   \frac{\mathrm{d}\Phi }{\mathrm{d}r} & = &
   - \frac{1 }{\rho c^2} \,
     \frac{\mathrm{d}P }{\mathrm{d}r} \,
     \left( 1 + \frac{P }{\rho c^2 } \right)^{-1},
\label{str-Phi}
\end{eqnarray}
where $P$ is the pressure and 
$\mathcal{E}\equiv \rho c^2$ is the energy density of the stellar matter. 
The quantity $\rho$, introduced instead of $\mathcal{E}$, is
called the  mass density of the matter. It includes baryon mass density
and mass defects produced by microscopic particle motion, by
strong, weak and electromagnetic interactions, but not by
the gravitational interaction.
The mass density in dense neutron star cores is noticeably
different from the traditional baryon mass density.

Equation (\ref{str-P}) is the Tolman-Oppenheimer-Volkoff
equation of hydrostatic
equilibrium \citep{1939Tolman,1939OV},
Equation (\ref{str-m}) describes mass balance, while
Eq.\ (\ref{str-Phi}) is a relativistic version of the
equation for the dimensionless gravitational potential $\Phi(r)$.
These equations
should be supplemented by the equation of state (EoS) 
that relates pressure to density and temperature, 
$P=P(\rho,T)$. In the layers where free leptons or nucleons are strongly
degenerate, the EoS is almost independent of temperature.
Then in the above equations one can use a barotropic EoS, that is
$P=P(\rho)$. Recalling that 
a neutron star is composed mostly of
strongly degenerate matter, we can conclude that
the neutron
star structure is largely independent
of its thermal state. Temperature
effects on hydrostatic structure
are important only near the surface (in the atmosphere
and the heat blanketing envelope). 

Equation (\ref{str-P}) can be
rewritten in the Newtonian form
\begin{equation}
  \frac{ \mathrm{d}P }{\mathrm{d}l} = -g \rho,
\label{str-Newton}
\end{equation}
where 
\beq
  g=\frac{ Gm \exp \lambda }{r^2 }\,
  \left( 1 + \frac{P }{\rho c^2} \right)
     \left( 1 + \frac{4 \pi P r^3 }{m c^2 } \right)
\eeq
 is a local gravitational acceleration.

Outside the star one has  
$P=0$ and $\rho=0$, so that
$m(r)=M$ is constant, which is the
total gravitational mass of the star.
In this case,
$\exp(2 \Phi)=\exp(-2\lambda)=1-r_\mathrm{g}/r$ meaning
the Schwarzschild metric outside the star, with $r_\mathrm{g}=2GM/c^2$ being the Schwarzschild radius.
At $r \gg r_\mathrm{g}$, the Schwarzschild
space-time becomes asymptotically
flat. Finally, for a non-relativistic star 
($P \ll \rho c^2$,
$P r^3 \ll m c^2$, $r_\mathrm{g} \ll R$) 
Eqs.\ (\ref{str-P})--(\ref{str-Phi}) reduce to the 
Newtonian equations of stellar equilibrium,
where $\Phi c^2$ plays role of the Newtonian
gravitational potential.
   
\subsection{Thermal structure and evolution}
\label{sub:evolution}

The thermal structure of a neutron star is characterized 
by the internal distribution of local temperature 
$T=T(r,t)$. In contrast to the hydrostatic structure 
that undergoes almost no evolution, the thermal structure may 
strongly evolve and 
affect observational manifestations of neutron stars. 
Let us consider not too hot 
(and not too young) stars,
with internal temperatures $T \lesssim 10^{10}$ K
(of age $\gtrsim 1$ min); they are fully
transparent for neutrinos (e.g., \citealt{1999PNS}). 

Generally, the thermal evolution is governed
by heat conduction within the star, with subsequent thermal
emission from the surface, and also by neutrino cooling from
the bulk of the star and possibly by some reheating from inside  
or from the surface. 

General relativistic equations of thermal evolution of a spherically
symmetric star were derived by
\citet{Thorne66,Thorne}. Basically, one needs to solve the two equations,
of thermal balance  and  thermal transport.

\emph{The thermal balance equation}
can be written as \citep{Richardson82},
\beq
    \frac{ 1 }{ 4 \pi r^2 \mathrm{e}^{2 \Phi+\lambda}} \,
    \frac{ \partial }{ \partial  r}
    \left( \mathrm{e}^{2 \Phi} L_r \right)
    = -Q_\nu + Q_\mathrm{h} - \frac{T }{ \mathrm{e}^\Phi} \,
           \frac{\partial S}{ \partial t},
\label{therm-balance}
\end{equation}
where $Q_\nu$ is the  neutrino cooling rate per unit volume,
$Q_\mathrm{h}$ is the heating power per unit volume (if any),   $S$ is
the entropy per unit volume, and $L_r$ is the ``local luminosity''
(non-neutrino energy transported through a sphere of radial coordinate
$r$ per unit time); all these quantities depend on $r$ and $t$.  The
thermal flux density associated with $L_r$ is $F=L_r/(4 \pi r^2)$.  It
may be convenient to include the entropy changes related to structural
modifications (such as phase transitions) into $Q_\mathrm{h}$. Then $T
\partial S / \partial t = C \partial T / \partial t$, where $C$ is the
heat capacity per unit volume at constant pressure. The heat capacities
at constant volume and constant pressure  are almost equal in the
strongly degenerate matter, that is almost everywhere throughout the
star  (e.g., \citealt{HPY07}); therefore we will not distinguish between
them.  The quantities $C$, $Q_\nu$, and $Q_\mathrm{h}$ have to be
determined from microscopic thermodynamic and kinetic theories and from
a model of internal heating (if available).  In the absence of the
latter ($Q_\mathrm{h}=0$), one deals with free (passive) cooling of the
star.  Typical microscopic scales (mean free paths, etc.) are much
smaller than space-curvature scales. If so, thermodynamic and kinetic
quantities can be calculated neglecting the effects of General
Relativity. 

\emph{The heat transport equation}
depends on the heat transport mechanism.  A stationary heat conduction
through non-moving matter in the local reference frame is governed by
equation
\beq
   \hat{\kappa}\nabla T = \bm{F}
\label{heatcond}
\eeq
where $\hat{\kappa}$ is the conductivity tensor and $\bm{F}$ is the heat
flux density. If the matter is isotropic, then the conductivity tensor
can be replaced by scalar $\kappa$.  We will assume it to be the case,
unless the opposite is stated.

If the transport is dominated by convection, then in the simplest
approximation \citep[e.g.][]{Schwarzschild58,Kippenhahn} 
\beq
  \frac{\partial T}{\partial l} = \frac{\partial P}{\partial l} \,
  \frac{T}{P}\,\nabla_\mathrm{ad},
\eeq
where 
\beq
  \nabla_\mathrm{ad} = \left(\frac{\partial P}{\partial T}\right)_S
\eeq
is the adiabatic temperature gradient.
The medium is stable against convection, if $\nabla_\mathrm{rad} <
\nabla_\mathrm{ad}$, where
\beq
\nabla_\mathrm{rad} = \frac{F P}{\kappa g \rho T}
\eeq
 is the ``radiative temperature gradient'', equal to the value that
$(\partial \ln T / \partial r)/(\partial \ln P / \partial r) =
\mathrm{d} \ln T / \mathrm{d} \ln P$ would have according to
Eqs.~(\ref{str-Newton}) and (\ref{heatcond}), were the convection
absent. 

In the strongly degenerate matter, as mentioned above, $P$ is almost
independent of $T$.  Therefore, in the strongly degenerate layers of a
neutron star, $\nabla_\mathrm{ad}$ is high and the convection is
suppressed. The convection may operate in  surface layers, where the
matter is less degenerate (or non-degenerate),  but its effects on
observables seem minor (e.g., \citealt{Z96a}; see also
Section~\ref{therm-env}). Hereafter we will focus on thermal conduction.

Using the equations of hydrostatic structure 
(Section~\ref{str-hydrostatic}) and
the heat transport equations in the local reference frame
\citep{Thorne}, one can rewrite a generally relativistic
Fourier equation of thermal
conduction in a neutron star as
\beq
    \frac{L_r }{ 4 \pi \kappa r^2} =
    - \mathrm{e}^{-\Phi-\lambda}
    \frac{\partial }{ \partial r} \left( T \mathrm{e}^\Phi \right).
\label{therm-Fourier}
\end{equation}
Thermal conduction 
is mainly provided by degenerate fermions
(electrons, muons, neutrons) almost everywhere in the star excluding
a very surface layer, where it becomes radiative (as reviewed, e.g.,  
by \citealt{PPP2015}).

Therefore,  one has to solve Eqs.~(\ref{therm-balance}) and
(\ref{therm-Fourier}) to determine $L_r(r,t)$ and $T(r,t)$.  These
equations should be supplemented by the initial and boundary conditions
which depend on a specific problem.  For an initially hot and passively
cooling neutron star the initial temperature profile $T(r,0)$ can be
taken rather arbitrary; the initial  temperature distribution relaxes on
a timescale of several months and does not affect further cooling (the
memory loss effect; see Section~\ref{sub:youngAndmergingNSs} for a brief
discussion of the cooling of neo-neutron stars at shorter timescales).
At the stellar center $T(0,t)$ should be finite and $L_r(0,t)=0$. The
boundary conditions at the surface are discussed in
Section~\ref{therm-blanket}.

Instead of $T(r,t)$ it is often convenient to introduce the redshifted 
internal temperature $\tilde{T}(r,t)$ which stops to depend on $r$ in 
an isothermal layer,
\begin{equation}
\widetilde{T}(r,t)=T(r,t)\, \mathrm{e}^{\Phi(r)}~~\to
\mbox{~independent~of}~r\mbox{~in~an~isothermal~layer}.
\label{eq:isothermal}
\end{equation}

\paragraph{Observables}
By solving a thermal evolution problem one calculates the effective 
surface temperature $\Ts$ of the star and 
the photon surface luminosity 
$ L_\gamma = 4 \pi \sigma_\mathrm{SB} R^2 T_\mathrm{s}^4(t)$ in a locally-flat reference frame at
the neutron star surface,
$\sigma_\mathrm{SB}$ being the Stefan-Boltzmann constant. 
A distant observer would register
the ``apparent'' (redshifted) luminosity $L_\gamma^\infty$,
``apparent'' effective surface temperature $T_\mathrm{s}^\infty$, and
``apparent'' radius $R_\infty$,
\begin{eqnarray}
&&   L_\gamma^\infty = L_\gamma (1 - r_\mathrm{g}/R) =
     4 \pi \sigma_\mathrm{SB} (T_\mathrm{s}^\infty)^{4} R_\infty^2,
\label{srt-L_gamma_infty}\\
&&   T_\mathrm{s}^\infty = T_\mathrm{s} \, \sqrt{1 - r_\mathrm{g}/R}, \quad
     R_\infty = R/ \sqrt{1 - r_\mathrm{g}/R}.
\label{therm-T_s_infty}
\end{eqnarray}
Calculated quantities
can be compared with observations. Typically, $\Ts^\infty/\Ts = R/R_\infty \sim 0.8$.

\section{Basic concepts of heat-blanketing envelopes}
\setcounter{equation}{0}
\label{therm-blanket}

\subsection{Outlook}
\label{sub:therm-outlook}

\paragraph{Heat blanket and internal region}

Direct calculation of $T(r,t)$ from the stellar surface to the center is
possible but time-consuming.  To facilitate calculations, one usually
divides the problem artificially into two parts by analyzing heat
transport in the outer heat-blanketing envelope ($R_\mathrm{b} \leq r
\leq R$) and in the interior ($r < R_\mathrm{b}$; the choice of the
boundary radius $R_\mathrm{b}$
is addressed below).  The full set of the dynamical equations for
$T(r,t)$ is solved in the internal region, while the heat blanket is
studied separately in a quasi-stationary and plane-parallel
approximation and serves as a boundary condition for the internal
solution. Here we focus on heat blankets.

\paragraph{Heat blanket: Formal definition. 
Mathematical and physical blankets}

The blanketing envelope is the layer under  the atmosphere (under the
radiative surface discussed later in this section) down to some
boundary. The choice of this boundary, characterized by radius
$R_\mathrm{b}$, corresponding mass density $\rhob$, or depth
$z_\mathrm{b}$, is conditional. It is chosen so as to optimize
computations and is subject to several requirements.  The blanketing
envelope should be thin (as compared to $R$) and contain negligibly
small mass; there should be no large sources of energy generation or
sink there; it should serve as a good thermal insulator of the internal
region; its thermal relaxation time should be sufficiently short to
treat the blanket quasi-stationary.  As a rule, these requirements are
satisfied by placing the bottom boundary at the density
$\rho_\mathrm{b}=\rho(R_\mathrm{b})$ between $10^8$ g cm$^{-3}$ and
$10^{11}$ g cm$^{-3}$.  Usually, following \citet{GPE83}, one sets
$\rho_\mathrm{b} \sim 10^{10}$ g cm$^{-3}$ (a few hundred meters under
the surface). A division into the interior and an envelope is often used
in stellar modeling (not only for neutron stars, but also, for example,
for white dwarfs -- e.g., \citealt{Koester_KI20}). Some requirements can
be relaxed as we discuss later.

In some cases it is possible to choose $\rho_\mathrm{b}$ in such a way
that the entire internal region be almost isothermal for the range of
$\Ts$ of study. Then the main temperature gradient occurs within the
heat blanket, and the modeling of the thermal evolution within the
internal region is greatly simplified because of  \req{eq:isothermal}.
In principle, one can introduce \emph{physical} heat blanketing
envelopes as insulating layers containing strongest temperature
gradients. However the bottom density of such envelopes would be very
sensitive to $\Ts$, as will be discussed in
Section~\ref{ssec:4:HeatBlanketsModel}. The smaller $\Ts$, the thinner
this physical heat blanket would be, which is inconvenient for
computations. Therefore, the artificial ``computational'' heat blankets
with fixed $\rho_\mathrm{b}$ are usually wider than their physical
counterparts. 

\subsection{Basic equations of heat blankets}
\label{subsub:basic_eqns}

Since a heat blanketing envelope is thin and light, 
the space-time curvature in the envelope
is nearly constant, so that the metric functions
are almost the same as at the surface (at $r=R$),
\begin{equation}
  \exp \lambda_\mathrm{s}= \exp(-\Phi_\mathrm{s})
  =\frac{1}{\sqrt{1-{r_\mathrm{g}/R}}}.
\label{eq-HB-metric}
\end{equation}
Therefore, the space-time is nearly flat there, although the time and
length scales are different than those for a distant observer. It is
convenient to introduce the proper depth $z=(R-r)\,\mathrm{e}^{\lambda_\mathrm{s}}$ from the surface (Section~6.9 of
\citealt{HPY07}). Equation  (\ref{str-P}) of hydrostatic equilibrium in
the envelope reduces to
\begin{equation}
     \frac{\mathrm{d}P}{\mathrm{d}z}=g_\mathrm{s} \rho,
\label{dP/dz}    
\end{equation}
where 
\beq
    g_\mathrm{s}=\frac{GM \mathrm{e}^{\lambda_\mathrm{s}}}{R^2}
\label{g_s}
\eeq
is the surface gravity.

Let the thermal relaxation in the blanket be sufficiently fast, so that
the heat transport problem can be treated as quasi-stationary, assuming
that $T(r)$  is explicitly independent of time  (although it can depend
on $t$ parametrically).  Then the heat flux density through the blanketing
envelope obeys \req{heatcond}, which in the absence of anisotropy
becomes
\beq
    \kappa \, \frac{\mathrm{d}T }{ \mathrm{d}z} = F_r.
\label{therm-Tcrust}
\end{equation}
Here, $F_r=L_r/(4\pi r^2)$ is the thermal flux density in the radial
(outward) direction in the local reference frame.
In the stationary envelope (i.e., $\partial S / \partial t=0$),
according to the energy conservation law,
\beq
\frac{\mathrm{d}F_r}{\mathrm{d}z} = Q_\nu - Q_\mathrm{h}.
\label{dFdz}
\eeq
In the absence
of any significant local energy sources and sinks
$Q_\nu - Q_\mathrm{h}=0$. In this case,
$L_r$ and $F_r$ are nearly constant, so that
\beq
L_r \approx 4 \pi R^2 F_r
\approx 4 \pi R^2 \sigma_\mathrm{SB} T_\mathrm{s}^4 \equiv L_\gamma.
\label{LrTs}
\eeq

Supplemented by the  EoS and by an appropriate thermal conductivity, 
Eqs.~(\ref{dP/dz}), (\ref{therm-Tcrust}), and (\ref{dFdz}) can be solved
to determine $\rho(z)$ and $T(z)$. The most important output would be 
temperature $\Tb$ and radial heat flux $F_\mathrm{b}$ at the bottom of
the heat blanket;  they depend on $\Ts$. By varying $T_\mathrm{s}$, one
can obtain the dependences  $F_\mathrm{b}(T_\mathrm{b})$, one of the
basic ingredients for the theory of neutron star  evolution. They are
used in boundary conditions at $r=R_\mathrm{b}$, 
\begin{equation}
  T(R_\mathrm{b})=\Tb, \quad  F_r(R_\mathrm{b})= F_\mathrm{b},
\label{boundarycond}  
\end{equation}
for solving the
thermal evolution equations (\ref{therm-balance}) and 
(\ref{therm-Fourier}) in the neutron star interiors 
($r<R_\mathrm{b}$). 
In the absence of internal energy sources and sinks in the envelope,
\req{dFdz} gives $F_r={}$constant.
This is the most common case, valid for not too hot neutron stars.
Then the relation $F_\mathrm{b}(T_\mathrm{b})$
is equivalent to the relation $T_\mathrm{s}(T_\mathrm{b})$, 
which is obtained by solving Eqs.~(\ref{dP/dz}) and (\ref{therm-Tcrust})
with $F_r=F_\mathrm{b}=\sigma_\mathrm{SB} T_\mathrm{s}^4$.

The most attractive feature of the heat-blanket
problem is its \emph{self-similarity}.
The structure of the blanket is largely independent of the internal
structure of the star, particularly, of specific 
values of mass and radius and of the EoS
of internal layers. The only global parameter of the
star which a heat blanket ``respects'' is the surface gravity $g_\mathrm{s}$. 
One can construct a model 
of the heat blanketing envelope for some assumed value of 
$g_\mathrm{s}$ and then rescale it for other values.
We will discuss this throughout the text.

Instead of the conductivity $\kappa$, one often introduces the opacity $K$,
\beq
   K = \frac{16\sigma_\mathrm{SB} T^3 }{ 3\kappa\rho}.
\label{K-kappa}
\end{equation}
Heat is transported through the envelope  mainly by radiation and
electrons,
\beq
 \kappa=\kappa_\mathrm{r}+\kappa_\mathrm{e},
  \qquad
  K^{-1} = K_\mathrm{r}^{-1} + K_\mathrm{e}^{-1},
  \label{op}
 \end{equation}
where $\kappa_\mathrm{r}$, $\kappa_\mathrm{e}$ and $K_\mathrm{r}$,
$K_\mathrm{e}$ denote the radiation and electron-conduction components
of the conductivity and opacity, respectively.  Specifically,
$K_\mathrm{r}$ is the Rosseland mean opacity
\citep[e.g.,][]{Mihalas1978}. Typically, the radiative conduction
dominates ($\kappa_\mathrm{r} > \kappa_\mathrm{e}$) in the outermost
non-degenerate layers of the envelope, whereas the electron conduction
dominates ($\kappa_\mathrm{e}>\kappa_\mathrm{r}$) in the deeper layers
of degenerate electrons.

\paragraph{Radiative boundary}

Above the heat blanketing envelope, there is a very thin neutron
star atmosphere that is usually neglected in calculations 
of such global parameters of neutron stars as total mass
and radius. It is the place where the spectrum
of thermal radiation, emergent from stellar interiors, is formed.
The {optical depth} $\tau$ is expressed
through the geometrical depth $z$ as
\beq
  \tau(z)=\int_{-\infty}^{z} K\big(\rho(z'),T(z')\big)\,\rho(z')\,\dd z'.
\label{tau(z)}
\end{equation}
With increasing $z$ within the heat blanket, $\tau(z)$
becomes very large.

The \emph{radiative boundary (radiative surface)} is  defined by the
condition $T=T_\mathrm{s}$. In the Milne-Eddington approximation to the
radiative transfer problem (e.g., \citealt{Mihalas1978}),  it is placed
at the Rosseland optical depth $\tau=\frac23$. Using this approximation
and assuming $K$ to be constant, from \req{tau(z)} one obtains a simple
relation
\beq
    K_\mathrm{s} P_\mathrm{s} = \frac{2 }{ 3} \, g_\mathrm{s},
\label{therm-Eddington}
\end{equation}
where $K_\mathrm{s}=K(\rho_\mathrm{s},T_\mathrm{s})$ and $P_\mathrm{s}$ are, respectively, the
radiative-surface opacity and pressure to be determined.

In reality, $K$ is not constant. However, it varies along the thermal
profile in the radiative zone much slower than $P$. This makes
\req{therm-Eddington} a good approximation, as we shall see in
Section~\ref{sect-analytic-non-degen}.

\subsection{The matter of heat blankets}
\label{sub:HBmatter}

\subsubsection{Electrons and ions}
\label{subsub:e-and-i}

The mass density $\rho$ in a heat blanketing envelope
varies in a wide range, from $\sim 0.1$ \gcc{} at the radiative boundary to 
$\sim 10^{10}$~\gcc{} at the bottom of
the envelope.
This is a plasma of electrons and ions whose properties
are reviewed, for instance, in
\citet{HPY07}. Near the stellar surface, depending on the
temperature, density, and composition, the plasma can
be partially ionized; its two (electron and ion) components
can be non-ideal. Deeper in the heat blanket, the ions become fully
ionized and the electrons constitute a nearly
ideal gas. With increasing $\rho$, the electrons become
degenerate, and at $\rho \gtrsim 10^6$ \gcc\ they become
relativistic. When the electrons are nearly free, they form a
slightly compressible negative charge background in which
the ions move. The ions constitute the so called Coulomb
ion plasma which can be in gaseous, liquid or solid (crystalline or
amorphous) 
state. The ion plasma can be one-component or contain ions of different species, 
$j=1,2,\ldots$
Let $A_j$ be the relative atomic weight and $Z_j$
the charge number of the ion species $j$.
The condition for 
electric neutrality of the plasma implies
\begin{equation}
    \nne = \sum_j Z_j n_j,
\label{e:ChargeNeutrality}
\end{equation}
where $\nne$ is the number density of electrons and $n_j$ is
the number density of ions $j$. The total number density of the
ions is  $\nnii =\sum_j n_j$. The mass density
of the matter is mostly contained in the ions, $\rho \approx
\sum_j m_j n_j$, where $m_j=A_j m_\mathrm{u}$, with $m_\mathrm{u}$ 
being the atomic mass unit. On the other hand, the pressure in
heat blankets is mainly provided by the electrons. In what follows
(unless the contrary is indicated), we assume full ionization.

A state of free electrons is conveniently characterized by  
the parameters
\begin{equation}
    p_\mathrm{F}=\hbar (3 \pi^2 \nne )^{1/3}, \quad
    \xr = \frac{p_\mathrm{F}}{\mel c}
    \approx 1.0088 \left(\frac{\rho_6\,\bar{Z}}{\bar{A}} \right)^{1/3},
\label{eq:xr}    
\end{equation}
where $p_\mathrm{F}$ is a measure of $\nne$ which has the meaning of
electron Fermi momentum if the electrons are strongly degenerate; $\xr$
is the relativity parameter of degenerate electrons, $\rho_6=\rho/10^6$
\gcc; $\bar{Z}$ and $\bar{A}$ are, respectively, the averaged values of
$Z_j$ and $A_j$. The averaging  is defined as $\bar{f}=\sum_j x_j f_j$
for any quantity $f$, where $x_j={n_j}/{\nnii}$ is the number fraction
of ion species $j$. In these notations, the electron degeneracy
temperature is
\begin{equation}
   T_\mathrm{F}=\frac{c}{\kB}\,\left( \sqrt{\mel^2 c^2+p_\mathrm{F}^2}
  - \mel c \right) =T_\mathrm{r}\,
  \left(\sqrt{1+\xr ^2}-1 \right),\quad
  T_\mathrm{r}=\frac{\mel c^2}{\kB} 
        \approx 5.930 \times 10^9\mbox{~K},
\label{degenerateT}  
\end{equation}
$\kB$ being the Boltzmann constant.

In a multicomponent ion plasma  it is convenient to introduce the
Coulomb coupling parameter for each ion species (e.g., \citealt{HPY07}),
\begin{equation}
    \Gamma_j = \frac{Z_j^2 e^2}{a_j k_\mathrm{B} T} = \frac{Z_j^{\frac{5}{3}} e^2}{a_\mathrm{e} k_\mathrm{B} T},
\label{e:Gammaj}
\end{equation}
where $e$ is elementary charge, $a_\mathrm{e} = \left(\slfrac{4\pi
\nne}{3}\right)^{-\slfrac{1}{3}}$is the electron sphere radius, and $a_j
= a_\mathrm{e} Z_j^{\slfrac{1}{3}}$ is the ion sphere radius for ions of
species $j$.  The charge of nearly free electrons within any ion sphere
compensates the ion charge. The parameter $\Gamma_j$ characterizes the
ratio  of electrostatic energy of an ion sphere to the thermal energy
$\kB T$. If $\Gamma_j \ll 1$ the Coulomb coupling of given ions is weak,
while at $\Gamma_j \gg 1$ it is strong.

It is also instructive to introduce the mean 
Coulomb coupling parameter for all ions
(e.g., \citealt{HPY07}),
\begin{equation}
    \bar{\Gamma} = \Gamma_0 \overline{Z^\frac{5}{3}}\,
     \bar{Z}^\frac{1}{3},
\label{e:GammaAv}
\end{equation}
where
\begin{equation}
    \Gamma_0 = \frac{e^2}{a_\mathrm{i} k_\mathrm{B}T}
\label{e:Gamma0}
\end{equation}
is a convenient notation, with $a_\mathrm{i}$ being a mean ion sphere
radius (a typical inter-ion distance) defined as
\begin{equation}
    a_\mathrm{i} = \left(\frac{3}{4\pi \nnii} \right)^\frac{1}{3}.
\label{e:IonSphereRad}
\end{equation}

At low enough  densities and high temperatures, where $\bar{\Gamma} \ll
1$,  the entire ion plasma is weakly coupled (resembles a mixture of
ideal gases). In  the opposite case of $\bar{\Gamma} \gg 1$ the ions 
are strongly coupled by Coulomb forces. The ions form a Coulomb liquid
at those temperatures at which $1 \lesssim \bar{\Gamma} \lesssim
\Gamma_\mathrm{m}$, where $\Gamma_\mathrm{m}$ corresponds to the melting
temperature $T_\mathrm{m}$. At $T < T_\mathrm{m}$
($\bar{\Gamma}>\Gamma_\mathrm{m}$) the liquid solidifies into a crystal;
the gas-liquid transformation at $\bar{\Gamma} \sim 1$ can be smooth
(without phase transition). In the presence of ions with strongly
different charges, the so called superionic structures are also
possible, where the ions with a larger $Z$ form a lattice, but the ions
with a smaller $Z$ do not (e.g., \citealt{Redmer_11}, and references
therein). If all the ions are of one and the same type, they are
described by the single parameter $\Gamma$. In the so called ``rigid
electron background'' model, $\Gamma_\mathrm{m}\approx 175$
\citep{2000PC}. This model can be sufficient for strongly degenerate
electrons, although even for them the allowance for electron
polarization can shift $\Gamma_\mathrm{m}$ value by tens percent
\citep{PC13}. Quantum effects of ion motion can substantially affect
crystallization of the plasma composed of light elements;
they can even preclude the crystallization for H or He (e.g.,
\citealt{Chabrier93,JonesCeperley96,BaikoYakovlev19}).

Many features of melting/crystallization for several 
ion species are still unclear. When the star cools, the layer
of liquid and gaseous ions (the ocean)
becomes thinner and shrinks to the surface.

Thermodynamic properties of the Coulomb plasma of ions,
its electric and thermal conductivities and
diffusion coefficients have been studied 
in many works. The
details on the EoS and thermodynamic properties
can be found, e.g., in the papers by
\citet{Hansen75,HTV77,2000PC,PC10}, as well as in a review
article by \citet{BausH80} and in the monograph by
\citet{HPY07}. Transport properties of Coulomb
plasmas are reviewed, for instance, by
\citet{PPP2015}. In a multicomponent
ion plasma it is important to know the diffusion coefficients
as we discuss in Section~\ref{sec:2}. 

\subsubsection{Chemical composition}
\label{subsub:chemicalcompos} 

The composition of the
heat blankets is generally unknown because it cannot be observed directly
being hidden for an observer by a
neutron star atmosphere. The composition 
may depend on the formation and evolution of the star.

Initially, it has been assumed that the envelopes as well as the
atmospheres of neutron stars consist of heavy elements (such as
iron) because the envelopes are formed in very young and hot 
stars where light elements are burnt out in thermonuclear reactions.

However, detailed studies of radiation spectra from 
neutron stars revealed that although some spectra are, indeed,
well described by the black-body model (or similar models of
atmospheres composed of iron) but other spectra are better
described by hydrogen or carbon atmosphere models (see
\citealt{COOLDAT}, and references therein). For example, spectra of 
neutron stars in supernova remnants
Cassiopeia A \citep{HoHeinke_09}, HESS J1731--347 \citep{Klochkov_etal13},
and G15.9+00.2 \citep{Klochkov_16} are well described by
carbon atmosphere models.

The compositions of underlying envelopes can also be
different. The envelopes may be affected by the
fallback of matter onto the stellar surface after a supernova
explosion, by accretion of hydrogen and/or helium from interstellar
medium or a companion star (if the neutron star enters or entered a
binary system, \citealt{Blaes92}), by diffusion and nuclear burning of
the matter in the envelope, and by other effects. For instance,
helium can be accreted directly or produced as a result of hydrogen
burning after accretion of hydrogen 
\citep[e.g.][]{ChiuSalpeter64,Rosen68,CB03,Wijngaarden_ea19}.
Some transiently
accreting neutron stars in low-mass X-ray binaries in quiescent
states (when accretion stops) contain hydrogen or helium layers as 
a leftover of active accretion stages (e.g., see \citealt{BBC02}).
Accordingly, it is instructive to study
different envelope models and their observational manifestations.

On the other hand, the chemical composition of heat blankets cannot be
absolutely arbitrary. There are important constraints which have to be
respected in theoretical models. The main constraint is imposed by 
gravitational stratification \citep{AI80,HameuryHB83}. There is a strong tendency
for such a stratification in neutron stars because of the very high
gravity. Lighter elements tend to be on top while heavier elements on
bottom (see Section~\ref{sec:2}). However, there could be processes
working in the opposite direction  (for instance, ion diffusion).
In addition, thermonuclear processes in the envelopes of
accreting neutron stars can instantaneously create 
complex ion mixtures 
\citep[see, e.g.,][for review and references]{Meisel_18}.

Finally, the densities and temperatures, at which light elements can
survive in a heat blanket, are naturally  restricted by nuclear physics,
particularly, by explosive or stable nuclear burning as well as by
electron captures. The density-temperature ranges  where different
elements survive for a sufficiently long time are not very certain and
depend on many factors, such as nuclear composition of the matter,
internal temperature of the star,  dynamics of mass accretion rate if
the star is accreting. The heavier the element, the wider its $\rho-T$
range.  Very roughly, hydrogen can survive at temperatures $T \lesssim
(4-7) \times 10^7$ K and  densities $\rho \lesssim (10^6-10^7)$ \gcc,
helium ($^4$He) at $T \lesssim (1-3) \times 10^8$ K and $\rho \lesssim
(10^8-10^9)$ \gcc, carbon ($^{12}$C) at $T \lesssim (3-7)\times10^8$~K
and  $\rho \lesssim 10^9 - 10^{10}$ \gcc{}
 (e.g., \citealt{Ergma86,Kippenhahn}; see also, e.g.,
\citealt{PiersantiTY14} for accreted helium, and
\citealt{PC12} for carbon envelopes). In the absence  of
light elements, a heat blanket could be mostly composed of iron.
Comprehensive reviews on nuclear burning in surface layers of neutron
stars have been given by \citet{2017Galloway} and by \citet{Meisel_18}.
In what follows, unless the contrary is indicated,  we will mainly
consider the $^1$H, $^4$He, $^{12}$C, and $^{26}$Fe isotopes, and we
will  drop isotopic indices, for brevity. Naturally, there could  be
many other elements and/or isotopes in the blanketing envelopes  which
can be included into consideration if necessary.

\subsection{Mass distribution in heat blankets}
\label{sec:basis}

\emph{A density profile within a heat blanket} is governed by
Eq.\ (\ref{dP/dz}). 
For simplicity, let the temperature effects be negligible ($T\to0$), 
and the pressure be provided by strongly degenerate electrons up to the surface $z=0$. 
We assume further that 
the ratio of the mean charge and mass numbers, 
$\bar{Z}$ and $\bar{A}$, is fixed. Then Eq.\ (\ref{dP/dz}) can be integrated with the result
(e.g., \citealt{HPY07}, Section~6.9)
\begin{equation}
    \xr^3 = \left[\frac{z}{z_0}
    \left(2+\frac{z}{z_0} \right)\right]^{3/2},
    \quad z_0=\frac{\mel c^2 \bar{Z}}{m_\mathrm{u} g_\mathrm{s} \bar{A}}
  =\frac{49.3 \bar{Z}}{ g_\mathrm{s14}\,\bar{A}}~\mathrm{m},
\label{rho(z)}    
\end{equation}
where $g_\mathrm{s14}$ is the surface gravity $g_\mathrm{s}$ in units
of $10^{14}$ cm~s$^{-2}$, and $\xr$ is given by Eq.\ (\ref{eq:xr}).
Since $\xr^3 \approx \rho_6 \bar{Z}/\bar{A}$ is determined by the density
$\rho$, Eq.\ (\ref{rho(z)}) gives the density profile $\rho(z)$ as
a function of depth $z$, $z_0$ being a depth at which
the electrons become relativistic ($\rho_6 \sim 1$).   
One has $\rho \propto z^{3/2}$ in the layer of non-relativistic
degenerate electrons and $\rho \propto z^3$ in the deeper layers where
the degenerate electrons are relativistic. 
Equation (\ref{rho(z)}) demonstrates  self-similarity of the structure
of outer layers of neutron stars advertised in Section~\ref{subsub:basic_eqns}. Note that the
equation is
inaccurate in a thin outermost layer of the star where 
the electrons are non-degenerate and the ions are not fully ionized.
It is qualitatively correct to the bottom of the outer crust, but
becomes invalid in the inner crust where free neutrons appear
and contribute to the pressure. 

\paragraph{Mass as a function of $z$}

 Integrating Eq.\ (\ref{str-m})
from the surface to a given depth $z$ using our plane-parallel
approximation, one derives a simple expression for the
gravitational mass $\Delta M(z)=m(R)-m(r)$ contained in the
the surface layer of depth $z$ (e.g., \citealt{GPE83}),
\begin{equation}
    \frac{\Delta M(z)}{M}=\frac{4 \pi G P(z)}{g_\mathrm{s}^2}.
\label{eq:dM/M}
\end{equation}
Therefore, $\Delta M(z)/M$ is determined by the pressure at a
given depth. This is another indication of self-similarity. 
In contrast to \req{rho(z)}, this expression is 
valid for any model of the pressure. It is convenient to
introduce the parameter 
\beq
   \eta \equiv g_\mathrm{s14}^2 \frac{\Delta M}{M} \approx 
  \frac{P(z)}{1.193\times 10^{34}\mbox{ dyn cm}^{-2}},
\label{eq:eta(z)}  
\end{equation}
and use $\rho(z)$ instead of $z$. Also, 
one often uses the column depth from the surface,
\begin{equation}
    y=\Delta M/(4 \pi R^2).
\label{e:columndepth}
\end{equation}    

At high depths $z$, where the electrons are strongly degenerate,
the pressure can be approximately (within
several percent) represented by the pressure of 
the ideal Fermi gas of completely degenerate electrons.
In this approximation, one has
\begin{equation}
        \eta = 1.51 \times 10^{-11}
        \left\{\xr  \gammar \left( \frac{2}{3} \xr^2 - 1 \right)
        + \ln \left( \xr  + \gammar \right)
        \right\},
\label{e:dM/M(rho)}
\end{equation}
where $\gammar \equiv \sqrt{1 + \xr^2}$ is the electron Lorentz factor
at the Fermi surface. In the non-relativistic limit ($\xr\ll1$), the
expression in curly brackets turns into $8\xr^5/15$; in the opposite
limit ($\xr\gg1$), it tends to $2\xr^4/3$.

For example, we can consider so called \emph{canonical neutron star
model} with $M=1.4 M_\odot$, $R=10$ km  ($g_\mathrm{s14}$=2.43) and the
envelopes composed of the iron.  Degenerate electrons become
relativistic ($\rho \sim 10^6$ \gcc)  at $z_0 \sim 10$ m, $\Delta M \sim
5 \times 10^{-13}~M_\odot$ and  the column density $y \sim  10^{8}$ g
cm$^{-2}$.  The heat blanketing envelope with $\rhob=10^{10}$ \gcc\
would have  the depth $z_\mathrm{b}\approx 160$ m, the mass $\Delta
M_\mathrm{b}\approx 1.9 \times 10^{-7}$ $M_\odot$ and $y_\mathrm{b}
\approx 3 \times 10^{13}$ g cm$^{-2}$.  If we assumed $\rhob=10^{9}$
\gcc, we would have  $z_\mathrm{b}\approx 75$ m, $\Delta
M_\mathrm{b}\sim 8.8 \times 10^{-9}$ $M_\odot$ and $y_\mathrm{b}=1.4
\times 10^{12}$ g cm$^{-2}$. The bottom of the outer crust 
($\rho_\mathrm{drip} \approx 4.3 \times 10^{11}$ \gcc)  would be reached
at $z_\mathrm{drip}\approx 560$ m, $\Delta M_\mathrm{drip}  \approx 2.9
\times 10^{-5}$~$M_\odot$ and  $y_\mathrm{drip}=4.6 \times 10^{15}$ g
cm$^{-2}$.   The latter example is a rough estimate because, actually, 
iron cannot survive to the neutron drip.

Using self-similarity relations one can easily rescale these results
to other values of $M$ and $R$. 
For instance, one can take the same $M$ but
larger $R=12$ km ($g_\mathrm{s14}=1.59$). Since $\Delta M\propto M/g_\mathrm{s}^2$,
at $\rho_\mathrm{b}=10^{10}$ \gcc\ one has $\Delta M_\mathrm{b}\approx 4.4
\times 10^{-7}$ $M_\odot$. At this density the electron gas is ultarelativistic
and $z \propto z_0 \rho^{1/3} \propto 1/g_\mathrm{s}$. Then
$z_\mathrm{b}\approx 240$ m and $y_\mathrm{b} \approx 4.9 \times 10^{13}$ g~cm$^{-2}$. 

\section{Analytic models of non-magnetic envelopes}
\setcounter{equation}{0}
\label{therm-analyt}

Analytic models of blanketing envelopes have been developed by
\citet{uy80c,HernqApple,VP}. 
Below we present a similar
analysis following mainly \citet{VP}. Contrary to the problem of
density distribution in an envelope 
(Section~\ref{sec:basis}), which has an exact and simple analytic
solution (\ref{rho(z)}), the problem of temperature distribution is more
complicated and, strictly speaking, cannot be solved in a closed
analytic form. Accurate solutions can be obtained numerically as
discussed in the next sections. In the present section, we will not
try to be as accurate as possible, but propose a simplified analytic
treatment of the temperature distribution which clarifies the main
features of the problem. We will focus on non-magnetic 
spherically symmetric envelopes.
Strongly magnetized envelopes will be analyzed in Section~\ref{therm-magn-analyt}.

One can subdivide the heat blanket
into two parts, the outer layer, where the heat is mostly carried
by photons, and the deeper layer, where the electron transport
dominates. We will assume, for simplicity, that the electrons are
non-degenerate in the former and degenerate in the latter layers and we
will check this assumption.

\subsection{Radiative layer}
\label{sect-analytic-non-degen}

Our consideration of the non-degenerate layer of a neutron star is
very close to the classical theory of non-degenerate envelopes of
white dwarfs (e.g., \citealt{Schwarzschild58}). Combining
Eqs. (\ref{dP/dz}) and (\ref{therm-Tcrust}), we obtain
\beq
   \frac{\mathrm{d} T }{ \mathrm{d} P} = \frac{F_r }{ g_\mathrm{s} \kappa \rho},
\label{therm-dT/dP}
\end{equation}
where $F$ is the thermal flux density
(see Section~\ref{sub:evolution}), and $\kappa$ is
the radiative conductivity, which will be taken in the form
\beq
      \kappa_\mathrm{r} = \kappa_0 T^\beta /\rho^\alpha,
\label{therm-kappa_rad}
\end{equation}
with constant $\alpha$, $\beta$, and $\kappa_0$. This relation 
approximates radiative conduction with the opacity given by the
Kramers's formula, $K \propto \rho/T^{3.5}$,  for $\alpha=2$ and
$\beta=6.5$. In a fully ionized, non-relativistic and non-degenerate
plasma, composed of electrons and ions with relative atomic weight $A$ 
and mass
number $Z$, where the opacity is provided by the free-free transitions,
\bea
    K_\mathrm{r} &\approx&
    75\,\bar{g}_\mathrm{eff}\,(Z^3/A^2)\,
    \rho\,T_6^{-3.5}       \mbox{~cm$^2$~g$^{-1}$},
\label{K_0}
\\
     \kappa_\mathrm{r} &\approx& \, 4 \times 10^{12} \,
     \frac{ T_6^{6.5} A^2 }{ \rho^2 Z^3
     \bar{g}_\mathrm{eff} } \mbox{~erg~cm$^{-1}$~s$^{-1}$~K$^{-1}$}.
\label{therm-kappa_0} 
\eea
Here,
$\rho$ is measured in g cm$^{-3}$, $T_6=T/10^6$~K, 
and $\bar{g}_\mathrm{eff} \sim 1$ is
an effective Gaunt factor, a slowly varying function of plasma
parameters (e.g., \citealt{Schwarzschild58,Mihalas1978});
it has much in common to a Coulomb logarithm 
for electron-ion collisions. For a colder plasma composed of heavy elements, where
bound-free transitions dominate over free-free ones, the Kramers's
formula  remains approximately valid, but the thermal conductivity
$\kappa$ is about two orders of magnitude lower. We will not analyze
this case, but the reader can easily study it by taking formally
$\bar{g}_\mathrm{eff} \sim 10^2$. According to \citet{VP},  \req{K_0}
gives an
order-of-magnitude approximation (within $\approx0.5$ in
$\log\kappa$) to the realistic Opacity Library (\textsc{opal}) opacities for hydrogen at
$T_6\sim10^{-1}-10^{0.5}$ and $\rho\sim(10^{-2}-10^1)\,T_6^3~\gcc$,
if we formally put $\bar{g}_\mathrm{eff}\approx\rho^{-0.2}$
(where $\rho$ is again in $\gcc$). An analogous
order-of-magnitude approximation to the \textsc{opal} opacities for iron at
$T_6\sim1-10^{1.5}$ and $\rho\sim(10^{-4}-10^{-1})\,T_6^3~\gcc$ is
given by \req{K_0} with $\bar{g}_\mathrm{eff} \approx 70 \,
\rho^{-0.2}$. Note that corresponding approximations for
$\kappa_\mathrm{r}$ also belong to the class of functions
(\ref{therm-kappa_rad}), but with $\alpha=1.8$.

Since the plasma is fully ionized, the pressure is produced by ideal
gases of electrons and ions, $P=(1+Z) \rho\, k_\mathrm{B}T
 /(A m_\mathrm{u})$, where $m_\mathrm{u}$ is again the atomic mass unit. Combining this
expression with Eqs.\ (\ref{therm-dT/dP}) and
(\ref{therm-kappa_rad}), we obtain
\beq
   \frac{\mathrm{d} T }{ \mathrm{d} P}=
   \frac{ F_r }{ g_\mathrm{s} \kappa_0} \,
   \frac{ P^{\alpha -1} }{ T^{\alpha + \beta -1} } \,
   \left( \frac{ A m_\mathrm{u} }{ (Z+1) k_\mathrm{B}} \right)^{\alpha -1}.
\label{therm-dT/dP1}
\end{equation}
Now let us employ the zero-order boundary condition $P(0)=T(0)=0$ at
the surface $z=0$ (Section~\ref{sub:evolution}) and integrate
\req{therm-dT/dP1} within the star. We get
\beq
   T^\beta = \frac{\alpha + \beta }{  \alpha} \;
        \frac{ F_r }{ g_\mathrm{s} \kappa_0} \,
        \frac{(1+Z) k_\mathrm{B} }{ A m_\mathrm{u} } \, \rho^\alpha.
\label{therm-T-rho}
\end{equation}
Using Eq.~(\ref{therm-kappa_rad}) and setting $\alpha=2$ and
$\beta=6.5$, we have
\beq
    \kappa = \frac{\alpha + \beta }{ \alpha} \,
      \frac{F_r }{ g_\mathrm{s}} \, \frac{(1+Z)k_\mathrm{B} }{ A m_\mathrm{u}}
      \approx 2.0 \times 10^{14} \; \frac{ 1+Z }{ A} \,
      \frac{ T_\mathrm{s6}^4 }{ g_\mathrm{s14}}
      \quad \frac{\mbox{erg}}{\mbox{cm~s~K}}.
\label{therm-kappa=const}
\end{equation}
Therefore, $T(z)$ increases within the
non-degenerate layer in such a way that the thermal conductivity
remains constant. Combining this equation with the conduction
equation $F_r=\kappa \, \dd T / \dd z$, we immediately obtain the linear
growth of the temperature with depth $z$,
\beq
    T(z) = \frac{ F_r }{ \kappa} \, z \approx 2.84 \times 10^5 \, g_\mathrm{s14}\;
           \frac{ A }{ 1+Z} \, z_\mathrm{cm} \mbox{~K},
\label{therm-T(z)}
\end{equation}
where $z_\mathrm{cm}$ is the depth $z$ measured in centimeters. The
constant thermal conductivity and the linear growth of $T$ are
well-known features of non-degenerate stellar envelopes.

Inserting Eq.\ (\ref{therm-T(z)}) into \req{therm-T-rho} we obtain
the density profile in the non-degenerate envelope,
\beq
    \rho \approx 0.0024 \, \frac{ g_\mathrm{s14}^{3.75} }{ T_\mathrm{s6}^2 } \,
     \left( \frac{A} { 1+Z} \right)^{3.75}
    \left( \frac{A^2}{ Z^3 \bar{g}_\mathrm{eff}}
    \right)^{1/2} z_\mathrm{cm}^{3.25} \mbox{~g~cm}^{-3}.
\label{therm-rho(z)}
\end{equation}
Therefore, $\rho \propto z^{3.25}$ and $P \propto z^{4.25}$. The
density dependence of the temperature is thus
\beq
  T_6\approx(50\,\bar{g}_\mathrm{eff}\,q)^{2/13}\,(\rho Z/A)^{4/13},
\label{nd-solution}
\end{equation}
where
\beq
  q\equiv [Z\,(1+Z)/A]\,T_\mathrm{s6}^4/g_\mathrm{s14}.
\label{nd-solution-q}
\end{equation}

\paragraph{Radiative surface}

Now we can check the accuracy of the approximation
(\ref{therm-Eddington}) for the radiative surface. From Eqs.\
(\ref{K-kappa}), (\ref{therm-kappa_rad}), and (\ref{therm-T-rho}), we
see that $K\propto P^\gamma$, where
$\gamma=(3\alpha-\beta)/(\alpha+\beta)$. Substituting this expression
for $K$ in \req{tau(z)}, we obtain the relation $\tau=KP/[g(\gamma+1)]$.
At $\tau=2/3$ it reproduces \req{therm-Eddington} with the left-hand
side multiplied by $(\gamma+1)$. The latter factor is nearly 1, because
$\gamma$ is small. For instance, $\gamma=-1/17$ at $\alpha=2$ and
$\beta=6.5$.

Using \req{K_0} and the ideal gas EoS 
$P=(\rho/m_\mathrm{u})\,[(Z+1)/A]\,k_B T$, we
obtain
\beq
    \rho_\mathrm{s} \approx 0.1\, \frac{A}{ Z}\,
    \left(\frac{A\,g_\mathrm{s14} }{ Z\,(Z+1)\,
\bar{g}_\mathrm{eff}}\right)^{1/2} T_\mathrm{s6}^{5/4}~\gcc.
\label{surface}
\end{equation}
Substituting $\bar{g}_\mathrm{eff}\approx1$ for hydrogen and
$\bar{g}_\mathrm{eff}\approx200$ for iron, we obtain, respectively,
$\rho_\mathrm{s}\sim0.07\,\sqrt{g_\mathrm{s14}}\,T_\mathrm{s6}^{5/4}\gcc$ and
$\rho_\mathrm{s}\sim0.004\,\sqrt{g_\mathrm{s14}}\,T_\mathrm{s6}^{5/4}\gcc$.

\paragraph{Degeneracy onset}

The solution given by \req{nd-solution} can be extended to a depth
where the electrons become degenerate ($T \sim T_\mathrm{F}$,
Eq.\ (\ref{degenerateT})). Let us estimate this depth
from the condition $\kB T_\mathrm{F}= p_\mathrm{F}^2/2\mel$,
because the electrons are still non-relativistic. We will label
the quantities at this depth by the subscript ``F''. We obtain
\bea
   z_\mathrm{F} \simeq \frac{30 }{ g_\mathrm{s14}} \,
   \frac{1+Z}{ A}\,(\bar{g}_\mathrm{eff} q)^{2/7}\mbox{~cm},
&\quad&
   \rho_\mathrm{F} \simeq  150 \,\frac{A}{ Z}\,(\bar{g}_\mathrm{eff} q)^{3/7} ~\gcc,
\label{rhoF}\\
   T_\mathrm{F} \simeq 8.5 \times 10^6 
      \,(\bar{g}_\mathrm{eff} q)^{2/7}\mbox{~K},
&\quad&
   x_\mathrm{rF} \simeq 0.053 \,(\bar{g}_\mathrm{eff} q)^{1/7},
\label{therm-degenerate} 
\eea
where $q$ is defined by \req{nd-solution-q} and $x_\mathrm{rF}$ is the
electron relativistic parameter (\ref{eq:xr}) at $z=z_\mathrm{F}$. Even
for very high effective surface temperatures $T_\mathrm{s} \sim 10^7$~K,
we have $x_\mathrm{rF} \lesssim 1$, i.e., the electrons are indeed
non-relativistic at the degeneracy boundary. The thickness of the
non-degenerate surface layer in such a hot star reaches several meters.
With decreasing $T_\mathrm{s}$, the quantities $z_\mathrm{F}$,
$\rho_\mathrm{F}$ and $T_\mathrm{F}$ decrease, i.e., the degeneracy
boundary shifts to the stellar surface. In a middle-aged 
neutron star, the typical surface temperature is
$T_\mathrm{s} \sim 10^6$ K, and the depth $z_\mathrm{F}$ is several
decimeters, while in an old and cold star, with $T_\mathrm{F}
\sim 10^5$ K, $z_\mathrm{F}$ is a few centimeters only.

\subsection{Electron-conduction layer}
\label{sect-analytic-degen}

The electron conductivity has been reviewed,
for instance, by \citet{PPP2015}. 
In the case of non-degenerate electrons, the conductivity
can be found, e.g., by the method of \citet{Braginskii},
which yields 
\beq
     \kappa_\mathrm{e}^\mathrm{nd} 
\approx 5\times10^{10}\,(F_Z/\Lambda)\, Z^{-1} \, T_6^{5/2}
\mbox{~erg cm$^{-1}$~s$^{-1}$~K$^{-1}$},
\label{kappa-ND}
\eeq
where $F_Z$ is a slow function of $Z$: for example,
$F_{26}=1.34$ and $F_1=0.36$,
whereas the Coulomb logarithm $\Lambda$ is $\sim1$
near the onset of degeneracy
and logarithmically increases with decreasing density. 

In degenerate matter (at
$z>z_\mathrm{F}$), the electron thermal conductivity 
is mostly limited by electron-ion scattering.
For this conduction mechanism (e.g., \citealt{PPP2015}, and references
therein),
\beq
     \kappa_\mathrm{e} = \frac{ \pi k_\mathrm{B}^2 T \mel c^3 \xr^3
             }{
             12 Z e^4 \Lambda \gammar^2}
            \approx 2.3\times10^{15}\,\frac{T_6}{\Lambda Z}\,
            \frac{\xr^3}{ \gammar^2}
            \mbox{~erg~cm$^{-1}$~s$^{-1}$~K$^{-1}$},
\label{therm-kappa_ei}
\end{equation}
where $\xr$ is the relativity parameter  (\ref{eq:xr}) and 
$\gammar^2=1+\xr^2$. The Coulomb logarithm $\Lambda$ is
close to unity in the liquid Coulomb plasma
 ($\Lambda\sim1$ at $T>T_\mathrm{m}$) and decreases to small
values in the crystalline matter ($\Lambda \sim T/T_\mathrm{m}$ at 
$T<T_\mathrm{m}$; see \citealt{1999PBHY}).
Equation (\ref{therm-kappa_ei})
transforms  into (\ref{kappa-ND}) if the dimensionless Fermi momentum
$\xr$ is replaced by an appropriate thermal average,
$\xr \to \sqrt{\kB T/(\mel c^2)}$.

\paragraph{Sensitivity strip}

\citet{GPE83} performed extensive numerical tests which
revealed that the accurate knowledge of the thermal conductivity is
particularly important in a certain ``sensitivity strip'' in the
$(\rho,T)$ plane. The $T_\mathrm{b}/T_\mathrm{s}$ ratio
changes appreciably if $\kappa$ is modified, say, by a factor 2
within this strip, while comparable changes of  $\kappa$ outside the
strip would leave the ratio almost unaffected. 
The strip lies near the
transition zone between the electron conduction
and radiative conduction. It is explained by the fact that,
as we see from Eqs.~(\ref{therm-kappa_0}) and (\ref{therm-kappa_ei}),
$\kappa_\mathrm{r}$ decreases while $\kappa_\mathrm{e}$ increases
with increasing density at a constant temperature.
Hence their crossover region presents a bottleneck 
for the  heat leakage from
the stellar interior.

The ``turning'' line in the $(\rho,T)$ plane, where
$\kappa_\mathrm{r}=\kappa_\mathrm{e}$, is easily determined from
Eqs.~(\ref{therm-kappa_0}) and (\ref{therm-kappa_ei}),
\beq
   \rho\approx 12\,(A/Z)\,\bar{g}_\mathrm{eff}^{-1/3}\, T_6^{11/6}~\gcc,
\label{turn1}
\end{equation}
where we set $(\Lambda\gammar^2)^{1/3} \approx 1$, for an
estimate. Using \req{therm-T-rho}, we can explicitly relate the
temperature $T_\mathrm{t}$ and the relativity factor $x_\mathrm{rt}$ at
the point, where the radiative conduction turns to the electron one,
\beq
   T_\mathrm{t}\approx 2.3\times10^7\,\bar{g}_\mathrm{eff}^{2/17}\,
     q^{6/17}\mbox{~K},
\quad
   x_\mathrm{rt}\approx 0.157\,\bar{g}_\mathrm{eff}^{-2/51}
     q^{11/51}.
\label{turn}
\end{equation}
Actually there is a turning zone rather than the turning
point, where both thermal conductivities are equally important. In
addition, the extrapolation of \req{nd-solution} to the turning
point is, strictly speaking, not justified, because the electron gas
becomes degenerate, $x_\mathrm{rt}>x_\mathrm{rF}$, for typical parameters.
Nevertheless, since $x_\mathrm{rt}$ and $x_\mathrm{rF}$ are not very
different, the segment of the temperature profile, where our
assumptions are violated, is relatively small, so that \req{turn}
provides a reasonable approximation. This is confirmed by a direct
comparison with numerical results \citep{PCY97}, which reveals a
discrepancy of a few tens percent at $T\gtrsim10^{5.5}$~K.

\subsubsection{Electron conduction solution}
\label{subsub:econdsolve}

An analytic temperature profile in the degenerate layers of a
neutron star envelope was first calculated by \citet{uy80c}. The
solution was based on the electron conductivity in the form of 
\req{therm-kappa_ei}. The hydrostatic equilibrium of the degenerate
surface layers is determined by Eq.\ (\ref{rho(z)}).
Using Eqs.~(\ref{therm-Tcrust}) and
(\ref{therm-kappa_ei}), one obtains
\beq
    T \, \frac{\dd T }{ \dd \xr}=\frac{12}{\pi}\,
    \frac{ F Z^2  e^4 \Lambda }{ m_\mathrm{u} \kB^2 A c g_\mathrm{s} }\,
    \frac{\gammar }{ \xr^2 }
    =(1.56\times10^7\mbox{~K})^2\,
    \frac{Z^2\Lambda T_\mathrm{s6}^4}{ A g_\mathrm{s14}}\,
    \frac{\gammar }{ 2\xr^2 }.
\label{dT/dx}
\end{equation}
Treating $\Lambda$, $A$, and $Z$ as constants, we can integrate this
equation from $x_\mathrm{rt}$ inside the star and obtain
\beq
   T^2(z)  = T_\mathrm{t}^2+
      (1.56\times10^7\mbox{~K})^2\,
      \frac{Z^2 \Lambda T_\mathrm{s6}^4}{ A g_\mathrm{s14}}
      \left[f(\xr)-f(x_\mathrm{rt})\right],
\label{therm-T1(z)}
\end{equation}
where $f(x) \equiv \ln\big(x+\sqrt{1+x^2}\big)- \sqrt{1+ 1/x^2}$.

Equation (\ref{therm-T1(z)}) describes the thermal structure of the
degenerate envelope. It shows that the largest temperature growth inside
the degenerate envelope takes place at lowest densities after the
turning point, as stated in the discussion of the sensitivity strip.
This is because the thermal conductivity $\kappa$ increases with
growing density, making the temperature profile flatter. Taking 
the decrease of the Coulomb logarithm with the density growth
into account,
one can show that in the deep layers the temperature tends to some
constant value $T(z)=T_\mathrm{b}$ which we treat as the
temperature at the heat blanket bottom.

\subsection{Internal temperature versus surface temperature}
\label{sub:Tb-vs-Ts}

Let us use the above solution to evaluate $T_\mathrm{b}$. Typically
$x_\mathrm{rt} \ll 1$, but at the inner boundary $\xr=x_\mathrm{b}\gg1$.
Under these conditions \req{therm-T1(z)} gives 
\beq
   T_\mathrm{b}^2 \approx T_\mathrm{t}^2+
      (1.56\times10^7\mbox{~K})^2\,
      \frac{Z^2 \Lambda T_\mathrm{s6}^4}{ A g_\mathrm{s14}
       x_\mathrm{rt}}
        \big[1 + x_\mathrm{rt}\ln(2x_\mathrm{b})\big]\Lambda.
\eeq
The term in the square brackets slowly grows with increasing density,
whereas $\Lambda$ slowly decreases. For a rough estimate we neglect 
their product and, using \req{turn}, obtain
\beq
   T_\mathrm{b} = \left(T_\mathrm{t}^2+T_{\Delta}^2\right)^{1/2},
   \quad
   T_{\Delta}\approx 4 \times 10^7 \left( \frac{ Z^2\,T_\mathrm{s6}^4
   }{ A\, g_\mathrm{s14}}
   \right)^{20/51}\mbox{~K},
\label{therm-T1}
\end{equation}
where we have also neglected some other factors close to unity, such as
$[Z/(Z+1)]^{0.1}$ and $\bar{g}_\mathrm{eff}^{1/51}$. More accurate
analytic approximations for $T_\mathrm{b}$ are obtained by fitting
the results of numerical calculations; they are described below (see
Section~\ref{sect-Tb-Ts} and \ref{app:TsTb}).

We see that the internal temperature $T_\mathrm{b}$ is determined by
the two temperatures, $T_\mathrm{t}$ and $T_\Delta$; they describe the
thermal insulating properties of the radiation- and electron-conduction
layers, respectively. The temperature growth takes place
in the very surface layers of the neutron star. Were the stellar
interiors in thermal equilibrium, the internal temperature would
actually be equal to $T_\mathrm{b}$ [corrected due to gravitational redshift,
\req{eq:isothermal}] everywhere in the internal region.

We also see that, for a typical surface temperature $T_\mathrm{s} \sim
10^6$ K, $T_\Delta$ is larger than $T_\mathrm{t}$, i.e., the main
thermal insulation is produced by the layers of degenerate electrons. 
The second expression in
\req{therm-T1}, being applied to iron matter, gives $ T_\mathrm{b}
\approx 1.06 \times 10^8 \, (T_\mathrm{s6}^4 / g_\mathrm{s14})^{0.39}$ K.
This formula is wonderfully close to \req{therm-GPE}
below, which was obtained by \citet{GPE83} by fitting  
numerical $\Tb$ values.

However, the $T_\Delta/T_\mathrm{t}$ ratio decreases with decreasing
$T_\mathrm{s}$. Therefore, the thermal insulation of the non-degenerate
layer becomes more important for a colder neutron star.

\begin{figure}[t]
\centering 
\includegraphics[width=.8\textwidth]{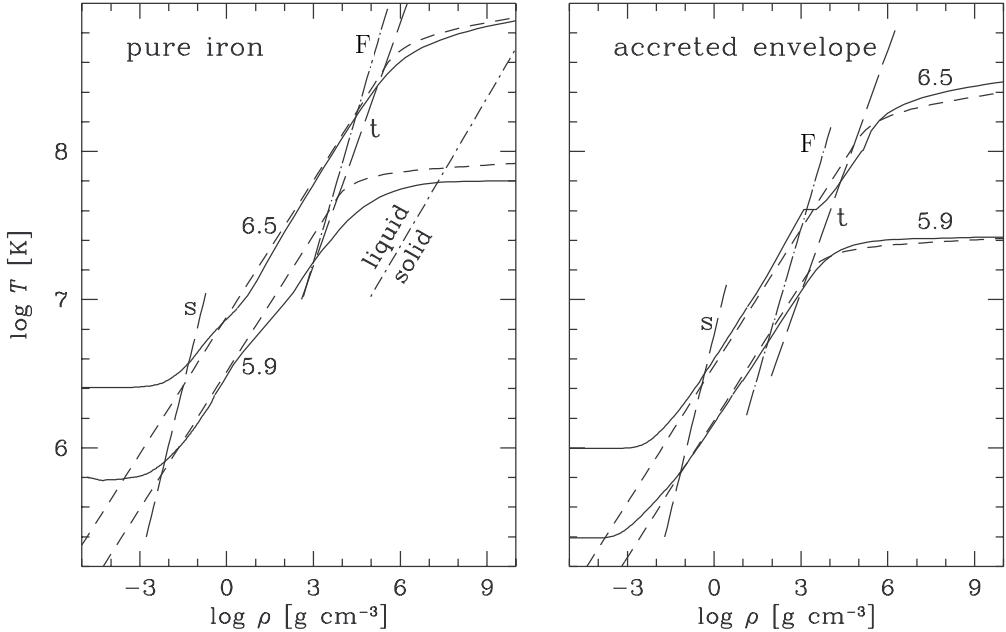}
\caption{Temperature profiles inside non-accreted
(left panel)
  and accreted (right panel) neutron-star
  envelopes at two effective temperatures,
  $\log T_\mathrm{s}\mbox{~[K]}=5.9$ and 6.5 (marked near the curves).
  Solid curves show numerical solution \protect\citep{SPYZ};
  dashed curves are analytic approximations (\protect\ref{nd-solution})
  and (\protect\ref{therm-T1(z)}).
  Straight lines marked ``s'', ``F'', and ``t'' give the values
  of $\rho$ and $T$ at which the various temperature profiles
  cross the radiative surface [\protect\req{surface}],
  the onset of electron degeneracy [\protect\req{therm-degenerate}],
  and the turning point  [\protect\req{turn}], respectively.
  The melting line of iron crystal is also shown.
  Adopted from \protect\citet{VP}.}
\label{fig-profiles}
\end{figure}

Fig. \ref{fig-profiles} illustrates the accuracy and limitations of the
analytic solution. The solid lines show
the temperature profiles for the canonical neutron star.
The profiles are obtained numerically as
described in Section~\ref{therm-env}.
The dashed curves depict the analytic approximations. The left panel
corresponds to an envelope composed of iron, while the right panel
refers to an accreted envelope (with the outermost shell composed
of hydrogen, and the deeper shells composed of heavier elements, He,
C, Fe, see Section~\ref{sect-therm-accret}). This shell structure is
responsible for the complex shape of the upper profile. The straight
lines show the points at which the temperature profiles at various
heat fluxes cross the radiative surface, the region of degeneracy
onset, the turning point $\kappa_\mathrm{e}=\kappa_\mathrm{r}$, and (on
the left panel) the bottom of the ocean (the ion crystallization point). 
The crystallization line is absent on the right panel, because freezing
of hydrogen and helium is suppressed by relatively large zero-point
vibrations of these light ions (e.g., \citealt{HPY07}, Section\ 2.3.4).

One can see that our analytic solutions correctly reproduce the thermal
structure of the envelope. Moreover, they provide a reasonable estimate
of the temperature at a given density. At low density
$\rho\lesssim\rho_\mathrm{s}$, the calculated profiles deviate from the
analytic approximation, because the atmosphere becomes optically thin
and isothermal.

Let us mention another important feature of our simplified solution.
Assuming $A/Z$ = constant and varying chemical composition of the
blanketing envelope, we approximately have $T_\mathrm{b} \propto
Z^{0.4}$. Thus, for a given $T_\mathrm{s}$, the stellar interior would
be cooler, if the star possessed an envelope made of light elements
\citep{CPY}.
This result is mostly explained by the $Z$-dependence of the thermal
conductivity of degenerate electrons \citep[e.g.,][]{1999PBHY}. This 
conductivity increases with lowering $Z$, which reduces the
temperature gradient and the internal temperature of the star.

\emph{How well should we know the thermal conductivity?} The answer
was given by \citet{GPE83} and \citet{HernqApple}. We can
come to the same conclusion by analyzing \req{therm-T1}. The
uncertainty of our knowledge of the radiative thermal conductivity
can be included into the Gaunt factor $\bar{g}_\mathrm{eff}$. We have
dropped this factor from \req{therm-T1} because it weakly affects
the temperature profiles, as can be seen from \req{turn}.
This is a consequence of the strong temperature dependence of the radiative
thermal conductivity, \req{therm-kappa_rad}. Even a large variation
of $\kappa_\mathrm{r}$ is easily compensated by a small variation of
$T$. The results are more sensitive to the thermal conductivity of
degenerate electrons in the sensitivity strip at not too strong
degeneracy. This
sensitivity strip coincides usually with the condition that the ions
constitute a strongly coupled liquid (the ion coupling parameter
$\Gamma$ ranges from $\sim 1$ to $\sim100$).

\subsubsection{Time resolution of heat blanketing models}
\label{sub:time_res}

Since the heat blanketing models are constructed
as quasi-stationary, the time 
resolution of the surface temperature 
variations, $\Ts(t)$, calculated by a cooling
code, is restricted. One cannot rely on the
variations which are shorter than the heat
diffusion time through a heat blanket.
 
A proper estimate of time over which a thermal perturbation
propagates in the heat blanketing envelope from a depth $z_2$ to a
depth $z_1$ is
\beq
   t_\mathrm{th} \sim \frac{1}{\sqrt{1 - r_\mathrm{g}/R}}
      \left[\frac12 \int_{z_1}^{z_2}\!\!
      \sqrt{\frac{C}{ \kappa}}\;
      \mathrm{d}z \right]^2,
\label{therm-relaxation-time}
\end{equation}
where $C$ is the heat capacity per unit volume and $\kappa$ is the
thermal conductivity.
The factor in front of the
square bracket describes gravitational dilatation of time
interval for a distant observer, and the integral itself should be taken
over a given non-perturbed thermal track. This estimate is a natural
consequence of the expression known from the classical studies of
thermal diffusion in stellar interiors \citep{hl69}. Let us estimate
$t_\mathrm{th}$ for a thermal wave propagating from a given depth
$z_2=z$ to the surface $z_1=0$. For this purpose we assume that the main
contribution into the integral comes from degenerate layers with the
electron thermal conductivity $\kappa$ given by \req{therm-kappa_ei} and
the heat capacity $C \approx 3 \kB n_\mathrm{i}$ appropriate to a
strongly coupled classical ion liquid or solid. In this case $C/\kappa
\approx 0.106 \, \Lambda \gammar^2/T_6$ s cm$^{-2}$. Let the
thermal wave be generated in the deep layer of the blanketing envelope
where the electrons are ultarelativistic ($\xr \gg 1$) and the
temperature is close to the internal temperature. The integration over
$z$ can be replaced by the integration over $\xr$ in the same
manner as in the derivation of \req{dT/dx}. Assuming further that the
main contribution into $t_\mathrm{th}$ comes from the layers, where
$T \approx T_\mathrm{b}$, $\xr\gg 1$, and the Coulomb logarithm
$\Lambda$ is constant, we obtain
\beq
   t_\mathrm{th} \sim
    \frac{ 2 \, \Lambda \, \xr^4
   }{ T_\mathrm{b6} \, \sqrt{1-r_\mathrm{g}/R} } \;
   \left( \frac{ Z }{ A g_\mathrm{s14} } \right)^2  \mbox{~days}.
\label{therm-tau-estimate}
\end{equation}
Taking the canonical neutron star model with 
 an iron heat blanket and setting $\Lambda=1$, we
arrive at $t_\mathrm{th} \sim 0.1 \, \xr^4/T_\mathrm{b6}$ 
days.  For example, if
$T_\mathrm{b} \approx 8.6 \times 10^7$ K
(appropriate for the surface temperature $T_\mathrm{s} = 10^6$ K), 
then a
thermal wave generated at $\rho=10^8$ g cm$^{-3}$ will travel to the
surface in $t_\mathrm{th} \sim 12$ hr, while a wave generated at
$\rho=10^{10}$ g cm$^{-3}$ will travel in $t_\mathrm{th} \sim 8$
months (also see Section~\ref{ssec:4:HeatBlanketsModel}).
The bottom of the heat-blanketing envelope is usually taken at
$\rho \sim 10^{10}$ g cm$^{-3}$, and the envelope solution derived
in the stationary approximation is implanted in the codes which
simulate neutron-star cooling (Section\ \ref{therm-blanket}). One
should not trust surface temperature variations over time scales of a few months
or shorter obtained using these cooling codes.

In a cold neutron star, the relaxation time can be determined by the
scattering of electrons off impurities rather than by the electron-ion
scattering (see, e.g., Appendix~A.4 of \citealt{PPP2015}). Numerical
calculations of $t_\mathrm{th}$ in the neutron star crusts were
performed, for instance, by
\citet{bbr98,Rutledge00,ur01,2009BC,2013PR,Yakovlev_21}
for the problem of thermal relaxation of transiently accreting neutron
stars in low-mass X-ray binaries. Generation of thermal disturbances in
the inner  neutron star crust and their emergence to the surface was
studied also for glitching neutron stars (e.g., \citealt{Hirano97}).

\subsection{Heat blankets of white dwarfs}
\label{sub:WD-HB}

White dwarf stars are ``close relatives'' of neutron stars.  They
consist of a massive and bulky core of degenerate electrons surrounded 
by a light and  relatively thin non-degenerate envelope (e.g.,
\citealt{ST83}). White dwarf masses are  comparable with neutron star
ones but white dwarf radii  are about three orders of magnitude larger.

White dwarfs, like neutron stars, possess heat blanketing envelopes
which keep their interiors sufficiently warm for a long time, comparable
with cosmological time-scales. Heat blankets of white dwarfs and neutron
stars are described by nearly the same physics, although the surface
gravity of white dwarfs is  smaller by about six orders of magnitude and
the composition of heat blankets may be different. Approximate  analytic
consideration of neutron star blankets in Section~\ref{therm-analyt} is
equally applicable to white dwarf blankets.

Analytic description of white dwarf thermal structure 
was developed in a seminal paper by \cite{1952Mestel}
(nicely summarized by \citealt{1971vanHorn}).
According to \citet{1952Mestel}, the white dwarf heat blanket essentially
coincides with the non-degenerate envelope.  
It was believed that high thermal conductivity of degenerate electrons 
should make the white dwarf
core isothermal. In our notations, Mestel's version of Eq.\  (\ref{therm-T1})
is $\Tb=T_\mathrm{t}$ (neglecting the contribution $T_\Delta$ of
degenerate electrons).  Note that, according to
our Eq.\ (\ref{turn}), $T_\mathrm{t} \propto \Ts^{24/17}$ while
Mestel obtained $\Tb \propto \Ts^{8/7}$. The difference in power-law indices 
is insignificant and stems from the fact that we estimate $T_\mathrm{t}$
at the turning line [\req{turn1}], whereas Mestel did so at the degeneracy
line [\req{therm-degenerate}]. 
  
Thus the Mestel's formula underestimates $\Tb$ for a given $\Ts$, and
the underestimate can be substantial. Anyway, people do not like
analytic formulas nowadays, and use computers instead. As a rule, the
white dwarf evolution is computed numerically (e.g.,
\citealt{WD92,Althaus_10}, and references therein) throughout entire
stars, without separate treatment of heat blankets. Nevertheless, 
analytic formulas are useful for insight and for benchmarking numerical
calculations. As will be seen in the next section, the relative
importance of thermal insulation of degenerate layers in a cooling  
star becomes lower and the turning point shifts to the degeneracy line.
This effect is more pronounced in cooling white dwarfs than in cooling
neutron stars. Therefore, as a white dwarf cools down, the Mestel's
approximation becomes more accurate.

\section{Basic non-magnetic heat  blanketing envelopes}
\setcounter{equation}{0}
\label{therm-env}

Now we turn to accurate calculations of the structure of non-magnetic heat
blankets of neutron stars. Magnetic envelopes will be analyzed in
Section~\ref{therm-magn-env}. 

\subsection{Historical remarks}
\label{sub:history}

Calculations of $T_\mathrm{b}$--$T_\mathrm{s}$ relations are being done
since the beginning of cooling simulations of neutron stars. Initially,
these relations were rather approximate, because of large theoretical
uncertainties of EoS and thermal conductivity in heat blankets.  The
first solid reliable relation was obtained in a classical paper by
\citet{GPE83}, who carried out a comprehensive study of blanketing
envelopes composed of iron using the best physical input available at
that time. These  authors considered the range of surface temperatures
$5.25 \leq \log \, T_\mathrm{s}\,\mbox{[K]} \leq 6.75$ and fitted their
numerical results by a remarkably simple formula,
\beq
     T_\mathrm{b} = 1.288 \times 10^8 \,
     (T_\mathrm{s6}^4/g_\mathrm{s14})^{0.455} \mbox{~~K}.
\label{therm-GPE}
\end{equation}
A simplified derivation of a similar expression was given in
 Section~\ref{sub:Tb-vs-Ts}.

Equation (\ref{therm-GPE}) has been used in numerous calculations.
It appears to be sufficiently accurate for not too cold
and not too hot iron blankets.

At the next step the problem was
reconsidered by PCY97 \citep{PCY97}, who
extended the results of \citet{GPE83} in two respects. First,
they studied the blanketing envelopes composed not only of iron but
also of lighter elements. Second, advanced
theoretical data on EoS and thermal conductivity implemented by
\citet{PCY97} allowed them to study colder neutron stars, with
$T_\mathrm{s}$ down to 50\,000~K. 
\citet{PY01} studied $T_\mathrm{s}-T_\mathrm{b}$ relations for magnetic
envelopes composed of iron. They depend on the
strength of the field $B$ and on its inclination to the surface.
\citet{PYCG03} obtained analogous
relations for the accreted envelopes and for a different value of
$\rho_\mathrm{b}$. 
We describe those results below.

\subsection{Physics input}
\label{sect-physinput}

PCY97 studied the blanketing envelopes composed, from surface to bottom, 
of hydrogen, helium ($^4$He), carbon ($^{12}$C),
and iron ($^{56}$Fe) shells (stratified onion-like structure). At any
given density the plasma contains ions of one chemical element that
can be in different ionization stages. The uncertainties in the composition have
been discussed briefly in Section~\ref{subsub:chemicalcompos}.
More details about different shells can be found in Section~\ref{sect-therm-accret}.

The EoSs of heat blankets are described, e.g., in \citet{HPY07} (Chapter
2). In the high-density domain (strongly degenerate
electrons,  almost full pressure ionization), PCY97 used an EoS of the
fully ionized electron-ion plasma. In the low-density domain (nearly
ideal plasma that can be partially ionized) one can employ the
\textsc{opal} \citep{OPAL} or another tabulated EoS. The intermediate
density domain (partially ionized, non-ideal plasma) is most
complicated. In this case, PCY97 used numerical tables of \citet{SCVH}
for H and He and an interpolation over the gap between the \textsc{opal}
tables and the domain of full ionization for the iron envelopes.

The electron heat conduction for partially ionized
plasmas was treated in the mean ion
approximation, using the formulas derived for fully ionized
degenerate plasmas. The effective ion
charge number $Z_\mathrm{eff}$
can be taken from
tables, whenever available. Otherwise PCY97 used an
interpolation procedure. The radiative thermal conductivity was
taken from the data of
\citet{OPAL}.

\subsection{Iron blanketing envelopes}
\label{sub:iron-blanket}

\begin{figure*}[t]
\centering
\includegraphics[width=\textwidth]{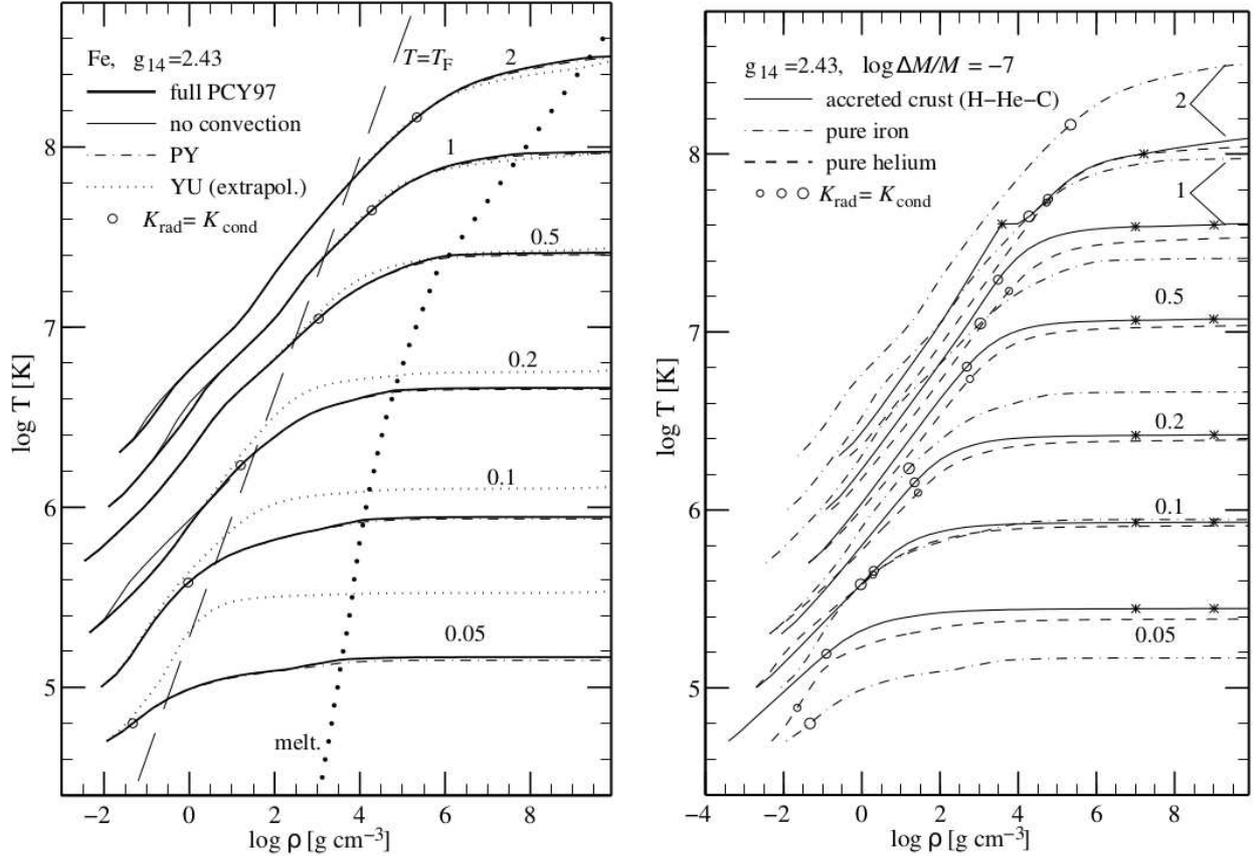}
\caption{\textit{Left panel}: Temperature profiles in an iron blanketing
envelope (thick solid lines) compared with the three approximations:
(i)  by extrapolating simplified $\kappa_\mathrm{e}$  of
\cite{YakovlevUrpin80} (YU) to $T > T_\mathrm{F}$;  (ii) by using
simplified $\kappa_\mathrm{e}$ of \cite{PY96} (PY); and (iii) neglecting
convection (thin solid lines). The curves are labeled by the values of
$T_\mathrm{s6}$. Circles show the points, where the radiative opacity
equals the conductive one; thick dots show the melting curve; long
dashes display the degeneracy line, $T = T_\mathrm{F}$. \textit{Right
panel}: Temperature profiles in a fully accreted envelope
(Section~\ref{sect-therm-accret}; solid lines) are compared to those in
the envelopes composed of pure iron (dot-dashed lines) and of pure
helium (dashed lines). The curves are labeled by the values of
$T_\mathrm{s6}$. The circles are turning points which separate the
regions of radiative and electron conduction; asterisks indicate the
H/He (lower $\rho$) and He/C (higher $\rho$) interfaces. (After PCY97.
See text for details.)
}
\label{fig-therm-pr}
\end{figure*}

%
\begin{figure*}[t]
\centering
\includegraphics[width=\textwidth]{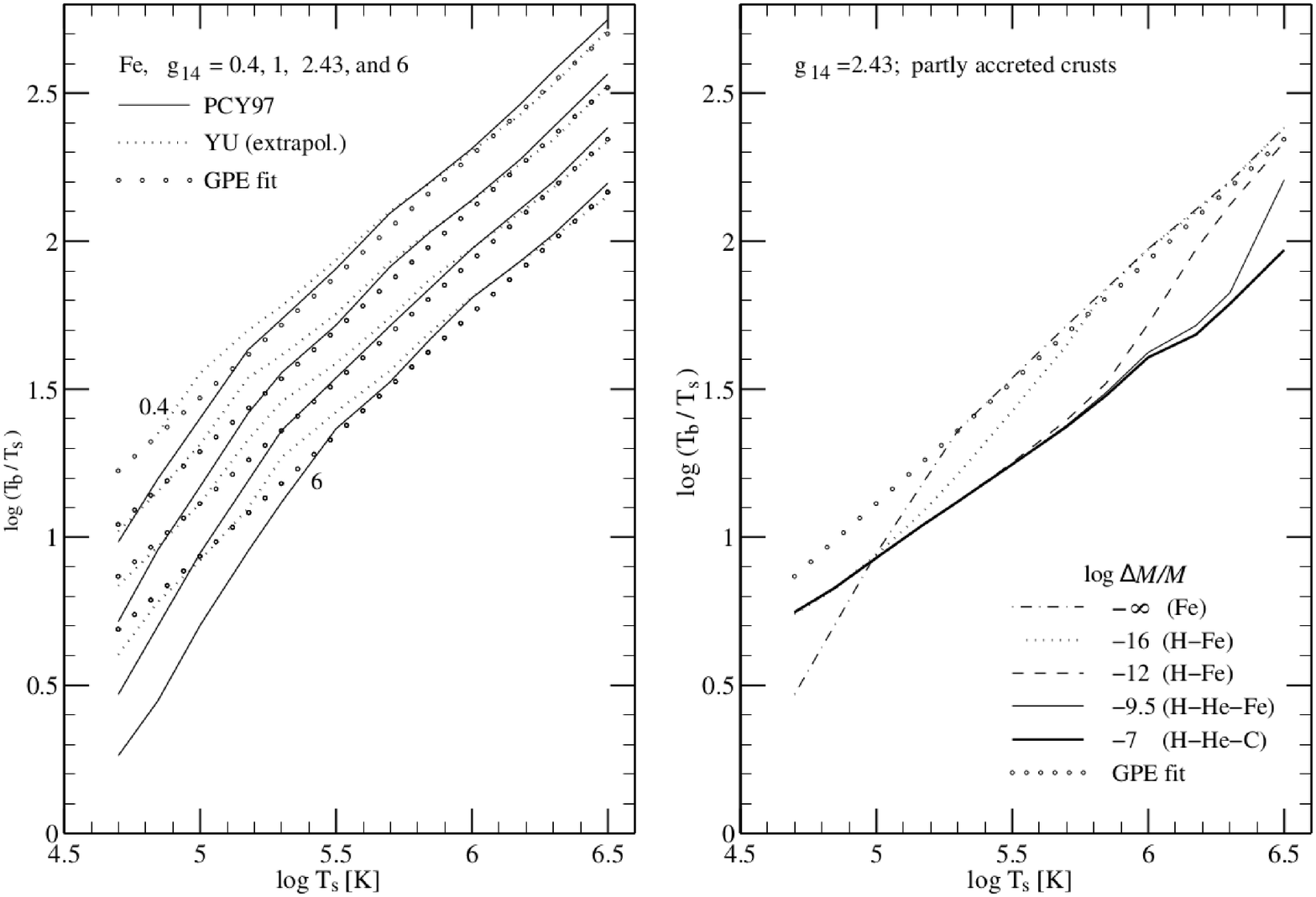}
\caption{Temperature ratio $T_\mathrm{b}/T_\mathrm{s}$ as
function of the effective surface temperature $\Ts$, compared with
the approximation of \citet{GPE83}, Eq.~(\ref{therm-GPE}), marked
as GPE.
\textit{Left panel}: Iron
envelopes with different surface gravities. 
\textit{Right panel}: Envelopes with different amounts of
accreted matter. 
 (After PCY97.
 See text for details.)
} 
\label{fig-therm-link}
\end{figure*}

The thermal structure of the envelope is studied by integrating
Eq.~(\ref{therm-Tcrust}) within the envelope. Fig.~\ref{fig-therm-pr} shows the dependence of temperature on density 
in the envelope at various $T_\mathrm{s}$. The integration is started
at the surface density $\rho_\mathrm{s}$, determined by the Eddington
boundary condition (\ref{therm-Eddington}).
In the left panel, the envelope is assumed to be composed of iron.
The integration is terminated at $\rho_\mathrm{b}=10^{10}$
g cm$^{-3}$. The value $g_\mathrm{s14}=2.43$ chosen in
Fig.~\ref{fig-therm-pr} corresponds to the canonical neutron star
model. Solid curves are calculated
using the physics input described above. Circles on the curves are
the turning points ($\kappa_\mathrm{r}=\kappa_\mathrm{e}$). Also shown are the
electron degeneracy curve and the melting curve.

In a wide range of $T_\mathrm{s}$, the outermost layers
can be convective
(see the left panel of Fig.~\ref{fig-therm-pr}). In
these layers, the energy is transported by convection rather than by
heat conduction. The convective energy flux is described
in the adiabatic approximation (see Section~\ref{sub:evolution}).

In order to check the effect of convection, calculations neglecting
 convection were performed. This extreme case is opposite to the
adiabatic one. In this approximation, one obtains
slightly higher temperatures inside the
convective part of the atmosphere
(the left panel of Fig.~\ref{fig-therm-pr}). The atmospheric temperature
profiles were also derived by \citet{Z96a} by numerically
solving the radiative transfer equation at moderate optical depths
and describing the convection using the mixing-length theory; they
lie between the two extremes mentioned above. In deeper layers, the
two extreme profiles tend to merge, because the thermal
conductivity $\kappa$ in \req{therm-Tcrust} increases inside
the envelope, thus reducing the temperature gradient at higher $\rho$. The
thermal structure of the blanketing envelope at $\rho\gtrsim 10$ \gcc\ is almost unaffected by convection.

The dotted and dot-dashed lines in the left panel of
Fig.~\ref{fig-therm-pr} show
temperature profiles calculated using simplified formulas for 
$\kappa_\mathrm{e}$. The dotted lines are obtained
with simplified expressions for 
$\kappa_\mathrm{e}$ derived by \citet{uy80a} and \citet{YakovlevUrpin80} for
strongly degenerate and fully ionized plasma; the expressions were
extrapolated into the domain of weak degeneracy and partial
ionization. It turns out that in a cold enough envelope the thermal
conductivity of non-degenerate or partly degenerate electrons
becomes important. A comparison with the tabular data of
\citet{HubbardLampe} reveals that a straightforward
extrapolation of the \citet{uy80a} formulae from their validity
domain (fully ionized, degenerate plasma) to the case of
non-degenerate matter may underestimate  
$\kappa_\mathrm{e}$ by orders of magnitude. As seen from
Fig.~\ref{fig-therm-pr}, this would significantly
overestimate the internal temperature at $T_\mathrm{s} \lesssim 2
\times 10^5$~K.

The dot-dashed profiles in the left panel of
Fig.~\ref{fig-therm-pr} were obtained
using the simplified thermal conductivity code \citep{PY96}, which
neglects contribution from electron-electron collisions and employs less accurate
Coulomb logarithms, but includes averaging of the effective
relaxation times with the electron Fermi-Dirac distribution at
partial electron degeneracy. The contribution 
of electron-electron collisions has been reconsidered
later \citep{2006Shternin} but in any case it seems
to be not very important for the conditions assumed 
in Fig.~\ref{fig-therm-pr}. The dot-dashed lines 
almost coincide
with the solid ones, indicating again that the temperature profiles
are most sensitive to the thermal conductivity of degenerate
electrons in Coulomb liquid of ions at $T_\mathrm{s} \gtrsim 10^5$ K
and to the thermal conductivity of mildly degenerate electrons at
lower $T_\mathrm{s}$.

This effect is also shown in Fig.~\ref{fig-therm-link}, which displays
$\log(T_\mathrm{b}/T_\mathrm{s})$ as a function of $\log T_\mathrm{s}$.
The left panel shows the case of iron envelopes. In this case, if $\log
T_\mathrm{s}\,[\mbox{K}]>5.5$, then the simple fit of \citet{GPE83},
Eq.~(\ref{therm-GPE}),  is fairly accurate. At lower $T_\mathrm{s}$ its
accuracy becomes worse. The scaling (self-similarity) relation,  $
T_\mathrm{b}=T_\mathrm{b}(T_\mathrm{s}^4/g_\mathrm{s})$, holds well in
the entire temperature--gravity range presented in this figure, except
for the lowest $T_\mathrm{b}$ and $T_\mathrm{s}$. In the last case, 
especially at high $Z$, radiative opacities are affected by bound-bound
transitions and strong plasma coupling effects. Therefore, they do not
obey the simple power law (\ref{therm-kappa_rad}) anymore. An
appropriate fit to the numerical results is given in
Section~\ref{sect-Tb-Ts}.

\subsection{Accreted envelopes}
\label{sect-therm-accret}

Here we describe PCY97 blanketing envelopes containing shells of light
elements (H, He, C; \citealt{PYCG03} supplemented this sequence by an
oxygen layer) and  possibly the iron shell at the bottom.  The iron
shell models the non-accreted part of the outer crust (which consists of
iron-group isotopes in its ground state). The light elements represent
the accreted matter and the products of its nuclear burning. The
interfaces between the shells of different light elements are placed at
the approximate limits of their stability against the burning. The
interface between the light elements and Fe is determined by the total
amount of the accreted matter. PCY97 called them  accreted envelopes;
the envelopes composed solely of the light elements were called
\emph{fully accreted}.  As outlined in Section~\ref{sub:HBmatter}, the
parameters of such shells are not free. In particular, lighter ions are
closer to the surface, owing to gravitational stratification
\citep{AI80}.  It is also important that lighter elements transform into
heavier ones at high enough temperatures  (via thermonuclear reactions)
and/or densities (via pycnonuclear reactions).

In heat blankets of different compositions,  hydrogen may be viewed    
as accreted, helium either as accreted or a product  of hydrogen
burning, and carbon as a result of nuclear burning. Iron may represent
either a primordial  composition of the stellar surface layers or a
final product of nuclear transformations of light   elements. In test
runs, the boundaries between the shells varied within wide limits.  In
final runs, the boundaries were varied within much more restricted
limits consistent with the models of nuclear burning existed by that
time. In their analysis, the authors took into account the results by
\citet{Iben,AI80,Pacz83,Ergma86,Miralda,Blaes92,Schramm92,Yak94}.  If
the temperature within a hydrogen, helium, or carbon shell exceeded a
certain limit, the nuclei within a given shell were replaced by heavier
ones (e.g., H$\to$He, etc.) reflecting thermonuclear burning.  Roughly,
it was assumed that hydrogen can survive at  $T \lesssim 4 \times 10^7$
K and/or $\rho \lesssim 10^7$ \gcc; helium -- at $T \lesssim 10^8$ K
and/or $\rho \lesssim 10^9$ \gcc, while carbon at $T \gtrsim 10^9$ K
and/or $\rho \lesssim 10^{10}$ \gcc. \citet{PC12} developed a more
accurate treatment of limiting boundaries between carbon, oxygen, and
iron-group substrate in the neutron-star envelopes. The positions of
other boundaries have also been updated (see
Section~\ref{subsub:chemicalcompos}). However, it was checked that
possible variations of these limiting boundary positions did not affect 
noticeably the resulting $\Ts-\Tb$ relations.

The right panel of Fig.~\ref{fig-therm-pr} displays the thermal
structure of a fully accreted envelope, where the accreted matter of
mass $\Delta M \sim 10^{-7} M$ extends to $\rho \simeq \rho_\mathrm{b}$ 
in a neutron star with $M=1.4\,M_\odot$ and $R=10$ km ($g_\mathrm{s} =
2.43 \times 10^{14}$ cm s$^{-2}$). The outer, intermediate, and inner
shells of this envelope, separated by asterisks, are composed of H, He,
and C, respectively.

One can observe significant differences from the iron envelope; they are 
explained below. For a not too cold neutron star ($T_\mathrm{s} \gtrsim
10^5$~K), the main temperature gradient occurs in a layer of
degenerate electron gas with ions in the liquid state. The thermal
conduction in this layer is mostly provided by the electrons, being
limited by the electron-ion scattering. The heavier the element, the
smaller the thermal conductivity, and the steeper is the temperature
growth inside the star. With decreasing $T_\mathrm{s}$, however, the
width of the heat-blanketing degenerate layer becomes smaller, and
the effect is less pronounced. In a cooler neutron star ($T_\mathrm{s}
\lesssim 10^5$~K), the main temperature gradient shifts into the
neutron star atmosphere, to the optical depths $\sim 1$. For heavier
elements, the atmospheric layers are denser at the same $T_\mathrm{s}$.
Then the internal temperature gradient is
weaker and the temperature grows slower inside the star. 
The effective surface temperature that
separates these two regimes is almost independent of the surface
gravity (see Fig.~3 of PCY97).

The right panel of Fig.~\ref{fig-therm-link} shows $\log(\Tb/\Ts)$ as a
function of $\log \Ts$ for various masses $\Delta M$ of H + He. The
dot-dashed line represents $\log(T_\mathrm{b}/T_\mathrm{s})$ for a
non-accreted (Fe) envelope from the left panel. Other lines are for
different  compositions at various $\Delta M$.  The effect of $\Delta M$
is seen to be quite pronounced.  Even a thin hydrogen or helium shell of
mass $\Delta M=10^{-16}M$, which extends only to $\rho \sim 10^3$ g
cm$^{-3}$, strongly modifies the $T_\mathrm{b}$--$T_\mathrm{s}$
relation.

\begin{figure}
\centering 
\includegraphics[width=.6\textwidth]{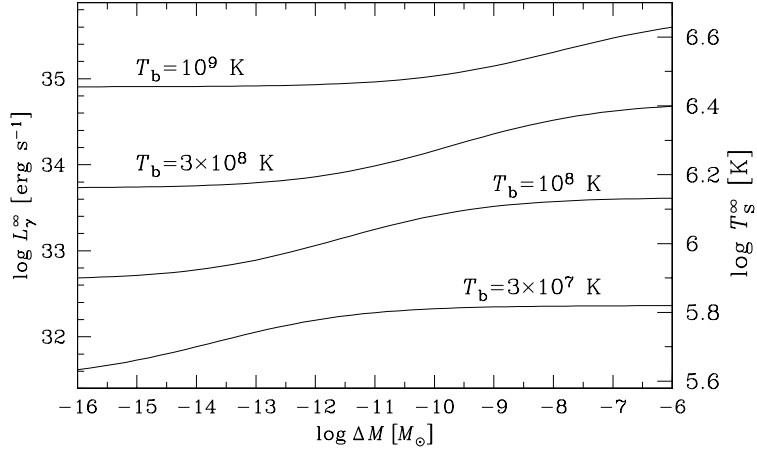} 
\caption{Photon surface luminosity (redshifted as
detected by a distant observer, left vertical axis) or redshifted
effective surface temperature (right vertical axis) of the canonical
neutron star model
for four values of the ``internal'' temperature
$T_\mathrm{b}$ (at $\rho_\mathrm{b}=10^{10}$ g cm$^{-3}$) versus the mass
 $\Delta M$ of H+He; after \citet{ygkp02}. }
\label{fig-therm-tedm}
\end{figure}

According to PCY97, the $T_\mathrm{s}$--$T_\mathrm{b}$ relation is
mostly determined by the total mass $\Delta M$ of H and He, contained in
the blanketing envelope, being rather insensitive to the boundary
density between the H and He shells and to a possible presence of the
carbon shell beneath the helium shell.

Fig.~\ref{fig-therm-tedm} shows the photon luminosity of the canonical
neutron star versus $\Delta M$ for four values of $T_\mathrm{b}=3\times
10^7$, $10^8$, $3\times 10^8$, and $10^9$ K. If the heat blanket is
fully accreted, the luminosity can increase by about one order of
magnitude, i.e., the surface temperature $T_\mathrm{s}$ can increase by
a factor of 2.

\subsection{Relation between internal and surface temperatures}
\label{sect-Tb-Ts}

PCY97 constructed a fitting formula for $\Ts$ as a function of
$T_\mathrm{b9}=T_\mathrm{b}/10^9$~K, $g_\mathrm{s14}$, and parameter
$\eta$, related to the accreted mass $\Delta M$ of light chemical
elements (H and He) by Eq.~(\ref{eq:eta(z)}). The fit was based on the
calculations of $\Ts-\Tb$ relations at $4.7 \leq \log \Ts \leq 6.5$,
$0.4 \leq g_\mathrm{s14} \leq 6$, and $0\leq\eta\leq10^{-7}$. The
boundaries between H, He and C  shells were varied in test runs but
fixed in the bulk of computations (unless the shells are not replaced by
Fe). For fixed $\Ts$ and $g_\mathrm{s14}$, the only physical parameter
that has been varied is $\rho_\mathrm{Fe}$, the upper boundary of the
iron shell; $\rho_\mathrm{Fe}\leq \rho_\mathrm{b}=10^{10}$ $\gcc$. 
According to PCY97, heat insulating properties of carbon and iron are
similar, so that $\rho_\mathrm{Fe}$  can be replaced (within a
reasonable accuracy) by $\rho_\mathrm{C}$, the upper boundary of the
carbon shell.  Recall that PCY97 found that insulating properties of H
and He plasmas are also similar. Therefore, the fit, albeit constructed
for H/He/C/Fe structure, can be used for other positions of H/He and
C/Fe interfaces with a similar (albeit somewhat lower) accuracy.

For a purely iron envelope, a crude estimate (with an error $\sim$
30\%) yields
\beq
   T_\mathrm{s6}=T_{\ast}\equiv
   \left(7\,T_\mathrm{b9}\sqrt{g_\mathrm{s14}}\right)^{1/2}.
\label{T*}
\end{equation}
According to \req{LrTs},
\beq
   F_\mathrm{20} \approx 0.567\, T_\mathrm{s6}^4,
\label{F-Ts}
\eeq
where $F_\mathrm{20}\equiv F_r/10^{20}$ erg cm$^{-2}$ s$^{-1}$.
Therefore, in spherical symmetry,
\beq
   L_r\approx 7.126\,(R/10\mbox{~km})^2\,T_\mathrm{s6}^4
   \times10^{32}\mbox{~erg~s}^{-1}.
\eeq

Let us define $\zeta \equiv T_\mathrm{b9}-(T_\ast /10^3)$. 
Then a refinement of the fit (\ref{T*}) by PCY97 reads
\beq
   F_\mathrm{20}^\mathrm{(Fe)}=0.567 g_\mathrm{s14}
    \, \left[ (7\zeta)^{2.25}+(\zeta/3)^{1.25}
   \right] .
\label{therm-t0}
\end{equation}
The typical fit error of $F_\mathrm{20}^\mathrm{(Fe)}$ is about 8\%, with
maximum 18\%, over the $T_\mathrm{s}-g_\mathrm{s}$ domain indicated
above.

For a fully accreted envelope, PCY97 had
\beq
   F_\mathrm{20}^\mathrm{(a0)}=0.567g_\mathrm{s14}
    \, (18.1\,T_\mathrm{b9})^{2.42},
\label{therm-t1}
\end{equation}
which is valid at not too high internal temperature, $T_\mathrm{b}
\lesssim10^8$~K.

Finally, for the partially accreted envelopes at any temperatures
within the indicated range, an interpolation formula of PCE97 was
\beq
   F_\mathrm{20}=\frac{a\,F_\mathrm{20}^\mathrm{(Fe)}
   +F_\mathrm{20}^\mathrm{(a0)}
   }{ a+1},
\label{therm-ttfit}
\end{equation}
where
\beq
   a=\left[1.2+(5.3\times 10^{-6}/\eta)^{0.38}\right]\,T_\mathrm{b9}^{5/3}.
\end{equation}
The typical fit error of Eq.\,(\ref{therm-ttfit}) for $F_\mathrm{20}$ is
about 12\%, with maximum $\sim20$\%, for all possible values of $\eta$
and any values of $g_\mathrm{s}$ and $T_\mathrm{s}$ within the indicated
ranges. For $\eta \gtrsim 10^{-7}$, the shell of light elements (H and
He) would formally extend beyond the heat blanketing envelope into the
zone where light elements cannot survive because of pycnonuclear
burning. In such cases, the actual mass of light elements will be lower
than the mass corresponding to the formal parameter $\eta$. However,
\req{therm-ttfit} remains valid even if  $\Delta M$ is formally
overestimated.

The dependence (\ref{therm-t0}) is realized not only at sufficiently low
accreted mass ($\eta\to0$), but also at sufficiently high
$T_\mathrm{b}$. The latter result reflects the fact that at high
$T_\mathrm{s}$ the thermal insulation is mostly produced by the electron
conductivity in the deep and hot layers of the envelope (within the
sensitivity strip), in which light elements (H, He) burn into heavier
ones. On the other hand, even at very low accreted mass, $\Delta M/M\sim
10^{-16}$, the approximation of fully accreted crust is still good
enough at sufficiently low temperature, because in this case the thermal
insulation is actually provided by the outermost accreted surface
layers.

\citet{PYCG03} noticed that thermonuclear burning of helium into heavier
elements leads to violation of the isothermality of the fully accreted
envelope at $\rho\gtrsim 10^{10}$ \gcc, if $\Tb\gtrsim10^8$~K. They
shifted $\rho_\mathrm{b}$ to the neutron-drip density, where the
isothermality is guaranteed, and obtained the $\Ts-\Tb$ relation for
this case.  They also extended the fit to higher temperatures. The fit
(\ref{therm-t0}) for the iron envelope remains valid with this increase
of $\rho_\mathrm{b}$ (within the indicated accuracy). For a fully
accreted envelope, the improved fit reads
\beq
   F_\mathrm{20}^\mathrm{(a)}=
   \frac{
   \left\{
     0.447+0.075\,\log(\Tb/\mbox{K})/[1+(6.2\,T_\mathrm{b9})^4]
     \right\} F_\mathrm{20}^\mathrm{(a0)} 
     + 3.2\,T_\mathrm{b9}^{1.67} F_\mathrm{20}^\mathrm{(Fe)}}{
     1 + 3.2\,T_\mathrm{b9}^{1.67}},
\end{equation}
where $F_\mathrm{20}^\mathrm{(a0)}$ is given by \req{therm-t1} and 
$F_\mathrm{20}^\mathrm{(Fe)}$ by \req{therm-t0}. The correction factor
accounts for thermonuclear burning of He at $T\gtrsim10^8$~K and for the
non-isothermality at high densities and temperatures. For a partially
accreted envelope, \citet{PYCG03} replaced the interpolation
(\ref{therm-ttfit}) by
\bea
   F_\mathrm{20}&=& a\,F_\mathrm{20}^\mathrm{(a)} +
   (1-a)\,F_\mathrm{20}^\mathrm{(Fe)},
\label{F20a}
\\
   a &=& \left[ 1 + 3.8\,(0.1\xi)^9 \right]^{-1}
        \left[ 1 + 0.171\,\xi^{7/2}\,T_\mathrm{b9} \right]^{-1} ,
\qquad
   \xi \equiv -\log(10^6\eta).
\eea

Since the results of PCY97 and \citet{PYCG03} have been obtained
neglecting neutrino emission in the blanketing envelope, the effective
surface temperature is determined by Eqs.~(\ref{F20a}) and (\ref{F-Ts}),
\beq
   \Ts = 1.1524\,F_{20}^{1/4}\times10^6 \mbox{~K}.
\label{TsF20}
\eeq
This assumption is acceptable, if $\Tb\lesssim10^9$~K. At still higher
temperatures,  neutrino cooling  within the heat blankets can be
important and the blanket models must be modified. In this case, 
\req{F-Ts} is valid only at the surface but not at the envelope bottom
$\rho=\rho_\mathrm{b}$, because the flux at the radiative surface is no
longer equal to $F_\mathrm{b}$ (see Section~\ref{subsub:basic_eqns}).
With increasing $\Tb$, energy from the blanket is progressively lost to
neutrino emission, while the photon emission levels off. In this case,
the boundary condition (\ref{boundarycond}) is not directly determined
by $\Ts$ (see Section~\ref{sub:2Dmagnet} below).

\section{Ion diffusion in heat blanketing envelopes}
\setcounter{equation}{0}
\label{sec:2}

So far we have considered the models of heat
blankets which contain ions of 
one species at any value of $\rho$. 
Evidently,  heat blankets 
may  contain mixtures of different
chemical elements. The parameters of ions in such  mixtures have 
been outlined in Section~\ref{subsub:e-and-i}. 
Generally, diffusion of the ions of different species
should be  
taken into account. Below, before constructing models of such envelopes, we focus 
on ion diffusion and its effect on the structure
and insulating properties of the envelopes.

\subsection{Diffusion currents in a dense plasma}
\label{ssec:2:DiffFluxEqs}

Diffusion processes in a mixture of rarefied gases and in a weakly coupled
multicomponent plasma (see Section\ \ref{subsub:e-and-i}) are well studied and described in the
classical monographs by \citet{CC52,Hirsh54}. Many studies of
diffusion in dense plasmas are based on the expressions for
diffusion currents taken for a mixture of rarefied (weakly Coulomb coupled)
particles, with the diffusion coefficients calculated for strongly
coupled plasmas. This approach does not take into account that the
Coulomb interaction affects not only the diffusion coefficients but
the diffusion currents themselves. 

Here, we present the  
derivation of the diffusion currents from first principles.
We will follow \citet{BY13,BY14a,BPY16} who
used the approach similar to that described by
\citet{LL_Fluid87}. We will derive the basic formulas in a general
form and apply them for heat blankets of
neutron stars. 

Let us consider a multicomponent Coulomb plasma,
which is out of equilibrium under the effects of external forces
$\bm{f}_\alpha$ (acting on particle species $\alpha$: 
electrons $\alpha = $ e and ions $\alpha = j =1,2,\ldots$), gradients of
number densities $\bm{\nabla} n_\alpha$ and the temperature
gradient $\bm{\nabla}T$. Here and hereafter, the gradient operator
$\nabla$ is assumed to act in the local frame of reference.
 All deviations from the equilibrium are
thought to be weak so that we can use the linear 
kinetics in which the diffusion currents are linear
with respect to corresponding thermodynamic forces. Let us
introduce generalized thermodynamic forces
\begin{equation}
  \widetilde{\bm{f}}_\alpha = \bm{f}_\alpha -
  \left(\bm{\nabla}\mu_\alpha -\left. \frac{\partial \mu_\alpha } {\partial T} \right|_P \bm{\nabla}T \right),
\label{e:GenForce}
\end{equation}
where $\mu_\alpha$ is the chemical potential of particles $\alpha$.

In the outer layers of neutron stars,
plasma particles are mostly affected by 
the gravitational force and the electric force,
\begin{equation}
\bm{f}_\alpha = Z_\alpha e \bm{E}+m_\alpha \bm{g}.
\label{e:Force}
\end{equation}
In this case, $Z_\alpha e$ and $m_\alpha$ are, respectively, the charge
and mass of particle species $\alpha$ ($Z_\mathrm{e}=-1$); $\bm{g}$ is
the gravitational acceleration, determined by \req{g_s},
 and $\bm{E}$ is the electric field in the local reference frame,
induced by a plasma polarization in the gravitational
field; this electric field ensures electric neutrality of the plasma,
\req{e:ChargeNeutrality}.

A deviation of the system from the state of diffusive equilibrium
is characterized by the quantities
\begin{equation}
\bm{d}_\alpha = \frac{\rho_\alpha}{\rho}
\sum_\beta n_\beta \widetilde{\bm{f}}_\beta- n_\alpha   \widetilde{\bm{f}}_\alpha,
\label{e:d-vect}
\end{equation}
where $\rho_\alpha = m_\alpha n_\alpha$ is the mass density of the
component $\alpha$ ($\rho$ being the total mass density).
Here we neglect the electron mass in conformity with the approximations
described after \req{e:d-alt} below.
Evidently, $\sum_\alpha
\bm{d}_\alpha = 0$. Using Eqs.\ \eqref{e:GenForce} and
\eqref{e:Force}, the Gibbs-Duhem relation 
\beq
\sum_\alpha n_\alpha
\bm{\nabla}\mu_\alpha = \bm{\nabla}P - S \,\bm{\nabla}T
\label{GibbsDuhem}
\eeq
 ($S$ being
the entropy density) and the electric neutrality condition
\eqref{e:ChargeNeutrality}, we obtain
\begin{equation}
\sum_\alpha n_\alpha \widetilde{\bm{f}}_\alpha =
\rho \bm{g} - \bm{\nabla}P.
\label{e:SumF}
\end{equation}
This is an important relation for the mechanical stability of the star.
Particle species $\alpha$ are in a state of mechanical equilibrium if
and only if  $\widetilde{\bm{f}}_\alpha = 0$. If, in addition, the
system is isothermal (i.e., $\bm{\nabla}T = 0$), then this expression
coincides with the condition of ``chemical'' equilibrium of particles
$\alpha$ \citep{CB10}. Furthermore, if the system as a whole is in the
state of hydrostatic equilibrium, then $\rho \bm{g} = \bm{\nabla}P$.
Recall that hydrostatic equilibrium in neutron stars is restored over
time-scales ranging from milliseconds to tens of seconds \citep{ST83}.
On the other hand, it takes from days to years  to reach diffusive
equilibrium in the envelopes of neutron stars (see
Section~\ref{ssec:4:NonEquilResults}). Therefore, if the system as a
whole is in hydrostatic equilibrium, then the diffusive equilibrium
implies also the mechanical equilibrium.

The outer layers of neutron stars are usually in 
hydrostatic equilibrium. Then the right-hand-side of Eq.\
\eqref{e:SumF} is zero, and Eq.\ \eqref{e:d-vect} is simplified,
\begin{equation}
\bm{d}_\alpha = - n_\alpha \widetilde{\bm{f}}_\alpha.
\label{e:d-short}
\end{equation}
Using Eqs.\ \eqref{e:GenForce} and \eqref{e:Force}, one can rewrite
\eqref{e:d-short} in the form
\begin{equation}
\bm{d}_\alpha = -\frac{\rho_\alpha}{\rho}\bm{\nabla}P - Z_\alpha n_\alpha e \bm{E}
+n_\alpha \left( \bm{\nabla}\mu_\alpha
- \left.\frac{\partial \mu_\alpha }{\partial T}\right|_P \bm{\nabla}T\right).
\label{e:d-alt}
\end{equation}

Because the electrons are much lighter than the ions, their
characteristic velocities are much higher (especially if they are
degenerate). We will be mostly interested in the ion transport, in which
case one can use the adiabatic Born-Oppenheimer approximation (e.g., 
\citealt{Schiff68}). This approximation implies that the electrons are
in mechanical quasi-equilibrium with respect to the ions (that is the
electrons instantaneously readjust themselves to ion displacements). 
Then $\bm{d}_\mathrm{e}=0$ and, according to \req{e:d-short}, 
$\widetilde{\bm{f}}_\mathrm{e}=0$. Therefore, in the limit
of $\mel \to 0$, we obtain
\begin{equation}
e\bm{E} = -\left(\bm{\nabla}\mu_\mathrm{e} - \left.\frac{\partial \mu_\mathrm{e}}
{\partial T} \right|_P \bm{\nabla}T\right).
\label{e:eE}
\end{equation}
Using standard relations of chemical equilibrium
(e.g., \citealt{LL_Stat93}), this expression
can be rewritten through chemical potentials of the ions. The
adiabatic approximation allows us to exclude the electrons from the
problem of ion transport (the diffusion currents of ions are mostly
determined by a non-equilibrium state of the ion subsystem,
\citealt{Paquette86}).

The chemical potentials are usually known as functions of
temperature and particle fractions. It is instructive
to express the derivative  $\partial \mu /
\partial T$ at constant $P$ and $x_j$ in terms of $\partial \mu /
\partial T$ at constant $n_j$,
\begin{equation}
\left.\frac{\partial \mu}{\partial T}\right|_{P,\{x_j\}} =
\left.\frac{\partial \mu}{\partial T}\right|_{\{n_j\}}
- \left.\frac{\partial P}{\partial T}\right|_{\{n_j\}}
\left(\sum_j n_j \left.\frac{\partial \mu}{\partial n_j}\right|_{T,\{n_k|k\neq j\}}\right)
\left(  \sum_j n_j \left.\frac{\partial P}{\partial n_j}\right|_{T,\{n_k|k\neq j\}} \right)^{-1}.
\label{e:MuPDeriv}
\end{equation}

The phenomenological expression for the mass density current can be
written as
\begin{equation}
\bm{J}_\alpha = \rho_\alpha \bm{v}_\alpha =  \frac{n m_\alpha}{\rho\kB T}
\sum_{\beta \neq \alpha} m_\beta D_{\alpha\beta}\bm{d}_\beta - D_\alpha^{T} \frac{\bm{\nabla}T}{T},
\label{e:J}
\end{equation}
where $\bm{v}_\alpha$ is the diffusion velocity of particles
$\alpha$, $D_{\alpha\beta}$ is a generalized diffusion coefficient
for particles $\alpha$ with respect to particles $\beta$,
$D_\alpha^{T}$ is a thermal diffusion coefficient for particles $\alpha$. The coefficient before
the sum is chosen in such a way for the expression to coincide with the
ordinary definition of diffusion coefficients in a mixture of
ideal gases  (e.g., \citealt{CC52,Hirsh54,LL_Kin}). By definition,
the diffusion fluxes should satisfy the relation
\begin{equation}
\sum_{\alpha}\bm{J}_\alpha=0,
\label{e:SumJ}
\end{equation}
which imposes certain restrictions on the diffusion and thermal
diffusion coefficients  (as described in 
\citealt{CC52,Hirsh54}). Equation \eqref{e:J} is strictly valid for
non-relativistic particles whereas the electrons in a dense plasma
can be relativistic. However, the adiabatic approximation can also be
valid for relativistic electrons (as long as they can be
treated as massless), so that the exclusion of electrons from the
ion diffusion problem is still possible.

A further use of Eq.\ \eqref{e:J} in the general form is complicated.
Hereafter, we restrict ourselves to particular cases that are most
appropriate to ion diffusion in heat blankets
of neutron stars. We will mainly consider binary ionic mixtures.
For them, \req{e:J} can be written as
\beq
   \bm{J}_1 = -\bm{J}_2 = -\frac{\nnii m_1 m_2}{\rho\kB T}\,
      D_{12} \left(\bm{d}_1+k_T\frac{\nabla T}{T}\right).
\label{k_T}
\eeq
Here, the last term is a correction due to thermal diffusion,
which is usually weak;
a dimensionless coefficient $k_T$ is called the thermal diffusion ratio.


\subsection{Diffusion in isothermal strongly coupled and
strongly degenerate plasmas}
\label{ssec:2:IsothermDiffFlux-SC}

To analyze the main features of \req{e:J} for the diffusion currents it
is sufficient to study an isothermal system. Moreover, as will be shown
below in this section, the terms associated with the temperature
gradient will disappear in the limit of a strongly non-ideal plasma.
Then general, non-isothermal expressions coincide with isothermal ones.
Let us consider a binary ion mixture (with two ion species $j=1,2$). To
be specific, we assume that  $Z_1 < Z_2$. Taking into account
quasi-equilibrium of electrons, $\bm{d}_\mathrm{e} = 0$, and also that
$\sum_{\alpha} \bm{d}_\alpha = 0$ (see above), we obtain $\bm{d}_1 = -
\bm{d}_2$. Therefore, it is sufficient to study only $\bm{d}_1$. Using
Eqs.~\eqref{e:d-alt} and \eqref{e:eE}, we obtain
\begin{equation}
{\bm{d}}_1=\frac{n_1 n_2}{\nne} \left( m_\mathrm{u} \left(Z_1 A_2-Z_2 A_1\right)
\frac{\bm{\nabla}P}{\rho} + Z_2 \bm{\nabla} \mu_1 - Z_1 \bm{\nabla}\mu_2 \right),
\label{e:d1}
\end{equation}
where $m_\mathrm{u}$ is again the atomic mass unit (i.e., $m_j = A_j m_\mathrm{u}$). Without
any loss of generality, the chemical potential of ions can be
presented as a sum of two terms, 
$\mu_j = \mu_j^{(\mathrm{id})}+\mu_j^{(\mathrm{C})}$,
where ``(id)'' labels the ideal gas contribution and ``(C)''
the contribution
of the
Coulomb interaction and other effects of non-ideality such as the
exchange interaction, polarizability of the electron background and
so on (see \citealt{PC10,PC13} for details). Under the conditions in
the envelopes of neutron stars, the main
contribution to  $\mu_j^{(\mathrm{C})}$ comes from the Coulomb interaction of
ions. As a result, the
vector $\bm{d}_1$ splits into the three terms, $\bm{d}_1 = \bm{d}_g
+ \bm{d}_{\nabla n} + \bm{d}_\mathrm{C}$, with \citep{BY13}
\begin{align}
&\bm{d}_\mathrm{g} = m_\mathrm{u} Z_1 Z_2\, \frac{n_1 n_2}{\nne}
\left( \frac{A_2}{Z_2} - \frac{A_1}{Z_1} \right) \frac{\bm{\nabla}P}{\rho}\,,
\label{e:d-grav} \\
&\bm{d}_{\nabla n} = \frac{n_1 n_2}{\nne}
 \left[ Z_2 \bm{\nabla}\mu^{(\mathrm{id})}_1
- Z_1 \bm{\nabla}\mu^{(\mathrm{id})}_2 \right]  = \frac{k_\mathrm{B} T}{\nne}
\left( Z_2 n_2 \bm{\nabla}n_1 - Z_1 n_1 \bm{\nabla}n_2 \right),
\label{e:d-gradn} \\
&\bm{d}_\mathrm{C}  = \frac{n_1 n_2}{\nne}
 \left[Z_2 \bm{\nabla}\mu^{(\mathrm{C})}_1
- Z_1 \bm{\nabla}\mu^{(\mathrm{C})}_2 \right].
\label{e:d-C}
\end{align}
In \req{e:d-gradn} we have used the relation $\bm{\nabla}\mu^{(\mathrm{id})}_j = k_\mathrm{B} T
\big(\bm{\nabla}n_j\big)/n_j$, which holds in an isothermal
system. Let us consider each term separately.
\begin{itemize}
  \item The term $\bm{d}_\mathrm{g}$ is responsible for the mechanism of
  gravitational separation of ions, provided their effective ``molecular
  weights'' $A/Z$ are different, which destroys the balance of gravity
  and electric forces. In the neutron-star envelopes and white dwarfs,
  this mechanism was studied previously, 
  e.g., by \citet{AI80,HameuryHB83,CB03,CB10}.
  
  \item The term $\bm{d}_{\nabla n}$ describes ordinary diffusion under
  the action of gradients of ion number densities; this is easily seen
  at $n_2 \ll n_1$ in which case $\nne \approx Z_1 n_1$ and
  $\bm{d}_{\nabla n} \approx -k_\mathrm{B} T\, \bm{\nabla} n_2$.
  In the neutron-star envelopes,
  this mechanism was also studied previously, e.g.,
   by \citet{HameuryHB83,CB10}.

  \item The term $\bm{d}_\mathrm{C}$ is responsible for a  Coulomb
mechanism of ion separation, which was put forward by \citet{DeBlasio00}
and studied by \citet{CB10} for equilibrium plasma configurations.  The
latter authors have located the domains of plasma parameters  where
$\bm{d}_\mathrm{C}$, $\bm{d}_\mathrm{g}$, or the nuclear mass defect of
ions with $A\approx2Z$ in $\bm{d}_\mathrm{g}$ are dominant. Later
\citet{BY13,BY14a} and \citet{BPY16} introduced the term
$\bm{d}_\mathrm{C}$ in the expressions for mass currents and studied its
effects for non-isothermal or non-equilibrium plasma states.

\end{itemize}

Let us study specific features of the
Coulomb mechanism of ion separation in a strongly coupled
and strongly degenerate plasma  ($\bar{\Gamma}\gg1$, $T\ll
T_\mathrm{F}$).  Within an accuracy of several percent,  thermodynamic
functions of such a plasma can be described in the ion-sphere
approximation using the linear mixing rule (see \citealt{PC10,PC13} and
references therein for details and for a more accurate description
beyond these simplified assumptions). Under these approximations,
\begin{equation}
\mu_j^{(\mathrm{C})} = -0.9\, \frac{Z_j^{\slfrac{5}{3}}e^2}{a_\mathrm{e}}, \quad
\bm{\nabla} \mu_j^{(\mathrm{C})} = -0.3\,\frac{Z_j^{\slfrac{5}{3}}e^2}{a_\mathrm{e}}
\frac{\bm{\nabla}\nne}{\nne}.
\label{e:muC-SC}
\end{equation}
Then
\begin{equation}
\bm{d}_\mathrm{C} = 0.3\, \frac{n_1 n_2}{\nne} \frac{Z_1 Z_2 e^2}{a_\mathrm{e}}
\left(Z_2^{\slfrac{2}{3}} - Z_1^{\slfrac{2}{3}} \right) \frac{\bm{\nabla}\nne}{\nne}.
\label{e:d-C-SC}
\end{equation}

The structure of $\bm{d}_\mathrm{C}$ resembles the structure of
$\bm{d}_\mathrm{g}$; $\bm{d}_\mathrm{C}$ describes a specific
`Coulomb' separation of ions in a gravitational field. The separation
occurs because the ions with different $Z$ have different ion-sphere
radii. Hence, the Coulomb energies of the ion spheres are different
(also see Section~\ref{ssec:2:Results}). A specific feature of the
Coulomb term is that it is present even for ions with 
$\slfrac{Z_1}{A_1} = \slfrac{Z_2}{A_2}$.

In order to illustrate this effect let us derive the final expressions
for the diffusive ion currents in a strongly coupled ion plasma, using
the ion-sphere approximation and assuming that the pressure is mainly
produced by degenerate electrons,  $P \approx P_\mathrm{e}
(\nne)$. These simplifying assumptions are sufficiently accurate
in the bulk of a typical neutron-star envelope. Then, using the condition
of hydrostatic equilibrium (\ref{dP/dz}), we obtain 
$\slfrac{\big(\bm{\nabla} \nne \big)}{\nne} =
\slfrac{\big(\bm{\nabla} P \big)}{\big(\gamma P \big)} = \slfrac{\rho
\bm{g}}{\big(\gamma P \big)}$, where  $\gamma=(\slfrac{\partial \ln
P}{\partial \ln \nne})_T$. Under the
above assumptions, the Coulomb contribution becomes
\begin{equation}
\bm{d}_\mathrm{C} = 0.3\, \frac{\rho n_1 n_2}{\nne}
                   \frac{Z_1 Z_2 e^2 \bm{g}}{\gamma a_\mathrm{e} P}
                    \left(Z_2^{\slfrac{2}{3}} - Z_1^{\slfrac{2}{3}}
                     \right).
\label{e:d-C-SC-Fin}
\end{equation}
In a binary ionic mixture, there is only one non-trivial
coefficient for binary diffusion, ${D}_{12}={D}_{21} \equiv D$.
With our treatment of electrons as massless fermions
 (Section~\ref{ssec:2:DiffFluxEqs}), 
while studying the ion transport, we can set $\bm{J}_\mathrm{e}
= 0$. Then Eq.\ \eqref{e:SumJ} yields that $\bm{J}_1 =
-\bm{J}_2$ and, taking into account $\bm{d}_1 = - \bm{d}_2$
(see above in this section), one can write down the diffusion
current as
\begin{equation}
\bm{J}_2=-\bm{J}_1=\frac{\nnii m_1 m_2}{\rho k_\mathrm{B}T}\,{D}{\bm{d}}_1. 
\label{e:J2-Gen}
\end{equation}
Substituting here Eqs.\ \eqref{e:d-grav}, \eqref{e:d-gradn}
and \eqref{e:d-C-SC-Fin}, we obtain
\begin{equation}
\bm{J}_2 = D\, \frac{m_1 m_2 \nnii}{\rho \nne}
\left( Z_2 n_2 \bm{\nabla}n_1 - Z_1 n_1 \bm{\nabla} n_2 \right)
+ \left(\bm{u}_g + \bm{u}_\mathrm{C} \right) m_2 n_2,
\label{e:J2-Vel}
\end{equation}
where $\nnii = n_1 + n_2$ and
\begin{align}
&\bm{u}_\mathrm{g}  =  \frac{\rho_1 \nnii D}{\rho \nne k_\mathrm{B} T}
Z_1 Z_2 m_\mathrm{u} \bm{g}\left( \frac{A_2}{Z_2} - \frac{A_1}{Z_1} \right),
\label{e:ug} \\
&\bm{u}_\mathrm{C}  =  \frac{\rho_1 \nnii D}{ \nne k_\mathrm{B} T} Z_1 Z_2 \bm{g}
\left(Z_2^{\slfrac{2}{3}}-Z_1^{\slfrac{2}{3}}\right)\,\frac{0.3 e^2}{a_\mathrm{e} P \gamma}
\label{e:uC-SC}
\end{align}
are the velocities of the gravitational (g) and Coulomb (C)
separations of ions, respectively.

If we consider a non-isothermal system in the limit of strong Coulomb
coupling, using the linear mixing rule and neglecting thermal diffusion,
then from Eqs.~\eqref{e:MuPDeriv} and \eqref{e:muC-SC} with  $P \approx
P_\mathrm{e} (\nne)$ we obtain that all the terms in
Eqs.~\eqref{e:d-alt} and \eqref{e:eE} related to $\nabla T$ vanish.
Accordingly, the non-isothermal expressions for the diffusion currents
coincide with the isothermal ones.

The diffusion separation of ions does not violate an overall
hydrostatic equilibrium, because the latter is established much
quicker, over hydrodynamic timescales. Therefore, the diffusion
of ions of species 2 inside a neutron star
envelope is accompanied by the diffusion
of ion species 1 toward the stellar surface. This is clearly seen
from the relation  $\bm{J}_1 = -\bm{J}_2$ discussed above.
This purely diffusive motion leads to the collisional generation
of the entropy (${\Sdot}_\mathrm{coll}$) and to the related
energy release with the rate $Q$ [erg~cm$^{-3}$~s$^{-1}$] (see, e.g.,
\citealt{CC52,Hirsh54}),
\begin{equation}
Q=T {\Sdot}_\mathrm{coll} = \frac{\rho}{\rho_1 \rho_2}
\bm{J}_2 \bm{\cdot} \bm{d}_1.
\label{e:Q}
\end{equation}

For practical applications of Eqs.\ \eqref{e:J2-Vel},
\eqref{e:ug} and \eqref{e:uC-SC} one needs the
mutual diffusion coefficient of ions, $D$. For a weakly coupled 
plasma it can be written as   \citep{CC52,Hirsh54}
\begin{equation}
D = \frac{3}{8} \sqrt{\frac{2 k_\mathrm{B} T}{\pi m_{12}}}
\left(\frac{k_\mathrm{B} T}{Z_1 Z_2 e^2} \right)^2 \frac{1}{\nnii \Lambda},
\label{e:D-Simple}
\end{equation}
where $m_{12} = \slfrac{m_1 m_2}{(m_1 + m_2)}$ is the reduced
mass of the ions 1 and 2, and $\Lambda$ is the Coulomb logarithm.
In order to apply this expression for a plasma with any coupling
(including a strong one) we will introduce a generalized Coulomb
logarithm suggested by \citet{Khrapak13} for calculating
the self-diffusion coefficient in a one-component ion plasma.
\citet{BY14b} extended this method to a binary ion mixture.
They introduced the effective Coulomb logarithm as
\begin{equation}
\Lambda_\mathrm{eff} = \frac{1}{2} \ln \left( 1 + X^{-2} \right), 
\quad
X = \left(1 + \left(3 \bar{\Gamma}
\right)^{\slfrac{3}{2}} \right)^{\slfrac{1}{3}} - 1.
\label{e:Leff-Simple}
\end{equation}
This formula allows one to estimate $\Lambda_\mathrm{eff}$ for Coulomb gas or liquid of 
ions. A more accurate approach will be presented in Section~\ref{sec:3}.

\subsection{Diffusion in isothermal weakly coupled plasma}
\label{ssec:2:IsothermDiffFlux-WC}

Now let us consider the opposite case of a weakly coupled plasma.
There are two main differences from the strongly coupled plasma.
First, Coulomb parts of the chemical potentials are different.
Second, the pressure is dominated by the ideal gas contribution,
$P \approx P^\mathrm{(id)} = (\nnii + \nne) k_\mathrm{B} T$. In the weakly
coupled plasma, such that the mean ion coupling parameter
(\ref{e:GammaAv}) is small
($\bar{\Gamma} = \Gamma_\mathrm{e} \overline{{Z}^{5/3}}
= \Gamma_0\,\bar{Z}^{1/3}\,\overline{{Z}^{5/3}} \ll 1$),
\begin{equation}
\mu_j^\mathrm{(C)} = -\frac{k_\mathrm{B} T}{2} Z_j \Gamma_\mathrm{e}^{\slfrac{3}{2}}
\sqrt{\frac{\overline{Z^2}}{3 \bar{Z}^3}} \left(3 Z_j \bar{Z}
 - \overline{Z^2} \right).
\label{e:muC-WC}
\end{equation}
Then the expression for the
Coulomb component of the vector  $\bm{d}_1$ can be presented
in the form [cf.\ \req{e:d-C-SC}]
\begin{equation}
\bm{d}_\mathrm{C} = \frac{n_1 n_2}{  \nne} \sqrt{\frac{\pi   \nne}{k_\mathrm{B} T}}\,
\frac{e^3 Z_1^2 Z_2^2 \left( Z_2 - Z_1 \right) }
{2 \bar{Z}^{\slfrac{3}{2}} \sqrt{\overline{Z^2}}}
\,\left(\bm{\nabla}\bar{Z}
  + \frac{\bar{Z}\,
   \overline{Z^2}}{Z_1 Z_2} \frac{\bm{\nabla}
      \nne}{  \nne}
        \right),
\label{e:d-C-WC}
\end{equation}
where the term $\bm{\nabla}\bar{Z}$ appears due to variations of ion
fractions with depth.

Further calculations are analogous to the strong-coupling limit in
Section~\ref{ssec:2:IsothermDiffFlux-SC}. Equation~\eqref{e:J2-Vel} has
the same form, but an expression for $\bm{u}_\mathrm{C}$ is different.
In this case,
\begin{equation*}
\frac{\bm{\nabla}   \nne}{  \nne} = \frac{\bm{\nabla} P}{P} +
\frac{\bm{\nabla} \bar{Z}}{\bar{Z}\left(\bar{Z}+1\right)}
= \frac{\rho \bm{g}}{P} + \frac{\bm{\nabla} \bar{Z}}{\bar{Z}\left(\bar{Z}+1\right)}.
\end{equation*}
As a result, we obtain [cf.\ \req{e:uC-SC}]
\beq
\bm{u}_\mathrm{C}  = \frac{\rho_1 \nnii D}{\rho
\left(k_\mathrm{B} T \right)^{\slfrac{3}{2}}} \sqrt{\frac{\pi}{ 
\nne}}\, \frac{e^3 Z_1^2 Z_2^2 \left( Z_2 - Z_1 \right) }{2
\bar{Z}^{\slfrac{3}{2}} \sqrt{\overline{Z^2}}}
\left[\bm{ \left(1+\frac{\overline{Z^2}}{Z_1 Z_2 \left(
\bar{Z} + 1 \right)} \right) \nabla}\bar{Z}
 + \frac{\bar{Z}\, \overline{Z^2}}{Z_1 Z_2}
      \frac{\rho \bm{g}}{P}  \right].
\label{e:uC-WC}
\eeq
Some numerical examples of the Coulomb separation velocities (after
\citealt{BY13,BY14a,BY14b}) will be given in Section~\ref{ssec:2:Results}.

Let us make one important remark. Although the mechanisms of the ion
separation are called gravitational \eqref{e:ug} and Coulomb
[\eqref{e:uC-SC}, \eqref{e:uC-WC}], both of them are associated with the
presence of the gravitational field. The gravitational mechanism is
directly responsible for the gravitational separation of ions provided
their buoyancy is different. The buoyancy of the ions in a plasma is
determined by their charge-to-mass ratio since the gravity acts toward
the stellar center while the electric field toward the surface. As
mentioned in Section~\ref{ssec:2:DiffFluxEqs}, this macroscopic electric
field appears as a response of the plasma to the external gravitational
field, to ensure  electric neutrality. As for the Coulomb mechanism, it
is related to the difference of Coulomb energies of ion spheres {in the
electric field} for the ions of different charges. Therefore, both
mechanisms are actually caused by the gravity.

The presented expressions for diffusion and Coulomb separation of ions
in ion mixtures (in various Coulomb coupling regimes) can be used in ion
liquids and gases. When the ions solidify, the diffusion is possible but
strongly suppressed \citep[e.g.,][]{Hughto11}. To the best
of our knowledge, for the fist time the component of the diffusion
current associated with ion-ion interaction in the presence of
gravitational field has been considered by \citet{BY13}. This component
becomes important for mixtures of ions with equal $A/Z$. Nevertheless,
we should stress that the classical monograph by \citet{Hirsh54}
presents general expressions which describe non-equilibrium processes in
the frame of thermodynamics of irreversible processes. Using those
expressions one can obtain the expressions for the diffusion currents
presented above. However, they have not been used in astrophysical
literature, where the Coulomb contribution in the diffusion current has
been neglected.

\subsection{Diffusion coefficients}
\label{sec:3}

\subsubsection{Methods of calculations}
\label{ssec:3:DiffCoeffReview}

In order to study the diffusion in a plasma,
aside of the diffusion currents, one needs the coefficients
of diffusion and thermal diffusion. Here
we outline the methods for calculating the diffusion coefficients
described in the literature. More details can
be found in \citet{BY14b}.

The main obstacle for calculating the diffusion coefficients in a
Coulomb plasma of ions is the long-range nature of the Coulomb
interaction. This diffusion is similar to that for particles interacting
via Debye potential\footnote{The same potential form appears in the
physics of dusty plasmas as well as in the nuclear physics, where it is
called the Yukawa potential.}  (statically screened Coulomb potential)
with a sufficiently large screening length,

\begin{equation}
\Phi_\mathrm{SSCP} = \frac{q_1 q_2}{r} \exp{\left(-\frac{r}{r_\mathrm{D}}\right)}.
\label{e:SSCP}
\end{equation}
In this case, $q$ is a particle charge,
\beq
   r_\mathrm{D} = \sqrt{4\pi \sum_j q_j^2 \nnii / \kB T}
\label{Debye}
\eeq
 is the Debye screening
length, SSCP means statically screened Coulomb potential.

The physics of diffusion processes is complicated. There are different
types of the diffusion coefficients:  the self-diffusion coefficients
$D_\mathit{ii}$, and the mutual diffusion coefficients $D_\mathit{ij}$, 
which determine diffusion currents (here $i, j=1, 2, \ldots$ label ion
species in a multi-component ion plasma).  Diffusion can be investigated
using different methods such as Chapman-Enskog theory, Green-Kubo
relations, molecular dynamic simulations, effective potential
theories, etc.

The most interesting case for us is the diffusion of ions in binary ion
mixtures which constitute a weakly  or a strongly Coulomb coupled liquid
(Section~\ref{subsub:e-and-i}). As mentioned above, the diffusion  in
gases is well studied and described in the famous monographs  
\citep{CC52,Hirsh54}, while the diffusion in liquids is less elaborated.
Our aim  is to choose a unified approach for calculating the diffusion
coefficients in gases and liquids. In a binary mixture, there is only
one independent mutual diffusion coefficient $D_{12}=D_{21}$  and two
self-diffusion coefficients, $D_{11}$ and $D_{22}$.

In a weakly coupled plasma ($\bar{\Gamma} \ll 1$),  the ions move more
or less freely and diffuse owing to relatively weak Coulomb collisions
with nearby ions. In this case, the diffusion coefficients are usually
expressed through a Coulomb logarithm $\Lambda$, which can be estimated
as a logarithm of sufficiently large ratio of the maximum-to-minimum
impact parameters of colliding ions. In this case one can use the
classical diffusion theory by \citet{CC52,Hirsh54}. In astrophysical
literature, this theory is often called the Chapman-Spitzer theory, which
means the application of the general diffusion theory to the Coulomb
interactions under astrophysical conditions, as described in the
classical monograph by \citet{Spitzer}. Earlier astrophysical
publications based on that theory were described, for instance, by
\citet{Paquette86}. With the growth of the Coulomb coupling, $\Lambda$
becomes smaller.  At $\bar{\Gamma} \sim 1$ one gets $\Lambda \sim 1$,
and the diffusion coefficient becomes $D \sim \omega_\mathrm{pi}
a_\mathrm{i}^2$,  where $\omega_\mathrm{pi}$ is the ion plasma frequency
given by \req{e:wp} below. Characteristic ion collision
frequencies reach the level of the plasma frequency, while the typical
ion mean-free-path becomes  comparable to inter-ion distances.

In a strongly coupled plasma  ($\bar{\Gamma} \gg 1$), the ions are
mostly trapped in their own potential wells (inside the appropriate
Wigner-Seitz cells) and constitute the Coulomb liquid or crystal. These
ions mainly oscillate around their quasi-equilibrium positions, and the
diffusion proceeds through thermally excited jumps from one
quasi-equilibrium position to another (neighboring) one. One can
distinguish the cases of classical  ($T \gtrsim T_\mathrm{pi}$) and
quantum ($T \lesssim T_\mathrm{pi}$) ion motion (where
$T_\mathrm{pi}=\hbar \omega_\mathrm{pi}/k_\mathrm{B}$ is the ion plasma
temperature; it is similar to the Debye temperature of the Coulomb
crystal). In the quantum case, the most important contribution comes
from collective oscillations.  As far as electrons are concerned, one
can consider approximations of the rigid or polarized electron
backgrounds. These cases usually lead to almost the same results.

\citet{FM79in} obtained approximate analytic expressions  for 
$D_{\mathit{ij}}$ through the Coulomb logarithm in the case of weak ion
coupling. They considered the cases of quantum and classical minimum
impact parameters in the Coulomb logarithm. In the case of weak
coupling, these results were further extended by \citet{IM85}.

\citet{Paquette86} calculated the mutual diffusion coefficients for a
binary ion mixture at weak and moderate Coulomb coupling using the
Chapman-Enskog formalism and statically screened Coulomb potential.  In
addition, they  analyzed previous molecular dynamics calculations of
self-diffusion coefficients at strong Coulomb coupling.

The first calculations of the self-diffusion coefficient $D_1$ by the
molecular dynamics were performed by \citet{Hansen75} who proposed
the following approximation at $\Gamma >1$ (in a one-component plasma,
where $\Gamma \equiv \bar{\Gamma}$),
\begin{equation}
D^*_1 = \frac{D_1}{\omega_\mathrm{pi} a_\mathrm{i}^2} \approx 2.95\, \Gamma^{-\nicefrac{4}{3}}.
\label{e:Hansen75}
\end{equation}
\citet{HJM85} calculated the diffusion coefficients $D_{12}$, $D_{11}$
and $D_{22}$ in a binary ion mixture at moderate and strong Coulomb
couplings. They derived the approximate relation,
\begin{equation}
D_{12} \approx x_2 D_{11} + x_1 D_{22},
\label{e:D12=D11+D22}
\end{equation}
where $x_{1}$ and $x_2$ are number fractions of ions  in an $^{1}$H --
$^{4}$He mixture ($x_2=1-x_1$). The authors  tabulated  $D_{12}$,
$D_{11}$ and $D_{22}$ for several values of the parameters $x_1$ and
$\Gamma_0$ [\req{e:Gamma0}].

Later the molecular dynamics simulations of the binary ionic mixtures
have been undertaken by many research groups. \citet{BP87}
calculated $D_{12}$ in a binary ion mixture using the molecular
dynamics and kinetic theory for strongly and weakly coupled plasmas.
Their results showed a good agreement with previous investigations.
\citet{Robbins88} studied the self-diffusion coefficient in a
one-component plasma.  \citet{RNZ95} modeled self-diffusion and
mutual diffusion in binary ion mixtures in a wide ranges of $A_2/A_1$
and $Z_2/Z_1$ ratios for strong, moderate and weak ion couplings. 
Extended calculations of self-diffusion coefficients in one-component
liquids, described by the Yukawa potential, were performed
by  \citet{OH00}, who used the Green-Kubo relation as well as the usual
expression for diffusion coefficients. They tabulated the
coefficient $D_1^*$ and approximated it by an analytic expression for
different screening parameters. 

\citet{DM05} calculated the self-diffusion coefficient for a
one-component ion system by the molecular dynamics using a
semi-empirical potential and approximated the results by an analytic
expression. Furthermore, \citet{Daligault06} analyzed the dynamics of a
liquid in a strongly coupled one-component plasma and studied the
transition from a free particle motion to the regime in which the ions
are ``confined'' in Coulomb potential wells. 

\citet{Daligault12} modeled self-diffusion in one- and two-component
strongly coupled ion systems by molecular dynamics and fitted the
numerical results at $\bar\Gamma\gtrsim25$ by the
expression 
\begin{equation}
D^* = \frac{D}{\omega_\mathrm{pi} a_\mathrm{p}^2}
 = \frac{A}{\Gamma}\exp{\left(-B \Gamma \right)},
\label{e:caging+jumps}
\end{equation}
where $A$ and $B$ are fit parameters. This expression can be derived in
the model of the ``confinement'' and thermally activated jumps of the
ions from one potential well to the nearest one. 
\citet{Daligault12} found that the numerical results are well
reproduced with $A=1.52$ and $B=0.0082$.
For the weak and
intermediate coupling regimes, $\bar\Gamma\lesssim25$,
\citet{Daligault12} proposed a model, which  extends the widely used
Chapman-Spitzer theory from the regime of weak coupling
to the regime of moderate coupling. According to this theory,
$
D = \kB T/(3m\nu),
$
where $\nu$ is a characteristic collision frequency
given by
\beq
    \nu= \frac43\sqrt{\frac{\pi}{m}}
    \frac{\nnii q^4}{(\kB T)^{3/2}}\,\Lambda,
\eeq
and $\Lambda$ is a Coulomb logarithm. In the Chapman-Spitzer theory
\citep{CC52,Spitzer,Paquette86},
$\Lambda = \ln(C r_\mathrm{D}/r_T)$,
$r_\mathrm{D}$ is the Debye screening length [\req{Debye}], which
characterizes the largest impact parameter,
$r_T=q^2/\kB T$ characterizes the smallest impact parameter,
and $C$ is a correction factor ($C=1$ is usually assumed).
\citet{Daligault12} replaced this expression by 
\beq
   \Lambda = \ln(1 + C r_\mathrm{D}/r_T)
\eeq
and found that it fits the numerical results from weak to moderate
coupling regimes and matches \req{e:caging+jumps} at $\Gamma\approx25$.
Also, he extended his results to the Yukawa
systems and to the mixtures.

 \citet{Khrapak13} considered self-diffusion coefficients in a
one-component plasma using the standard Chapman-Enskog theory of a
weakly-coupled plasma and molecular dynamics results by different
authors at strong coupling. Based on these results, he
suggested a simple and convenient analytic approximation for a Coulomb
coupling of any strength. This was done by introducing a
generalized Coulomb logarithm, $\Lambda_\mathrm{eff}$.

 \cite{BD13} suggested that both cases (of weak and strong coupling)
can be described within one and the same formalism of effective
potential of ion-ion interaction and traditional Chapman-Enskog theory.
They considered several effective potentials obtained from  radial
distribution functions (RDFs, also called pair correlation functions) of
the ions, $g(r)$, which were calculated either by molecular dynamics or
by hypernetted chain technique. The effective potential allows one
not only to describe the screening effect (that could have been done
using a statically screened Coulomb potential), but also take into
account ion correlations, including strong ones. This method treats the
screening and ion-ion correlations in a self-consistent manner, without
introducing any external screening length. The authors stressed the
convenience to express the diffusion coefficients through a generalized
Coulomb logarithm. 

\citet{BY14b} followed the same strategy and extended the method
to binary ionic mixtures (see Section~\ref{ssec:3:HNC}). They
expressed $D^*$ through an effective Coulomb logarithm and constructed a
new fit to it, which we reproduce in \ref{app:Lambda_Eff}.

Although we will not consider diffusion in Coulomb crystals, we remark
that the problem was investigated by \citet{DeBlasioLazzari96} using
macroscopic relations from \citet{Haase}. Later \citet{Hughto11}
simulated the self-diffusion in the Yukawa solid using the
molecular dynamics. They found that diffusion in the solid phase is
strongly suppressed as compared with the liquid.
For example, according to Tables II and IV of \citet{Hughto11}, 
$D^*$ decreases by more than an order of magnitude as $\Gamma$ grows
from 175 to 200, which implies $B\sim0.1$ in \req{e:caging+jumps}.

Calculations of  diffusion coefficients in magnetized Coulomb plasmas
were performed, e.g., by \citet{Bernu81,RJW03}. These authors obtained
self-diffusion coefficients along and across the field lines. The former
coefficient is larger than the latter; both of them decrease
with increasing field strength.

Hereafter we will neglect the quantum-mechanical effects
on ion diffusion and consider the cases or rigid and slightly
compressible electron gas. We will restrict ourselves only to the
classical ion-ion scattering in the presence of strongly degenerate
electrons.

\subsubsection{Calculation of the effective potentials with the
  hypernetted chain method}
\label{ssec:3:HNC}

Let us outline the results by \citet{BY14b} who calculated  the diffusion
coefficients, using the effective potentials of ion-ion interaction in
binary ions mixtures. As already discussed in
Section~\ref{ssec:3:DiffCoeffReview} (and see \citealt{BD13}), the effective
potentials are determined by the RDFs of
ions. These functions can be calculated by different techniques,
particularly, by molecular dynamics or by the hypernetted chain method. 
\citet{BY14b} have used the hypernetted chain method, which requires less
computational resources. As will be shown below, a choice of the  method
for calculating RDF to obtain the diffusion coefficients is not of
principal importance.

In this particular subsection lengths are measured in units of the ion
sphere radius, $a_\mathrm{i}$, Eq.\ \eqref{e:IonSphereRad}, and the
potentials are measured in units of  ${k_\mathrm{B} T}/{e}$.

A state of a binary mixture of ions is determined by the mass and charge
numbers of ions and also by two dimensionless parameters $x\equiv x_1$,
the relative number fraction of ions 1, and by $\Gamma_0$,
\req{e:Gamma0}.

Let $g_\mathit{ij}(r)$, $h_\mathit{ij}(r)$ and $c_\mathit{ij}(r)$
($i,j=1,2$) be the RDFs, total and
direct correlation functions, respectively  (e.g., \citealt{Croxton74}). All
these functions are symmetric with respect to their subscripts
 [e.g., $g_\mathit{ij}(r) =
g_\mathit{ji}(r)$]. By definition, one has $h_\mathit{ij}(r) =
g_\mathit{ij}(r)-1$. The effective potential (also called mean field
potential) $\Phi(r)$ in a one
component plasma is defined by the relation $g(r)=\exp{[-\Phi(r)]}$
(e.g., \citealt{BD13,Croxton74}). An expansion of this formalism to
a binary mixture is quite evident,
\begin{equation}
g_\mathit{ij}(r) = \exp{\left[-\Phi_\mathit{ij}(r)\right]}.
\label{e:RDF-phi-TCP}
\end{equation}
For calculating the mutual diffusion coefficients one needs,
first of all, the potential $\Phi_{12}(r)$, responsible
for the interaction between ion species 1 and 2.

Generally, all these functions cannot be determined in analytic form.
The hypernetted chain approximation (e.g., \citealt{HTV77,SPS73,Ng74}) 
consists in a joint solution of two types of the equations. They are (i)
Ornstein-Zernike relations, which connect direct and total correlation
functions and  (ii) a hypernetted chain closure. The
Ornstein-Zernike equations are exact, whereas hypernetted chain closure
is an approximation. This approximation reads
\begin{equation}
  g_\mathit{ij}(r) =  h_\mathit{ij}(r) + 1 = \exp{\left[h_\mathit{ij}(r) - c_\mathit{ij}(r) - \phi_\mathit{ij}(r)\right]},
\label{e:HNC-closure-base}
\end{equation}
where $\phi_\mathit{ij}(r)$ is a \emph{non-screened (bare)}
 Coulomb potential,
\begin{equation}
  \phi_\mathit{ij}(r) = \frac{Z_i Z_j \Gamma_0}{r},
\label{e:bare-Coulomb}
\end{equation}
the electron screening being neglected. Notice that the ion-ion
interaction potential enters only hypernetted chain closure relations
and that no ion-ion screening is employed here. As will be shown
later, the ion screening is obtained automatically during the solution
of hypernetted chain equations.

This system cannot be solved directly because of the long-range nature
of Coulomb interaction. For a one-component plasma, this problem has
been solved by \cite{SPS73} and \citet{Ng74} by introducing short-range
potentials and correlation functions. A similar method has been used by
\citet{HTV77} for binary mixtures.  Here we will not go into the details
of the calculations. Technical details and numerical schemes are
discussed in \citet{BY14b}, where the authors followed the methodology of
\citet{SPS73,Ng74,HTV77}.

\begin{figure*}
  \centering
  \includegraphics[width=.48\textwidth]{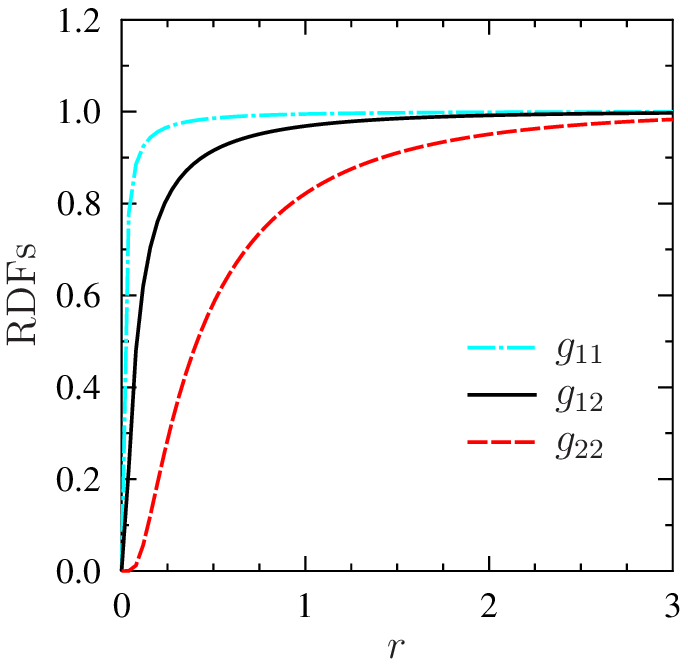}
  \includegraphics[width=.48\textwidth]{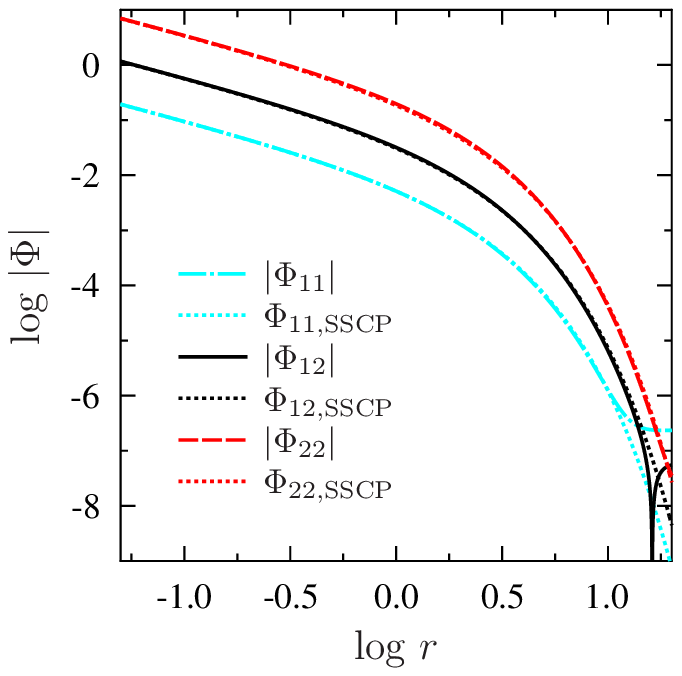}
  \caption{Radial distribution functions (left panel) and absolute
values of effective potentials (right panel) for H\,--\,C
mixture in a weakly coupled plasma with $x_\mathrm{H} = 0.6$ and
$\Gamma_0 = 0.01~ (\bar{\Gamma}\approx 0.12)$. Dotted lines show the
comparison with the corresponding statically screened Coulomb potential,
Eq.\ \eqref{e:SSCP-WC}. See the text for details.
}
  \label{fig:RDF-Phi-WC}
\end{figure*}

Having completed the calculations, \citet{BY14b} 
compared the calculated Coulomb excess energy with
the results of \citet{HTV77} and found an agreement 
 to five-six significant digits.

\begin{figure*}
  \centering
  \includegraphics[width=.48\textwidth]{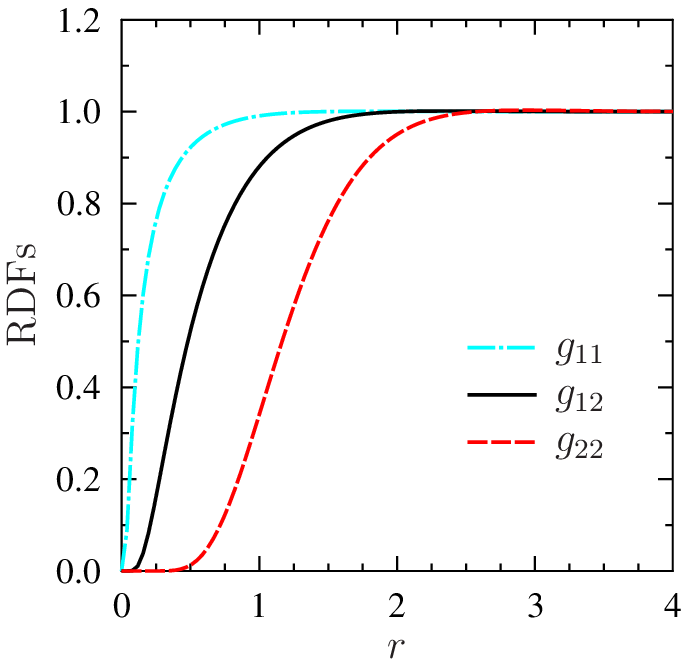}
  \includegraphics[width=.48\textwidth]{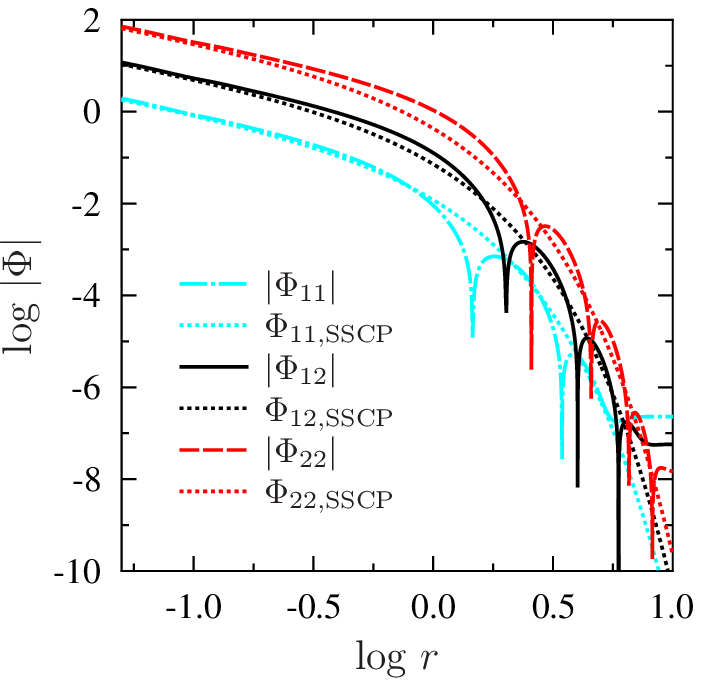}
  \caption{The same as Fig.~\ref{fig:RDF-Phi-WC}, but for an
intermediate coupling regime at $\Gamma_0 = 0.1~ (\bar{\Gamma}\approx
1.2)$.}
  \label{fig:RDF-Phi-Inter}
\end{figure*}

Now let us consider some examples of the obtained RDFs and corresponding
effective potentials for the cases of weakly, intermediate and strongly
coupled plasmas. We also want to compare the effective potentials with
the statically screened Coulomb potentials valid for weakly coupled plasmas.
To this aim we rewrite \req{e:SSCP} in a dimensionless form and
rewrite the Debye screening length in our dimensionless units,
\begin{equation}
        \Phi_{\mathit{ij},\mathrm{SSCP}} = \frac{Z_i Z_j \Gamma_0}{r}
\exp{\left(-\frac{r}{r_\mathrm{D}}\right)}, \quad
\frac{1}{r_\mathrm{D}}  = \sqrt{3 \Gamma_0 \overline{Z^2}}.
\label{e:SSCP-WC}
\end{equation}
%

\begin{figure*}
  \centering
  \includegraphics[width=.48\textwidth]{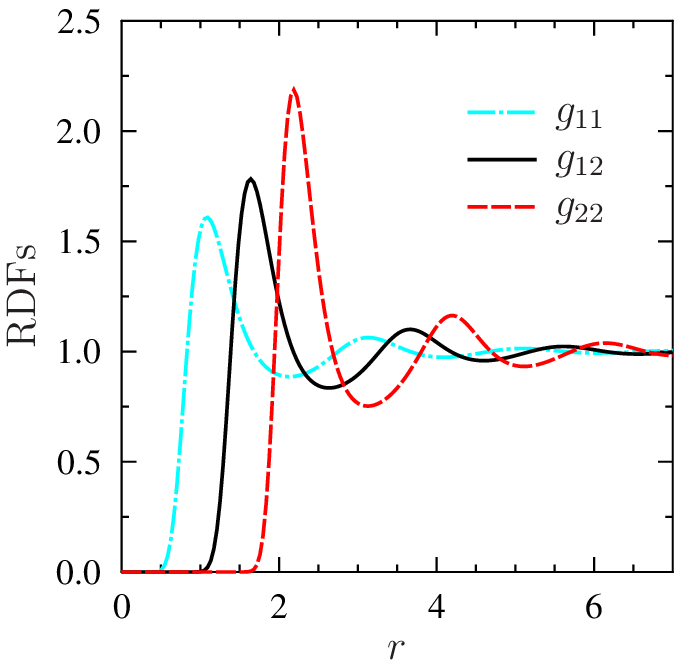}%
  \includegraphics[width=.48\textwidth]{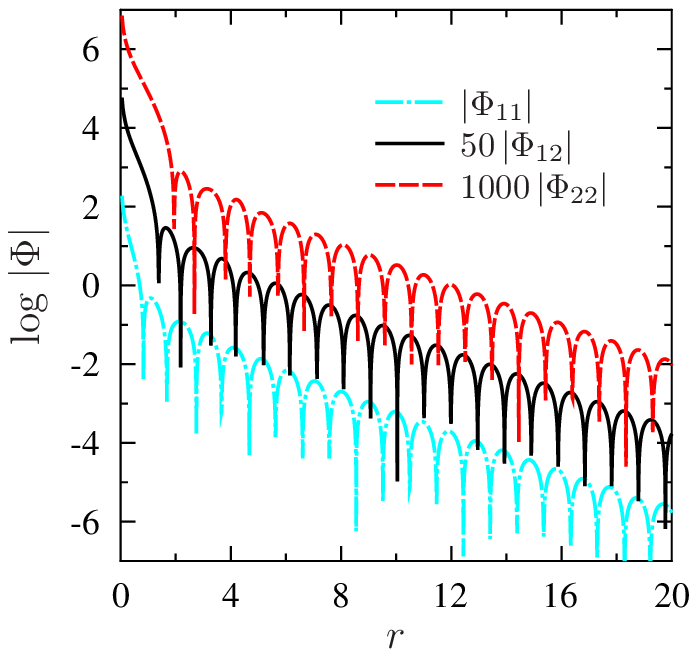}
  \caption{The same as Fig.~\ref{fig:RDF-Phi-WC}, but for a strongly
coupled plasma regime at $\Gamma_0 = 10~ (\bar{\Gamma}\approx 123)$. For
better visual representation $\Phi_{12}$ is multiplied by 50 and
$\Phi_{22}$ by 1000.}
  \label{fig:RDF-Phi-SC}
\end{figure*}

The results are demonstrated on Figs.~\ref{fig:RDF-Phi-WC},
\ref{fig:RDF-Phi-Inter} and \ref{fig:RDF-Phi-SC} for H/C mixture with
$x_\mathrm{H} = 0.6$ and $x_\mathrm{C} = 0.4$. Fig.~\ref{fig:RDF-Phi-WC}
shows weakly coupled plasma regime with $\Gamma_0 = 0.01$ which
corresponds to $\bar{\Gamma}\approx 0.12$. The RDFs are displayed on the
left panel and corresponding absolute values of effective potentials on
the right panel (effective potentials can change signs as a result of
strong ion correlations in non-ideal plasmas; we plot their absolute
values). The dotted lines show the comparison with the corresponding
statically screened Coulomb potentials [Eq.\ \eqref{e:SSCP-WC}].  In the
weak coupling case (Fig.~\ref{fig:RDF-Phi-WC}), the RDFs demonstrate a
behavior typical for an almost ideal gas. The effective potentials 
agree well with the statically screened Coulomb potentials (some
deviations at $r > 10$ are due to numerical issues, but they are 
unimportant, as the absolute values of the potentials in this region are
very small, $\left| \Phi \right| \lesssim 10^{-6}$).  We emphasize that
hypernetted chain calculation employed only the bare Coulomb potential. Yet,
the obtained effective potentials demonstrate the ``correct''  screening
for weakly coupled plasma. In other words, this method allows one to
calculate the ion screening from first principles.

Fig.~\ref{fig:RDF-Phi-Inter} shows the same as Fig.~\ref{fig:RDF-Phi-WC}
but for an intermediate coupling regime ($\bar{\Gamma}\approx 1.2$). The
RDFs still demonstrate gas-like behavior, but the effective potentials
start to deviate from the Debye-screened Coulomb potentials and 
oscillate at $r\gtrsim1$. 

The case of strong coupling (Coulomb liquid) is shown on Fig.\
\ref{fig:RDF-Phi-SC}. The system is the same as in Figs.
\ref{fig:RDF-Phi-WC} and \ref{fig:RDF-Phi-Inter} but with $\Gamma_0 =
10$  ($\bar{\Gamma}\approx 123$). For better visual representation
$\Phi_{12}$ is multiplied by 50 and $\Phi_{22}$ by 1000. A comparison to
Debye-screened Coulomb (Yukawa-like) potentials is not displayed, since
the Debye approximation fails for strongly coupled plasmas. The behavior
of the RDFs  is characteristic to condensed matter (i.e., to strongly
correlated systems such as liquids or solids). The effective potentials
oscillate as the results of ion-ion correlations.

\subsubsection{Computing diffusion coefficients}
\label{ssec:3:DiffCoeff}

Here we return to the ordinary physical units.

The standard Chapman-Enskog procedure gives the following expression
for the leading order approximation to the mutual diffusion coefficient in a binary mixture
\citep{CC52,Hirsh54,Paquette86}:
\begin{equation}
D_{12} = \frac{3}{16} \frac{k_{\mathrm{B}} T}{m_{12} \nnii}
 \frac{1}{\Omega_{12}^{(1,1)}}.
\label{e:D-general-def}
\end{equation}
where $m_{12}$ is again the reduced mass of colliding ions and 
$\Omega_{12}^{(1,1)}$ is an effective average product of  the cross
section and relative velocity, which is  related to the transport cross
section after integrating over a Maxwellian velocity distribution. It is
given by
\beq
\Omega_{12}^{(\xi,\zeta)} = 
\sqrt{\frac{\kB T}{2\pi m_{12}}}
\int_0^{\infty}\! \exp{\left(-y^2\right)}
y^{2\zeta+3} Q_{12}^{(\xi)}(y) \, \dd y,
\label{e:Omega-dim-def}
\eeq
where
\beq
Q_{12}^{(\xi)}(u) = 2 \pi \int_0^{\infty} \!
\left[1 - \cos^\xi{\left(\chi_{12}(b,u)\right)} \right] b\,
 \dd b
\label{e:Q-dim-def}
\eeq
is an effective cross section at a given energy (a given relative
velocity), $b$ is an impact parameter, $u$ is a dimensionless relative
particle velocity at infinity (in the units of $\sqrt{\slfrac{2
k_\mathrm{B} T}{m_{12}}}$),
\beq
\chi_{12}(b,u) = \left| \pi- 2b \int_{r_{12}^{\mathrm{min}}}^{\infty} \!
\frac{\dd r}{r^2 \sqrt{1 - \frac{b^2}{r^2} - \frac{\phi_{12}}{u^2}}}
\right|
\label{e:chi-dim-def}
\eeq
is the scattering angle,
$\phi_{12}(r)$ is the interaction potential
between the ions  $1$ and $2$, and $r_{12}^{\mathrm{min}}$ is the
distance of the closest approach,
that is the root of the denominator
under the integral \eqref{e:chi-dim-def}.

Let us introduce the ``hydrodynamic'' plasma frequency of the ion
mixture (see, e.g., \citealt{HJM85}),
\begin{equation}
\omega_\mathrm{pi} = \sqrt{\frac{4 \pi \nnii \bar{Z}^2 e^2}{\bar{A}m_\mathrm{u}}}.
\label{e:wp}
\end{equation}
We will express the mutual diffusion coefficients in units of
$\omega_\mathrm{pi} a_\mathrm{i}^2$,
\begin{equation}
D_{12}^{*} = \frac{D_{12}}{\omega_\mathrm{pi} a_\mathrm{i}^2}.
\label{e:D-dim-def}
\end{equation}
For a weakly coupled binary mixture,
the dimensionless 
diffusion coefficient \eqref{e:D-dim-def} is calculated
analytically [\citealt{CC52,Hirsh54}; cf.~\req{e:D-Simple}],
\begin{equation}
D_{12}^{*} =  \sqrt{\frac{\pi}{6}} \frac{1}{\Gamma_0^{\nicefrac{5}{2}}}
\sqrt{\frac{\bar{A}\left(A_1 + A_2 \right)}{\bar{Z}^2 A_1 A_2}}
\frac{1}{Z_1^2 Z_2^2 \Lambda^{\mathrm{(cl)}}},
\label{e:D-WC-def}
\end{equation}
where $\Lambda^{\mathrm{(cl)}}$ is the ``classical'' Coulomb
logarithm for a weakly coupled plasma,
\begin{equation}
\Lambda^{\mathrm{(cl)}} =
 \ln{\left( \frac{1}{\sqrt{3} \Gamma_0^{\nicefrac{3}{2}}
    Z_1 Z_2 \sqrt{\overline{Z^2}}} \right)}\,, 
\quad
 \Lambda^{\mathrm{(cl)}} \gg 1.
\label{e:Lambda-WC-def}
\end{equation}

The algorithm of calculating $D_{12}^*$ for a plasma of arbitrary
Coulomb coupling strength consists of three steps. First, one uses the
hypernetted chain method (Section~\ref{ssec:3:HNC}) to compute the
RDFs. Then one determines the effective
potential $\Phi_{12}(r)$ from Eqs.\ \eqref{e:RDF-phi-TCP} and
substitutes it instead of $\phi_{12}(r)$ in the integral
\eqref{e:chi-dim-def}. Finally, $D_{12}^*$ is calculated from
Eqs.\ \eqref{e:D-dim-def}, \eqref{e:Omega-dim-def} and
\eqref{e:Q-dim-def}.

Such calculations of the mutual diffusion coefficients have been
performed by \citet{BY14b} for the $^{1}$H -- $^{4}$He, $^{1}$H -- $^{12}$C, $^{4}$He
-- $^{12}$C, $^{12}$C -- $^{16}$O and $^{16}$O -- $^{79}$Se mixtures
at different values of $\Gamma_0$ and $x_1$. For convenience of
applications, the calculated values
of  $\Lambda_{\mathrm{eff}}$ have been approximated by an analytic expression. The Coulomb
logarithm is defined through $D_{12}^{*}$ as
\begin{equation}
\Lambda_{\mathrm{eff}} = \sqrt{\frac{\pi}{6}} \frac{1}{D_{12}^* Z_1^2 Z_2^2
  \Gamma_0^{\nicefrac{5}{2}}} \sqrt{\frac{\bar{A}\left(A_1 + A_2 \right)}
  {\bar{Z}^2 A_1 A_2}}.
\label{e:Lambda-eff-def}
\end{equation}
Then $D_{12}^{*}$ is given by \req{e:D-WC-def} at any coupling, but with
the classical Coulomb logarithm $\Lambda^{\mathrm{(cl)}}$
replaced by the effective Coulomb
logarithm $\Lambda_{\mathrm{eff}}$.
This approach is in line with that by   \citet{Khrapak13,BD13}
(see Section~\ref{ssec:3:DiffCoeffReview}). 
The diffusion coefficient is again given by \req{e:D-Simple},
although $\Lambda_\mathrm{eff}$ is now more refined. 
It is convenient, because $\Lambda_{\mathrm{eff}}$ is a slowly varying 
function of plasma parameters (especially of particle fractions).
It has been
approximated by a universal expression 
\eqref{e:Lambda-fit} (\ref{app:Lambda_Eff}). 
The expression contains five fit parameters for any
binary mixture (listed in Table \ref{tab:DiffCoeff-FitParam}
of \ref{app:Lambda_Eff}). This gives a unified 
description of mutual diffusion coefficients for
 binary mixtures. 

\begin{figure}[!]
  \centering
  \includegraphics[width=.48\textwidth]{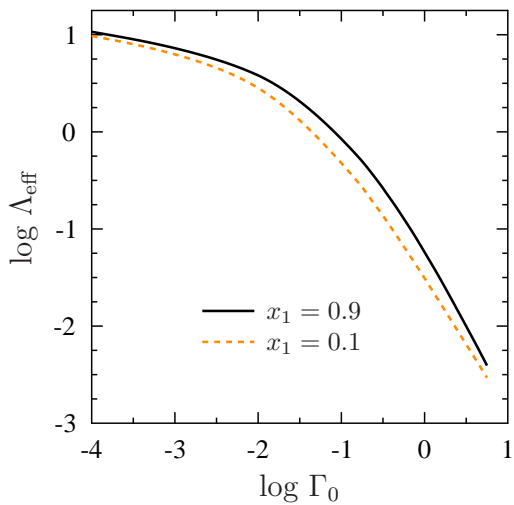}
  \caption{Effective Coulomb logarithm $\Lambda_{\mathrm{eff}}$ versus $\Gamma_0$,
    calculated from Eq.\ \eqref{e:Lambda-eff-def}, for an H\,--\,C
     mixture.
}
  \label{fig:Lambda}
\end{figure}

Fig.~\ref{fig:Lambda}
 shows $\Lambda_{\mathrm{eff}}$ as a function
of $\Gamma_0$ for an H\,--\,C mixture. One can easily
identify the regions of weak and strong
Coulomb couplings (\ref{app:Lambda_Eff}), as well as the transition region of moderate coupling.

\subsubsection{General features of diffusion coefficients}
\label{ssec:3:Discussion}

There is no rigorous proof of the existence of an effective
ion-ion potential, which would properly include all many-body effects (correlations)
between the ions in a strongly coupled Coulomb plasma (or, generally, between
particles in a liquid). Moreover, it is likely that such a potential does not 
exist. Nevertheless, the method of effective potentials is a promising
tool for solving many problems of physics of strongly
coupled plasmas with reasonable accuracy (see
\citealt{BD13}).
  
We have employed the standard hypernetted chain method
of radial distribution functions. Although there exist
modified versions of this method (e.g., \citealt{II83}), the
accuracy of the standard method is sufficient for calculating the diffusion
coefficients.  As seen from Fig.~2 of \cite{BD13}, even the use of
`exact' RDFs, calculated by molecular dynamics, 
gives the diffusion coefficients that are 
close to those determined by the Chapman-Enskog method.

\begin{figure}[!]
  \centering
  \includegraphics[width=.48\textwidth]{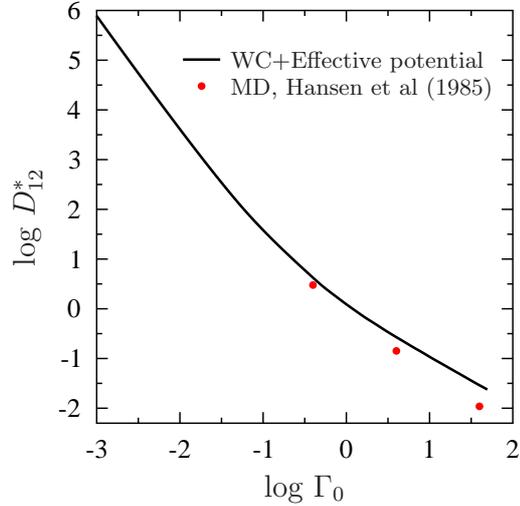} 
\caption{Reduced diffusion
  coefficient $D_{12}^*$ in an H/He mixture ($x_\mathrm{H} =
  x_\mathrm{He} =0.5$) as a function of $\Gamma_0$ (the solid curve), as
  compared with the molecular dynamics (MD) calculations by
  \cite{HJM85}. Note that here $\bar\Gamma=2.48\,\Gamma_0$.
  }
  \label{fig:DiffCoef}
\end{figure}

As seen from Fig.~ \ref{fig:DiffCoef}, 
the diffusion coefficients  $D_{12}^*$,
computed via the effective potentials, are systematically
higher than the values of $D_{12}^{*\mathrm{MD}}$, calculated by
\citet{HJM85} via the molecular dynamics; the difference increases
with growing $\Gamma_0$. The same behavior was
mentioned by  \citet{BD13} (their Fig.~2).  
A comparison of the results of molecular dynamics simulations
\citep{RNZ95,BP87} with  the results of \citet{BY14b} described here
leads to similar conclusions. Apparently the 
discrepancy at strong coupling arises from the
approximate character of the effective potentials method.
    
There are two ways to improve  the accuracy of the effective potentials
method: either by using second-order corrections to the diffusion
coefficients in the framework of the standard Chapman-Enskog method
(Sonine polynomials expansion, see \citealt{CC52, Hirsh54} and also
Eqs.\ (40)--(45) in \citealt{BY14b}) or by improving the standard
Chapman-Enskog method itself. The latter possibility was employed by
\cite{BD15}, who proposed to apply the modified Enskog theory of dense
gases to strongly coupled Coulomb plasmas. This allowed them to somewhat
reduce the discrepancy between the effective potential method and
molecular dynamics results.

The effect of the second order correction was considered by
\citet{BY14b} and in more details by \citet{SBD17} who demonstrated that
with the increase of $\bar{\Gamma}$ the second order correction quickly
vanishes (see Fig.~3 of \citealt{SBD17}). Therefore, it is reasonable
to omit these corrections; the accuracy of the results is limited by the
accuracy of analytic approximations and by the method of effective
potentials itself. 

A comparison of sparse calculations of the self-diffusion
coefficients in binary ionic mixtures with those
 in one-component ion plasmas reveals  that
the method of effective potentials is more accurate for 
one-component plasmas than for binary mixtures. Relations similar to
Eq.\ \eqref{e:D12=D11+D22} can be also derived for the diffusion
coefficients obtained via effective potentials (see \citealt{BY14b} for
details).

It would be important to confirm the validity of the effective
potential  approach and formulate the conditions at which it is
reasonably accurate. We have already demonstrated that the approach
becomes less accurate with increasing Coulomb coupling. When the
temperature drops to the melting temperature $T_\mathrm{m}$, the quantum
effects in ion motion may become important for many properties of the
matter (e.g., \citealt{HPY07}). In particular, they can be important for
diffusion. This effect has not been studied in the literature in detail.
As long as the quantum effects are neglected, the method of effective
potentials appears to be quite adequate (although the inclusion of
quantum effects would be desirable).

The main advantage of the presented results is their simplicity,
uniformity, and convenient fit expressions. Another advantage is that
the method of effective potentials can be easily generalized for
calculating other kinetic properties of strongly coupled ion plasmas,
for instance, diffusion and thermal diffusion coefficients in
multicomponent mixtures of ions which are often needed for applications
but almost not explored in the astrophysical literature. This is
especially important for thermal diffusion coefficients. In strongly
coupled systems, they are usually calculated using non-equilibrium
molecular dynamics (e.g., \citealt{SDFR98} and \citealt{EM07}), which is
a more complicated numerical problem than the modeling based on standard
(equilibrium) molecular dynamics.

\cite{KBD17} developed further 
the idea of effective Coulomb logarithms and applied the
 effective potential method to the
calculation of thermal diffusion coefficients (as
proposed by \citealt{BY14b}). The state of the art
of the effective potentials approach is described by \cite{BD19}.

The above calculations have been conducted
assuming a rigid (incompressible) electron background.
The results can be generalized for the case of compressible
electron background of any degeneracy and relativity, but the 
effects of electron polarization in dense matter of neutron stars 
are expected to be weak.

\subsection{Estimates of diffusive velocities}
\label{ssec:2:Results}

The diffusion coefficients   can be used for  estimating diffusion
velocities in heat blanketing envelopes of neutron stars.    Although
the diffusive current  \eqref{e:J2-Vel} has the standard form, it
contains a new Coulomb term [Eqs. \eqref{e:uC-SC} and \eqref{e:uC-WC} in
the limits of strong and weak couplings]. The effects of Coulomb forces
on ion separation were first described by \citet{DeBlasio00} and later
studied by \citet{CB10}, who considered equilibrium isothermal
configurations of ion mixtures with account for the Coulomb effects (the
method of ``chemical'' equilibrium). Now this result can be extended to
the case of non-equilibrium and/or non-isothermal systems.

As mentioned in Section~\ref{ssec:2:IsothermDiffFlux-SC}, the Coulomb
contribution is especially important for the ion mixtures with the same
charge-to-mass ratio, such as He, C, and O. The gravitational
contribution \eqref{e:ug} for such ions is non-vanishing only owing to
the mass defect. In strongly coupled plasmas, typical for the
neutron-star envelopes, it is about one order of magnitude smaller than
the Coulomb contribution. On the other hand, for the mixtures of ions
with different charge to mass ratio, the gravitational contribution
dominates and is typically one order of magnitude larger than the
Coulomb contribution. For these mixtures, the Coulomb contribution can
be neglected.

The estimates show \citep{BY13} that although the velocities of the
Coulomb separation of ions in the neutron star envelopes are not
negligible, the diffusive energy release (\ref{e:Q}) is small and cannot
reheat cooling middle aged (ages $\lesssim 10^5-10^6$ yr) neutron stars.

\begin{figure*}[!]
  \centering
  \includegraphics[width=.48\textwidth]{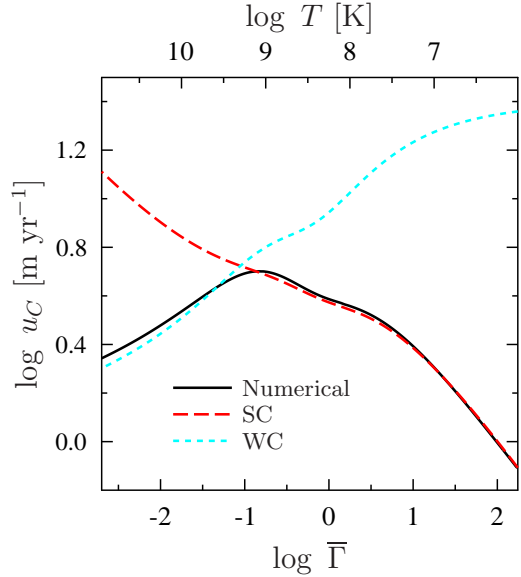}
  \caption{Coulomb separation velocities of ions in the
    $^{4}$He\,--\,$^{12}$C mixture in the  blanketing envelopes of
    neutron stars as a function of the mean ion coupling parameter 
    \eqref{e:GammaAv} (the lower axis) or temperature (the upper axis);
    $x_{\mathrm{He}} = 0.6$, $x_{\mathrm{C}} = 0.4$, $\rho = 10^6$ \gcc,
    and $g_\mathrm{s} = 2\times 10^{14}$~cm s$^{-2}$. SC  refers to a
    strongly coupled plasma, Eq.\ \eqref{e:uC-SC}; WC to a weakly
    coupled plasma, Eq.\ \eqref{e:uC-WC}. (After 
    \citealt{BY14a}.)}
  \label{fig:UC-Gamma}
\end{figure*}

Fig.~\ref{fig:UC-Gamma} shows the velocity of the Coulomb separation of
ions in the He\,--\,C mixture as a function of the average Coulomb
coupling parameter \eqref{e:GammaAv} or temperature. The estimates are
performed using the formulae presented in
Section~\ref{ssec:2:IsothermDiffFlux-SC}.  In particular, we use a
simplified diffusion coefficient  (\ref{e:D-Simple}), which gives nearly
the same results as the more refined theory described in
Section~\ref{sec:3}.  The plasma density and gravitational acceleration
are typical for the heat blankets of neutron stars, $\rho = 10^6$~\gcc\
and $g_\mathrm{s} = 2\times 10^{14}$~cm s$^{-2}$. The composition
($x_{\mathrm{He}} = 0.6$, $x_{\mathrm{C}} = 0.4$) is assumed to be
uniform. Note that this figure is for illustrative purposes only as the
temperatures needed to get $\bar{\Gamma} \lesssim 0.01$ are too high and
neither helium, nor carbon can actually exist at these temperatures (due
to nuclear burning).

Specifically, Fig.~\ref{fig:UC-Gamma} exhibits  the mean (diffusive)
velocity of carbon ions  $\bm{u}_\mathrm{C}$ with respect to the matter
as a whole  (which is at rest because of the overall hydrostatic
equilibrium). In practice, one often introduces the relative diffusion
velocity  $\bm{u}=\bm{u}_1 - \bm{u}_2$ of one ion species with respect
to the second one. It is easily recovered from $\bm{u}_1$  with the use
of relation $m_1 n_1 \bm{u}_1+ m_2 n_2 \bm{u}_2=0$; it is of the same
order of magnitude.  The solid line is calculated numerically from the
general expression \eqref{e:d-C} and from the expressions for 
$\mu_j^{(\mathrm{C})}$ given by \citet{PC10} (see
Section~\ref{ssec:4:DiffEquilEnvelopes} for more details). The
long-dashed line corresponds to the limit of strongly coupled plasma,
Eq.~\eqref{e:uC-SC}; the short-dashed line is for the weak-coupling
limit, Eq.\ \eqref{e:uC-WC}. One can see that the transition  between
the regimes of weak and strong coupling takes place at $\bar{\Gamma}
\sim 0.1$. The limiting cases of weak and strong coupling appear in good
agreement with numerical calculations.  In the transition region, the
velocity $u_\mathrm{C}$ reaches maximum at nearly the same 
$\bar{\Gamma} \sim 0.2$. This maximum occurs because $J ^{(\mathrm{C})}
\sim Dd ^{(\mathrm{C})}$. With increasing  $\bar{\Gamma}$, the quantity
$d ^{(\mathrm{C})}$ grows, and the diffusion coefficient $D$ becomes
lower. The competition between the diffusion coefficient $D$ and the
Coulomb contribution to deviations from equilibrium (to $d
^{(\mathrm{C})}$) creates the maximum in the velocity curve. Therefore,
although the Coulomb effects are most noticeable in the regime of strong
coupling, the velocity of the Coulomb separation reaches maximum at
intermediate coupling, $\bar{\Gamma} \sim 0.2$.

According to Fig.~\ref{fig:UC-Gamma}, the velocity of the Coulomb
separation in the heat blankets of neutron stars
can be as high as $\sim 5$ m~yr$^{-1}$. Taking
into account characteristic depths of the heat blankets, 
a strong separation there is
expected to occur during decades.  The Coulomb separation in a $^{4}$He -- $^{12}$C mixture may 
also be efficient in white dwarfs \citep{BY13,BY14b}.  

Taking $\bar{\Gamma}=1$ (i.e., $T=1.78 \times 10^8$ K),
from Fig.~\ref{fig:UC-Gamma} we have $u_\mathrm{C} \sim 3$ m yr$^{-1}$. 
Note, however,
that direct gravitational separation in a 
plasma, if allowed, would proceed much faster. If we
took the same $\rho=10^6$ \gcc, $T=1.78 \times 10^8$ K, and
$g_\mathrm{s} = 2 \times 10^{14}$ cm s$^{-2}$ but consider the
$^{12}$C--$^{26}$Fe mixture (where $Z_1/A_1 \neq Z_2/A_2$
and the direct separation operates) and put, for instance, $x_\mathrm{C}=0.4$, we would have 
much larger separation velocity $u_\mathrm{Fe}\sim 15$ m yr$^{-1}$. 
Further applications of Eq.\ \eqref{e:J} 
to heat blankets of neutron stars are discussed below.

\section{Diffusive heat blanketing envelopes of neutron stars}
\setcounter{equation}{0}
\label{sec:4}

\subsection{Constructing diffusion-equilibrium envelopes}
\label{ssec:4:DiffEquilEnvelopes}

In this section we follow \citet{BPY16} and directly focus on diffusion in 
neutron star heat blankets with the aim to 
determine $\Ts-\Tb$ relations. Here we
study blanketing envelopes made of binary ion
mixtures in diffusive equilibrium. They are
not isothermal (not in a state of full 
thermodynamic equilibrium) because of the heat flux from
stellar interiors to the surface.

We will adopt the same assumptions as in Sections~\ref{ssec:2:DiffFluxEqs} and \ref{ssec:2:IsothermDiffFlux-SC}.
In addition, we neglect the effect of
thermal diffusion. The validity of these assumptions will be
discussed later. With the formulated assumptions, we come
to \req{e:J2-Gen} for the diffusion currents of ions.
Now, however, we do not  restrict ourselves
to the approximations of strongly or weakly coupled plasma,
but consider a plasma with arbitrary Coulomb coupling in the presence of
temperature gradients. 

The diffusion equilibrium implies the absence of diffusion currents.
According to Eq.\ \eqref{e:J2-Gen} this is equivalent to the
condition $\bm{d}_1 = 0$. In addition, since 
$\bm{d}_2 = -\bm{d}_1$ and the electrons are in
the state of quasi-equilibrium, $\bm{d}_\mathrm{e} = 0$ (see Sections~\ref{ssec:2:DiffFluxEqs} and \ref{ssec:2:IsothermDiffFlux-SC}),
we come to $\bm{d}_1 = \bm{d}_2 =
\bm{d}_\mathrm{e} = 0$. Furthermore, assuming an overall hydrostatic
equilibrium of the envelope and Eq.\
\eqref{e:d-short}, we obtain
$\widetilde{\bm{f}}_1 = \widetilde{\bm{f}}_2 =\widetilde{\bm{f}}_\mathrm{e} =
0$. These are the equations of the diffusion equilibrium.

Using Eqs.\  \eqref{e:GenForce},
\eqref{e:Force} and \eqref{e:MuPDeriv} [cf.{} \req{e:eE}],
we arrive at the basic system of equations 
\begin{equation}
\widetilde{\bm{\nabla}}\mu_\mathrm{e} = -e\bm{E},
\quad
\widetilde{\bm{\nabla}}\mu_j = m_j \bm{g} + Z_j e\bm{E} ,
\label{e:SysEquilEq}
\end{equation}
where $\widetilde{\bm{\nabla}}$ is defined as
\begin{equation}
\widetilde{\bm{\nabla}}\mu_\alpha \equiv \sum_j \frac{\partial \mu_\alpha}{\partial n_j} \bm{\nabla} n_j + \frac{\partial P}{\partial T} \left( \sum_j n_j \frac{\partial \mu_\alpha}{\partial n_j} \right) \left(\sum_k n_k \frac{\partial P}{\partial n_k}\right)^{-1} \bm{\nabla}T.
\label{e:CorrGrad}
\end{equation}
The indices $j,k$ take the values 1 and 2, while the index $\alpha$ can
be ``e'', 1, or 2. The chemical potentials and  the pressure are assumed
to be known, together with their derivatives, as functions of
temperature and number densities of ions. The quantities which are
unknown include $\bm{\nabla}n_j$ and $e\bm{E}$. Interestingly, one does
not need to know the diffusion coefficients themselves, under the
formulated assumptions. However, generally, without neglecting the
thermal diffusion, one needs the diffusion and thermal diffusion
coefficients to determine an equilibrium configuration.

To close the system of Eqs.\ \eqref{e:SysEquilEq} and
\eqref{e:CorrGrad} one should add the expressions 
(\ref{dP/dz}) and (\ref{therm-Tcrust}) 
for the hydrostatic equilibrium  and for the radial
thermal heat flux density $F_T$ 
in the local plane parallel approximation,
\begin{equation}
\frac{\dd P}{\dd z}=g_\mathrm{s} \rho,
\quad
\kappa \frac{\dd T}{\dd z}= F_T.
\label{e:FlatEnvelope}
\end{equation}
These are the equations for calculating diffusively equilibrated envelopes.
Their integration should be carried out from the radiative
surface ($T=\Ts)$ inside the envelope to its bottom, 
$\rho = \rhob$. In this way one can calculate all physical parameters
in the envelope  (particularly, $T$, $P$ and
$n_\alpha$) as a function of $z$ or, equivalently, as a
function of $\rho$. The calculations give the required 
$\Ts - \Tb$ relations.

The EoS and thermodynamic functions have
been taken from \citet{PC10}
with the improvements mentioned in \citet{PC13}.\footnote{A
corresponding Fortran code is accessible at
\url{http://www.ioffe.ru/astro/EIP/}.
\label{EIP}
}
The thermal conductivity $\kappa$ has been taken as
a sum of the electron conductivity 
$\kappa_\mathrm{e}$ and the radiative one $\kappa_\mathrm{r}$.
These conductivities have been determined using analytic
approximations described in Appendix~A of
\citet{PPP2015}\footnote{A corresponding Fortran code
  is accessible through
\url{http://www.ioffe.ru/astro/conduct/}};
the radiative thermal conductivity is constructed using
Rosseland spectral opacities presented in 
the Opacity Library \citep{OPAL},\footnote{Available
at the web page
\url{http://mesa.sourceforge.net/index.html}
of the \textsc{mesa} project (\citealt{MESA19} and references therein).
}
or within the Opacity
Project (\textsc{op}, \citealt{OP}).\footnote{Available at
extrapolation along the opacity tables has been carried out
by the method described by PCY97.\footnote{The 
extrapolation method beyond these tables has been
improved by \citet{PC18}.}

Equations
\eqref{e:SysEquilEq}  are analogous to the conditions
of chemical equilibrium presented by  \citet{CB10}. The difference
consists in the presence of terms containing $\bm{\nabla} T$
in Eqs.\ \eqref{e:GenForce} and
\eqref{e:CorrGrad}, which were neglected
by \citet{CB10}. 

Before discussing the results, let us describe the parameters of
heat blanketing envelopes.

\subsection{Models of heat blanketing envelopes}
\label{ssec:4:HeatBlanketsModel}

\citet{BPY16} have constructed blanketing
envelope models, which consist of binary ion mixtures of
H\,--\,He, He\,--\,C, or C\,--\,Fe. As mentioned in 
Section~\ref{subsub:basic_eqns}, the envelope
models are self-similar,
being dependent on the surface gravity $g_\mathrm{s}$. 
The results obtained at one value of $g_\mathrm{s}$ can be easily rescaled
to another value.  \citet{BPY16}
have used $g_{s0}=2.4271\times10^{14}$ cm s$^{-2}$,
which corresponds to the canonical neutron star model with
$M=1.4\,M_\odot$ and $R=10$ km. For two realistic EoSs
of neutron star interiors\footnote{One should not confuse this
EoS with the EoSs in heat blanketing
envelopes} APR \citep{APR98} and BSk24 \citep{Pearson_ea18},
 such surface gravity corresponds to stars
with   $M=1.73\,M_\odot$, $R=11.3$
km and with $M=2.00\,M_\odot$, $R=12.3$ km, respectively.

All 
analyzed diffusively equilibrated envelopes demonstrate
stratification of elements. Lighter ions concentrate
in the upper layers of the envelopes while the heavier ions
are localized near the envelope bottom.
An essentially binary mixture (a transition layer) is formed
between the upper and lower layers. A schematic plot of a heat blanketing
envelope is presented in Fig.\
\ref{fig:SchemeEnvelopeGeom}, which shows also the directions
of electric field and heat flux.

The thickness of the transition layer depends on many parameters,
including the types of ion species, temperature and the depth from the
stellar surface. The mixtures under consideration are significantly
different. In the  H\,--\,He and C\,--\,Fe mixtures, the
effective ``molecular weights'' of the ions are noticeably
different; accordingly, the gravitational separation of the ions
is leading there (see Section~\ref{sec:2}). In contrast, in the He\,--\,C 
mixtures the ``molecular weights''
of the ions are equal, and the ions are separated by the Coulomb
mechanism which is weaker than the gravitational one. Therefore 
the transition layer in the  He\,--\,C mixture should be much wider
than in the H\,--\,He and
C\,--\,Fe mixtures (as confirmed by the calculations described
below).

\begin{figure}[!]
  \centering
  \includegraphics[width=.48\textwidth]{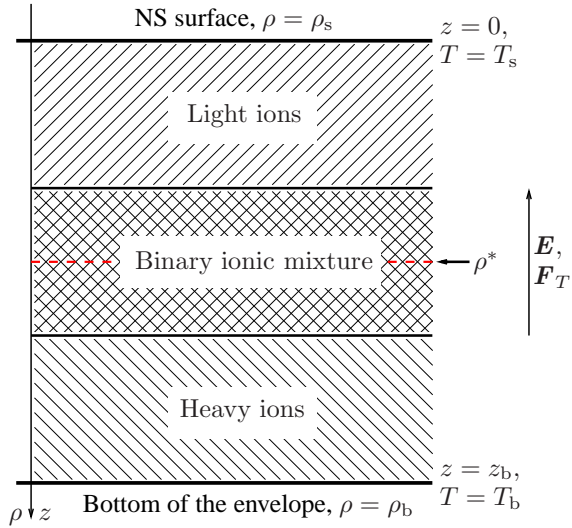}
  \caption{Scheme of a neutron star blanketing envelope in
    the plane-parallel approximation (see the text for details).}
  \label{fig:SchemeEnvelopeGeom}
\end{figure}

For further analysis one needs to introduce a parameter which would
characterize the amount of lighter and heavier ions in the envelope.
PCY97 used the parameter $\eta$,  \req{eq:eta(z)}, which is directly
related to the pressure $P^*$ at the bottom of the outer layer with
mass  $\Delta M$, setting $\Delta M$ equal to the total mass of light
ions in the envelope (see Sections~\ref{therm-blanket}
and~\ref{therm-env}). Instead,  \citet{BPY16} used an equivalent
parameter $\rho^*$, related to $\eta$ through 
\begin{equation}
\xr(\rho^*) \equiv \xr^*= 1.0088 \,
 \left(\rho_6^* {Z}/{A}\right)^{{1}/{3}},
\label{e:RelParam}
\end{equation}
where $\rho_6^*\equiv\rho^*/10^6$ \gcc{}  and $\xr^*$ is the solution of
\req{e:dM/M(rho)} for the given $\eta$. If $P^*$ corresponds to a
strongly degenerate layer of the envelope ($T_\mathrm{F}\gg T$), then
$\rho^*$ is almost equal to the mass density at $P=P^*$, that is at the
bottom of the outer layer whose mass  $\Delta M$ equals the total mass
of the lighter ions in the considered binary mixture. In the
non-degenerate matter, $\rho^*$ does not have a straightforward physical
meaning.

For a given chemical composition and the surface gravity $g_\mathrm{s}$,
the heat blanket is characterized by the surface temperature $\Ts$ and
by the amount of lighter ions (i.e., by $\Delta M$, $\eta$, or
$\rho^*$), and also by the bottom density $\rhob$ (see below).
Naturally, these parameters are restricted
(Section~\ref{subsub:chemicalcompos}).  Another restriction implies
$\rho^* \lesssim \rhob$; otherwise the heat blanket would contain only
light ions. Besides, it would be meaningless to consider $\Delta M$
values smaller than the mass of the neutron star atmosphere, which
implies $\Delta M\gtrsim 10^{-18}-10^{-16}$ $M_\odot$. The reported
calculations have been mostly conducted at those temperatures and
densities where given elements can survive over long time. 

The surface temperature $\Ts$ has been varied from $\sim 0.3$ MK to
$\sim 3$ MK, a typical range of observable surface temperatures of
isolated neutron stars \citep{COOLDAT}. In accordance with the above
restrictions,  $\rho^*$ has been varied up to $\sim 10^6$ \gcc\ for
H\,--\,He mixtures;  up to  $\sim 10^8$ \gcc\ for He\,--\,C mixtures; 
and  up to $\sim 10^9$ \gcc\ for C\,--\,Fe mixtures.

The choice of $\rhob$ requires special comments (Section~\ref{sub:therm-outlook}).  From physical point of view, $\rhob$ can be
chosen at such $\rho=\rhob^*$, that at higher  $\rho$  the envelope
becomes nearly isothermal. However in practice, this choice is
inconvenient since such $\rhob^*$ depends on many parameters, first of
all on  $\Ts$, as demonstrated in Fig.~\ref{fig:Rho_b}. 
 In that
figure, we show the dependence of $\rhob^*$ (determined as the density
at which $T$ is only several percent below its limiting value) on $\Ts$
for the envelope which contains pure iron (the solid curve) and pure
carbon (the dashed curve). One sees that when $\Ts$ drops by one order
of magnitude, $\rhob^*$ drops by $\sim 5$ orders.  

\begin{figure}[!]
  \centering
  \includegraphics[width=.48\textwidth]{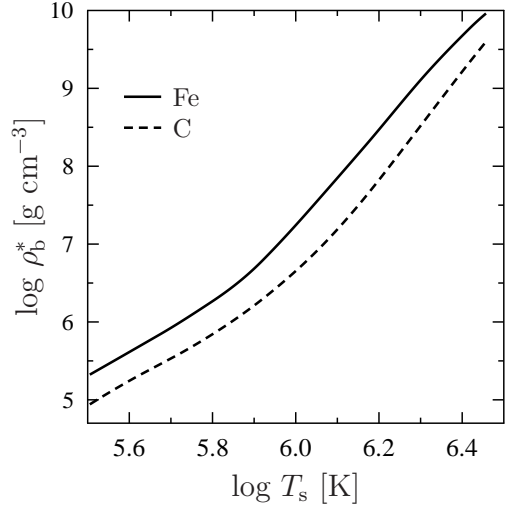}
  \caption{Effective density $\rhob^*$
    at the ``physical'' heat blanket bottom, where plasma becomes 
    isothermal with an accuracy within 10\%,
    as a function of $\Ts$ for the canonical neutron star model.
    The solid line corresponds to the blanket made of pure iron while
    the dashed line is for pure carbon. After \citet{BPY16}.
 }
  \label{fig:Rho_b}
\end{figure}

The figure clearly shows that the {real physical} heat blanket  becomes
thinner when the star cools. To simplify further use of blanket models,
instead of $\rhob^*$ one usually \emph{assumes} a fixed value of
$\rhob$. The most used (standard) value is $\rhob = 10^{10}$ \gcc{}
(e.g., \citealt{GPE83}). This value can be varied depending on a
specific problem. Calculations of thermal evolution of neutron stars
employ fixed  $\Tb-\Ts$ relations as a boundary condition for finding
the temperature distribution in  stellar interiors
(Section~\ref{sub:evolution}).  The higher $\rhob$, the simpler the
solution of the latter problem. On the other hand, the higher $\rhob$, 
the larger is the heat diffusion time $t_\mathrm{th}$ through  this
envelope. Clearly, one cannot rely on evolutionary simulations  over
timescales shorter than  $t_\mathrm{th}$. An estimate for the canonical
neutron star model with $\Ts = 1$ MK and the iron blanketing envelope
yields $t_\mathrm{th}\sim10^{-3}\xr^4$ days
(Section~\ref{sub:time_res}). With $\rhob = 10^{10}$ \gcc, it gives
$t_\mathrm{th}$ about a year (Section~\ref{sub:time_res}).  Therefore,
if one needs to model faster processes, one should decrease $\rhob$
(complicating operation of numerical algorithms). For instance, under
the same conditions but with $\rhob = 10^{8}$~\gcc{} one has
$t_\mathrm{th}$ within a day,  and for $\rhob \lesssim 10^{6}$~\gcc{}
(used, e.g., by \citealt{PC18,BeznogovPR20,Yakovlev_21}) one has
$t_\mathrm{th}$ within a few minutes.

To allow for different  possibilities, the calculations \citep{BPY16} have been
carried out at  $\rhob = 10^{8}$, $10^{9}$ and
$10^{10}$ \gcc{} (except for the H\,--\,He mixtures, for which
the density $10^{10}$ \gcc\ is unrealistic and has been 
excluded). This allows users to choose
suitable $\rhob$ for their specific problem.

Note that under certain conditions (e.g., in the presence of
sufficiently strong magnetic fields), the envelope can  reach
isothermality at larger density, $\rhob>10^{10}$ \gcc{} (e.g.,
\citealt{PYCG03}; see Section~\ref{sect-Tb-Ts}).  One should also
remember that in the case of the carbon envelope, the temperature may
noticeably grow at higher densities, immediately beyond the transition
to the iron-group elements, so that $\rhob^*$ for a pure carbon may mark
a false physical bottom of the heat blanket.  In these cases, one may
need to increase $\rhob$ to reach a desired isothermality.  Then
$t_\mathrm{th}$ may increase (up to $\sim100$ years at the neutron-drip
density).

\subsection{Heat blankets in diffusive equilibrium}
\label{ssec:4:DiffEquilResults}

Here we describe the results by \citet{BPY16}.
Fig.~\ref{fig:DemoEnvelopeProfiles} presents distributions of ions and
the temperature profiles $T(\rho)$ in the envelopes composed of the
He~--~C and C~--~Fe mixtures assuming $\rhob=10^{10}$~\gcc. Calculations
have been performed for two effective temperatures, $\Ts=$ 0.8 and 1.5
MK (the solid and dashed curves, respectively). The mass of lighter
elements corresponds to $\rho^*=10^6$ \gcc{} for the He\,--\,C envelopes
(the dotted and dot-dashed curves) and $\rho^*=10^8$ \gcc{} for the
C\,--\,Fe envelopes (the solid and dashed curves). The chosen value of 
$\rho^*$ for the He\,--\,C mixture corresponds to the depth of the
transition layer $z^* \approx 3$ m and to the total depth of the heat
blanketing envelope $z_\mathrm{b} \approx 161$ m; for the C\,--\,Fe
mixture we have, $z^* \approx 28$ m and  $z_\mathrm{b} \approx 145$ m,
respectively (for the canonical neutron star model).

\begin{figure*}[!]
  \centering
  \includegraphics[width=.48\textwidth]{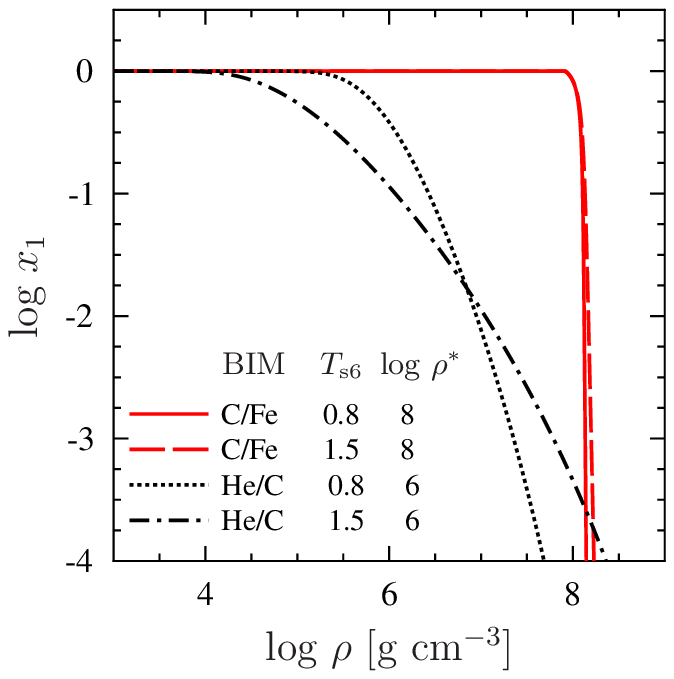}%
  \includegraphics[width=.48\textwidth]{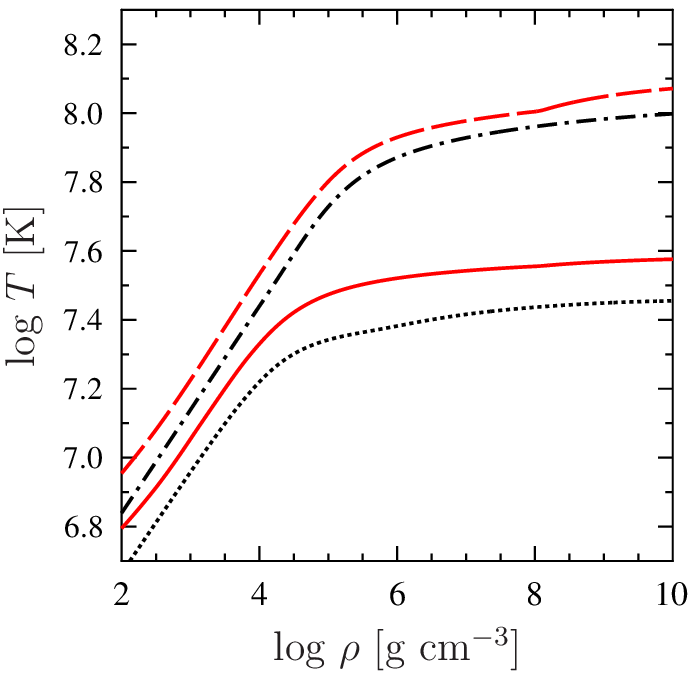}
  \caption{Fraction of lighter ions as a function of $\rho$ (left) and
    the  $T(\rho)$  dependence (right) in the heat blanketing envelopes
    composed of He\,--\,C and C\,--\,Fe mixtures,  for the canonical neutron
    star. Calculations are performed for $T_\mathrm{s6}=\Ts/10^6$ K
    = 0.8 (solid lines) and 1.5 (dashed lines) at $\rho^*=10^6$ \gcc{} 
    (He\,--\,C; dotted and dot-dashed lines) and $10^8$ \gcc{} (C\,--\,Fe; solid and dashed
    lines). See
    the text for details. After Fig.~1 of \citet{BPY16}.
}
  \label{fig:DemoEnvelopeProfiles}
\end{figure*}

The left panel of Fig.\ \ref{fig:DemoEnvelopeProfiles} shows the density
profiles of lighter ions. One can see that the transition layer for the
He\,--\,C mixture is much wider than for the C\,--\,Fe mixture (typical
relative depths $\slfrac{\delta \rho}{\rho^*}$ are different by about
one order of magnitude). This confirms the expectations discussed in
Section~\ref{ssec:4:HeatBlanketsModel}; there is good agreement with the
results of Section~\ref{sec:2} and with the predictions by \citet{CB10}
on the difference of gravitational and Coulomb mechanisms of ion
separation in the mixtures with different and equal effective molecular
weights. Because of the wide transition regions in the He\,--\,C
mixture, the diffusive ``tail'' of helium ions extends to densities much
higher than $\rho^*$, and contributes significantly to the total helium
mass $\Delta M_\mathrm{He}$. For the  C\,--\,Fe mixture, the diffusive 
``tail'' of carbon is much shorter, so that almost entire mass of carbon
is contained in the region  $\rho \lesssim \rho^*$. The difference in
the behaviors of diffusive ``tails'' can be most important for diffusive
nuclear burning. In addition, as seen in the figure, the width of the
transition layer increases with the growth of $T$, especially for the
He\,--\,C mixtures.

The right panel of Fig.~\ref{fig:DemoEnvelopeProfiles} displays the
temperature profiles. It is seen that at high enough $\rho$ the
temperature approaches a constant, meaning that the the envelope
becomes nearly isothermal. As shown in the previous section, the
isothermal layers are reached at lower densities with the decrease of
$\Ts$. The plasma composed of lighter ions has higher thermal
conductivity (e.g., PCY97; see also \citealt{PPP2015} and references
therein). Therefore, the thermal
conductivity of the He\,--\,C mixture is overall higher than of the C --
Fe mixture. Accordingly, at a given $\Ts$, the $T(\rho)$ curves for the
He\,--\,C mixture are lower than for the C\,--\,Fe mixture.

\begin{figure*}[!]
  \centering
  \includegraphics[width=.48\textwidth]{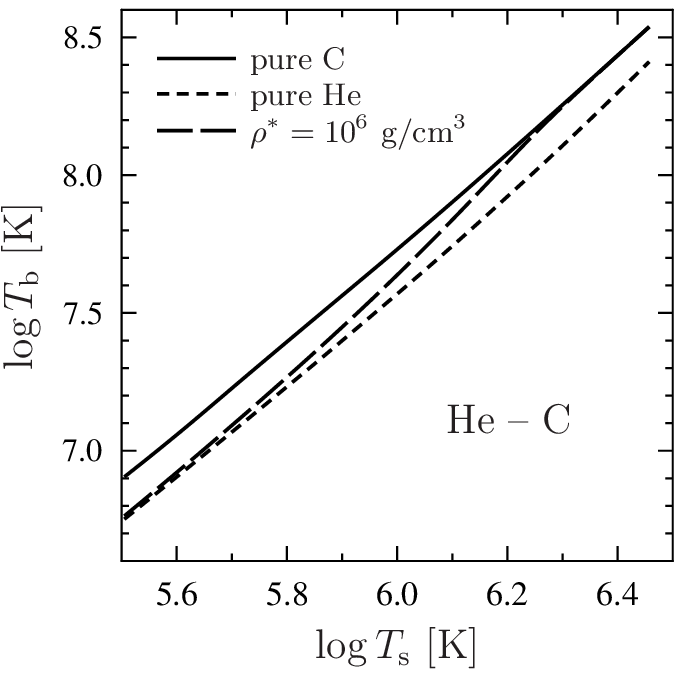}%
  \includegraphics[width=.48\textwidth]{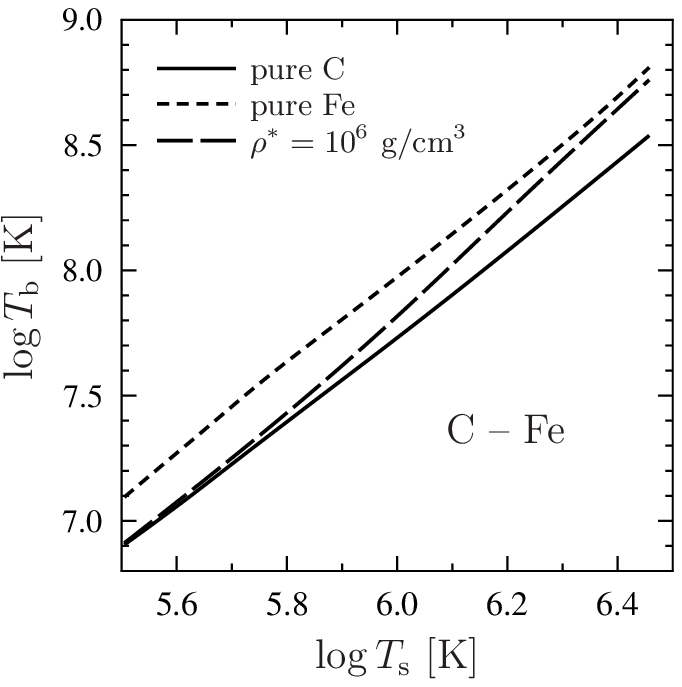}
  \caption{$\Tb - \Ts$ relations for the envelopes composed of pure He
    or C (left) and C or Fe (right) and their mixtures at $\rho^*=10^6$
    \gcc{} and $\rhob = 10^{10}$ \gcc{} for the  canonical neutron star.
    The solid lines correspond to pure carbon on both panels; the
    short-dashed lines are for pure helium (left) and iron (right); the
    long-dashed lines are for the He\,--\,C (left) and C\,--\,Fe (right)
    mixtures. See the text for details. After Fig.~3 in \citet{BPY16}.}
  \label{fig:Tb-Ts}
\end{figure*}

Fig.~\ref{fig:Tb-Ts} presents typical $\Tb(\Ts)$ relations for the
He\,--\,C envelopes (left panel) and C\,--\,Fe ones (right panel) at $\rhob =
10^{10}$ \gcc. The curves correspond for the
envelopes consisting of pure elements (He or C on the left panel; C
or Fe on the right panel) and their mixtures at fixed $\rho^*=10^6$ \gcc. 
As in Fig.~\ref{fig:DemoEnvelopeProfiles}, lighter elements
have lower $\Tb$ for a given $\Ts$, than heavier elements. For ion
mixtures, the curves are intermediate between the curves for pure
elements. Variations of $\rho^*$ change thermal insulation of the
heat blanketing envelope and, hence, change $\Tb$ (as detailed
below).

\begin{figure*}[!]
  \centering
  \includegraphics[width=.48\textwidth]{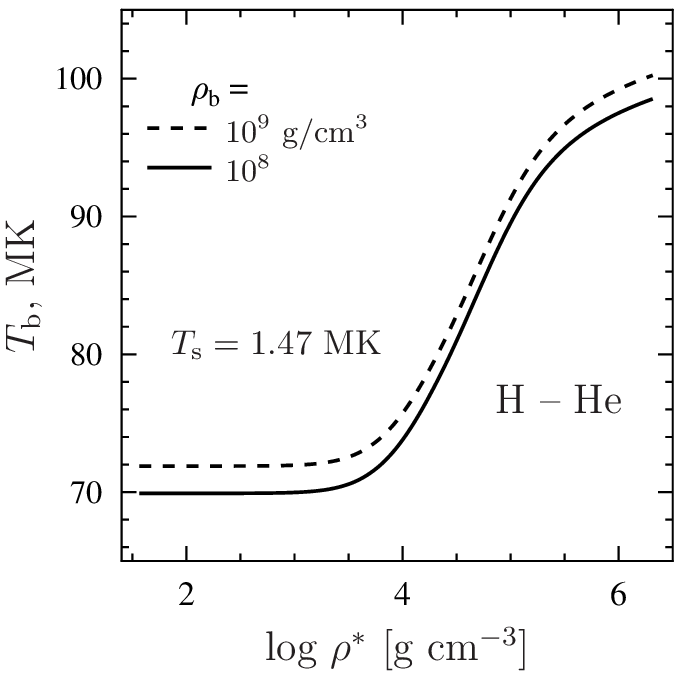}
  \includegraphics[width=.48\textwidth]{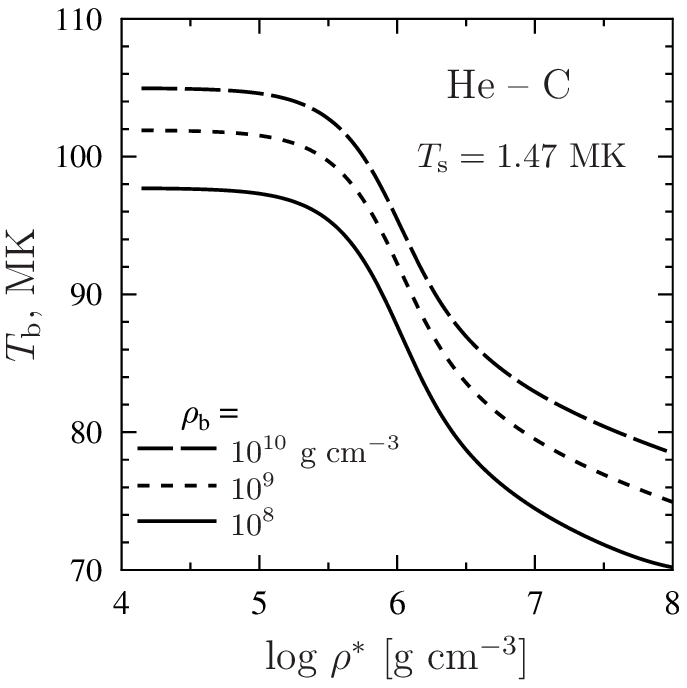}
  \caption{Internal temperature $\Tb$ as a function of $\rho^*$ for the
    canonical neutron star with heat blanketing envelopes made of H --
    He mixtures at $\rhob=10^8$ or $10^{9}$ \gcc{} (left) and He~--~C
    mixtures at $\rhob=10^8$, $10^9$ or $10^{10}$ \gcc{}  (right). The
    surface temperature is $\Ts=1.47$ MK. One can see the transition
    from the envelopes made predominantly of heavier ions (low $\rho^*$)
    to the envelopes consisting mainly of lighter ions (large $\rho^*$).
    See the text for details. After Fig.~4 of \citet{BPY16}.}
  \label{fig:Tb-Rho*}
\end{figure*}

Fig.~\ref{fig:Tb-Rho*} demonstrates the dependence of the internal
temperature  $\Tb$ on the effective transition density $\rho^*$ at fixed
surface temperature $\Ts=1.47$ MK for the H\,--\,He (left panel) and  He
-- C (right panel) heat blankets. The solid lines correspond to
$\rhob=10^8$ \gcc, the short-dashed lines to $\rhob=10^9$ \gcc, and the
long-dashed lines are for the He\,--\,C envelope with $\rhob=10^{10}$
\gcc{} (hydrogen cannot survive at such high densities, but carbon can).
All the curves show a characteristic transition from the envelope made
mostly of heavier ions (small values of $\rho^*$), to the envelope that
consists mainly of lighter ions (higher values of $\rho^*$). The
intermediate range of $\rho^*$, where both ion components are of
principal importance, is seen to be sufficiently wide. It is worth to
notice the different transition behavior of the $\Tb(\rho^*)$ curves for
different mixtures. The behavior of the H\,--\,He mixture is ``special'':
while increasing the amount of lighter  (hydrogen) ions, $\Tb$ grows up,
whereas for the He\,--\,C and C\,--\,Fe mixtures the behavior is the
opposite. This effect has been first noticed by \citet{BPY16}. The
special case of hydrogen occurs because of two reasons. First, charge to
mass ratio for protons (hydrogen ions) $Z/A\approx1$ strongly differs
from $Z/A\approx0.5$ for other ions. Second, helium has low
radiative opacity.

As discussed in Section~\ref{sect-analytic-degen},
the region of densities and temperatures, where $\kappa_\mathrm{e}
\approx \kappa_\mathrm{r}$, constitutes the sensitivity strip 
which gives the main contribution
to $\Tb-\Ts$ relations.
As long as the transition layer (i.e. $\rho^*$) does not
fall in the sensitivity strip,
$\Tb$ is nearly independent of $\rho^*$. In contrast, when the
transition range falls into the sensitivity strip, then
the dependence of $\Tb$ on $\rho^*$ is the strongest.

With growing $\Ts$, the sensitivity strip
shifts inside the heat blanket. 
This explains the behavior of long
dashed curves in Fig. \ref{fig:Tb-Ts}. At low $\Ts$ the transition
region is deeper than the sensitivity strip. Therefore, the mixture
behaves as pure lighter component. With increasing $\Ts$, the
sensitivity strip moves deeper and reaches the transition region,
where the mixture demonstrates its two-component nature. At
larger  $\Ts$ the sensitivity strip appears deeper than the
transition region, and the mixture behaves as pure heavier
component.
This behavior  was further explored by \cite{Wijngaarden_ea19} who
investigated the sensitivity of $\Tb$ to $\rho^*$ in the $\Ts-\rho^*$
plane considering not only the diffusion and the position of the
sensitivity strip, but also diffusive nuclear burning (see their
Figs.\ 4 and 5). Yet, it is important to note that published works on
diffusive nuclear burning treated diffusion rather approximately
(trace ion approximation, no thermal diffusion) which might seriously
affect the nuclear burning (see below).

\begin{figure}
  \centering
  \includegraphics[width=.48\textwidth]{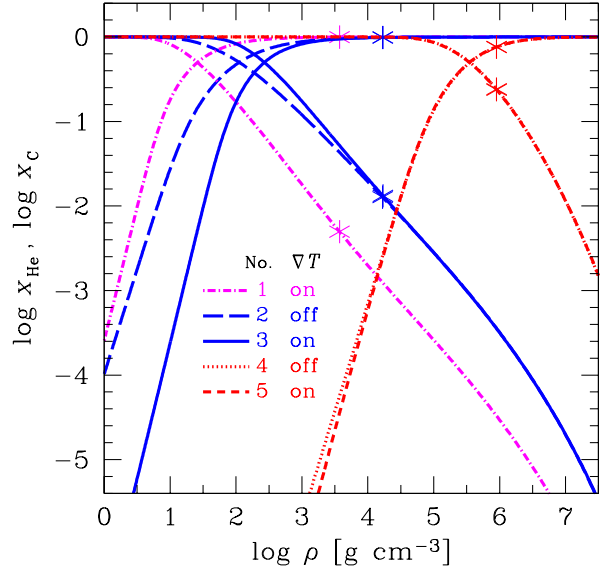} 
  \caption{Profiles of helium fraction (decrease) and carbon fraction
    (increase) as functions of density $\rho$ for five envelope models
    of He\,--\,C mixtures in the canonical neutron star with $\Ts = 1.1$
    MK. Models 1~--~5 are shown by curves of different types. They
    either include or exclude the $\bm{\nabla}T$ term in 
    \req{e:GenForce} (respectively, ``on'' or ``off'' in the legend). An
    asterisk at each curve indicates the position of the effective
    transition density $\rho^*$. See text for details. After Fig.~5 in
    \citet{BPY16}.}
  \label{fig:GradT}
\end{figure}

Fig.~\ref{fig:GradT} demonstrates the effect of the terms containing
$\bm{\nabla}T$ in Eqs.\ \eqref{e:GenForce} or \eqref{e:CorrGrad} on the
properties of He\,--\,C envelopes. The figure shows the fractions of
helium, $x_\mathrm{He}(\rho)$, and carbon, $x_\mathrm{C}(\rho)$,
calculated in five cases (curves 1\,--\,5) for one and the same surface
temperature $\Ts = 1.1$ MK.  The left part of this figure (at
$\log\rho\lesssim1.5$) should be taken with caution, because the matter
at these densities is non-degenerate  (according to \req{rhoF},
$\rho_\mathrm{F}\sim300$ \gcc), so that  the involved assumptions (such
as the linear mixing rule),  which are applicable for strongly
degenerate plasmas, are no longer valid there. Curves 1, 3 and 5 are
computed including the contribution of $\bm{\nabla}T$ terms, whereas
curves 2 and 4 neglect this contribution (which is equivalent to the
approximation made in Section~\ref{sec:2} and in the paper by
\citealt{CB10}). Curves 2 and 3 are computed for one value of 
$\rho^*\approx1.7\times10^4$ \gcc, whereas curve 1 assumes the same
fraction of carbon at the radiative surface (from which one integrates
the equations), as curve 2, $x_\mathrm{C}(z=0) = 2 \times10^{-6}$. This
boundary condition leads to a different value of the accumulated mass of
helium, and therefore to a different $\rho^*\approx 3.7\times10^3$ \gcc.
Nevertheless, the difference between curves 1, 2 and 3 has almost no
effect on the  $\Tb-\Ts$ relation. The values of $\Tb$ for these
relations differ by less than 1\% because the corresponding values of
$\rho^*$ lie out of the sensitivity strip. In contrast, curves 4 and 5
have $\rho^* \approx 9 \times 10^5$ \gcc{} inside the sensitivity strip.
However, in this case the effects of  $\bm{\nabla}T$ are weak owing to
stronger electron degeneracy (as already mentioned in Section~\ref{ssec:2:IsothermDiffFlux-SC}, in the approximations of linear
mixing, strongly non-ideal ion plasma, and strongly degenerate
electrons, all the $\bm{\nabla}T$ terms disappear). That is why curves 4 and
5 are close to each other and the $\bm{\nabla}T$ terms, again, do not
affect the  $\Tb-\Ts$ relation.

As seen from curves 1, 2 and 3, the contribution of the  $\bm{\nabla}T$
term depends, among other things, on the statement of the problem. It is
important which quantity is fixed as a boundary condition -- the
accumulated mass or the fraction of ions at the surface. According to
calculations, the $\bm{\nabla}T$ terms have the strongest effect on the
ion fraction profiles if the transition region coincides with the region
of moderate coupling of ions. This situation occurs at sufficiently high
$\Ts$ in the outer regions of the envelopes ($\rho\lesssim 10^7$ \gcc)
which consist of light elements (such as hydrogen, helium, carbon).
However even in these cases the effect of  $\bm{\nabla}T$ on the
$\Tb-\Ts$ relations, profiles of pressure, temperature and density is
weak.

\subsection{Envelopes out of diffusive equilibrium}
\label{ssec:4:NonEquilResults}

In addition to diffusively equilibrated heat blanketing envelopes,
\citet{BPY16} considered the envelopes out of diffusive equilibrium.
Since ion diffusion is relatively slow (see below), a non-equilibrium
state can exist for a long time without violating a global hydrostatic
equilibrium. By way of illustration, let us study a fixed ion
distribution,  $x_j(\rho)$, ignoring the equations of diffusive
equilibrium. The structure of the envelope can be calculated by
integrating Eqs.\ \eqref{e:FlatEnvelope}.

The results are presented in Fig.~\ref{fig:NonEquil}. The left panel
shows three envelope models for H\,--\,He mixtures. The right panel
presents three models for He\,--\,C mixtures. The figures demonstrate the
helium number fraction as a function of mass density for the canonical
neutron star model with surface
temperature $\Ts=10^6$~K. All three models for the H\,--\,He mixtures have
the same amount of hydrogen ($\log \rho^*=5.06$), while all three models
for the He\,--\,C mixture have the same amount of helium ($\log
\rho^*=7.18$). The helium fraction increases with $\rho$ on the left
panel, because helium is heavier than hydrogen, but it decreases on the
right panel since helium is lighter than carbon. The solid lines refer
to diffusion-equilibrated configurations, while the dashed lines refer
to non-equilibrated configurations with wider (long dashes) and narrower
(short dashes) transition regions, than in the equilibrated case.

\begin{figure*}[!]
  \centering
  \includegraphics[width=.48\textwidth]{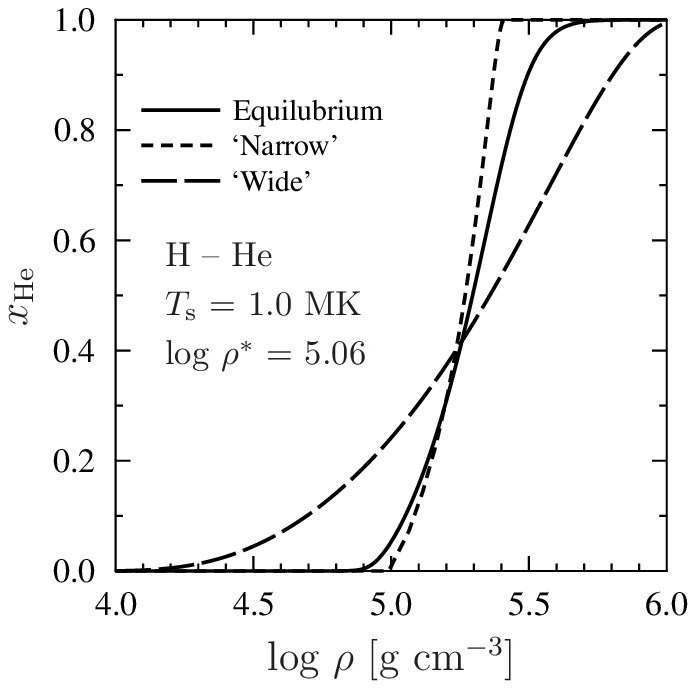}%
  \includegraphics[width=.48\textwidth]{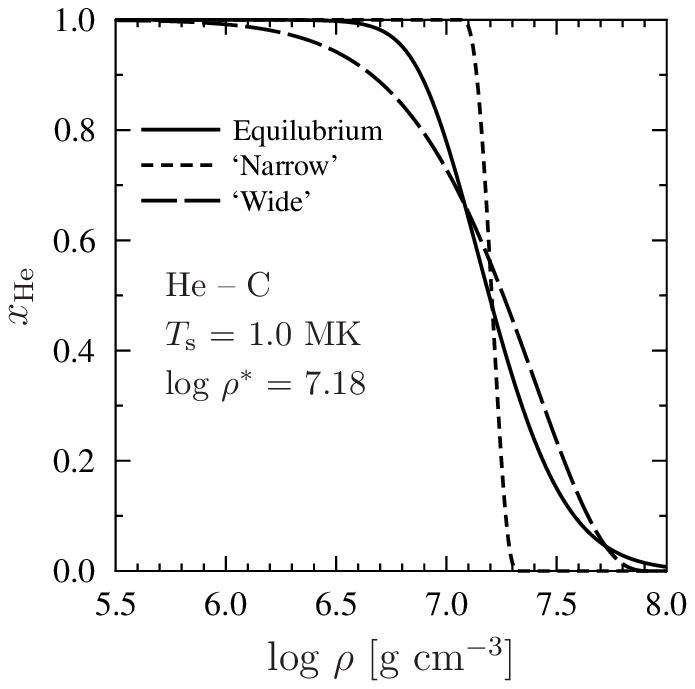}
  \caption{Helium fraction as a function of density in
    heat blanketing envelopes composed of  H\,--\,He mixtures (left
    panel: $\Delta M=5.09 \times 10^{-14}$ $M_\odot$, $\log
    \rho^*=5.06$) or He\,--\,C mixture (right panel: $\Delta M=3.04 \times
    10^{-11}$~$M_\odot$, $\log \rho^*=7.18$) 
    at $g_\mathrm{s14}=2.43$ and $\Ts=1$ MK. The solid lines correspond to
    diffusively equilibrated envelopes, whereas long-dashed and
    short-dashed lines refer to diffusely non-equilibrated envelopes
    with wider (long dashes) and narrower (short dashes) transition
    layers. See text for details. After Fig.~7 in \citet{BPY16}.}
  \label{fig:NonEquil}
\end{figure*}

All three models give nearly the same values of $\Tb$. For instance, for
the He~--~C mixture, one gets the temperature $\Tb=4.00 \times 10^7$ K
at $\rhob=10^{10}$ \gcc. For the H~--~He mixture at $\rhob=10^{9}$
\gcc{} one has $\Tb= 4.64 \times 10^7$ K for the equilibrium and
``narrow'' profiles and $\Tb=4.54 \times 10^7$ K for the ``wide''
profile. Therefore, the resulting  $\Tb-\Ts$ relations are weakly
sensitive to a heat blanket configuration. The main parameter which
regulates the $\Tb(\Ts)$ relation is the accumulated mass of light
elements $\Delta M$ divided by $g_\mathrm{s}^2M$ (or, equivalently, $\eta$ or
$\rho^*$). This is true at least as long as the ion distributions are
not too wide, as seen for the H\,--\,He mixture, for which the ``wide''
profile gives a slightly different value of $\Tb$. On the other hand,
strong deviations from an equilibrium configuration cannot exist for a
long time (see below).

The insensitivity of the $\Tb-\Ts$ relations to the distribution of ion
fractions is helpful for understanding the importance of thermal
diffusion effects. Although thermal diffusion may change ion fractions,
these changes will not affect the resulting $\Tb(\Ts)$ relations.
However, thermal diffusion can be important for the processes that are
sensitive to the distribution of the ion fractions (e.g., diffusive
nuclear burning). \cite{BPY16} demonstrated this by making several
estimates, assuming a constant thermal diffusion ratio $k_T=0.1$ in
\req{k_T}, which is the conservative upper limit obtained in
calculations with the effective potential method. For the H\,--\,He
mixture with $x_{\mathrm{H}}=x_{\mathrm{He}}=0.5$, the thermal diffusion
rate does not exceed  3\% of the ordinary diffusion rate, and for the
He\,--\,C mixture ($x_{\mathrm{He}}=x_{\mathrm{C}}=0.5$) it does not
exceed 6\%.

Using Eq.\ \eqref{e:J2-Gen}, one can calculate the diffusion velocity of
ions for diffusively non-equilibrium configurations considered in
Fig.~\ref{fig:NonEquil}.  Then, introducing a typical width $\Delta z$
of the diffusively non-equilibrium layer and taking characteristic
diffusive velocities $v$, one can estimate typical
diffusion-equilibration time $t_\mathrm{eq}\sim \Delta z /v$ for these
configurations. For the H~--~He envelopes, the estimate gives $\Delta z$
about a few meters, the diffusive velocity $v \sim 10^{-4}-10^{-3}$ cm
s$^{-1}$ and the diffusion-equilibration time $t_\mathrm{eq}$ of a few
days or weeks. For the He~--~C envelopes, $\Delta z$ is also about a few
meters but the diffusion velocity is much slower, $v \sim
10^{-7}-10^{-6}$~cm s$^{-1}$. Accordingly, $t_\mathrm{eq} \sim
10-100$~yr. The equilibration in the He~--~C mixture lasts much longer
as a result of the slow Coulomb separation of ions. Therefore, the
diffusive equilibration takes from a few days to a century, depending on
the chemical composition of heat blankets.

\subsection{Diffusive and the onion-like heat blanketing envelopes}
\label{ssec:4:Concl}

Let us compare the main properties of the diffusive
(Section~\ref{sec:4}) and onion-like (PCY97, Section~\ref{therm-env})
blanketing envelopes. For illustration, we consider a canonical neutron
star with the effective surface temperature $\Ts=0.89$ MK. Then  the
redshifted surface temperature is  $\Ts^\infty=0.68$ MK. This choice
corresponds to the magnetic hydrogen atmosphere plus power-law fits to
the Vela pulsar spectrum from \textit{Chandra} observations by
\citet{Pavlov_etal01} and from $XMM-Newton$ observations by
\citet{ManzaliDLC07}. Recently \citet{2018VELA} obtained a similar
value $\Ts^\infty=0.700\pm0.005$ MK for \textit{Chandra} observations. 
It is remarkable that the attempts to improve the estimate using wide
ranges of $M$ and $R$ give  almost the same $\Ts^\infty$ in all these
ranges (e.g., \citealt{2018VELA}). Using a $\Tb - \Ts$ relation, one can
determine the non-redshifted temperature $\Tb$ at the envelope bottom and
the redshifted internal temperature of isothermal stellar interiors 
$\widetilde{T} = \Tb \sqrt{1-r_\mathrm{g}/R}$. In reality, the Vela
pulsar possesses the surface magnetic field $B \sim 3 \times 10^{12}$~G.
Therefore, the $\Ts$ distribution over its surface is non-uniform (see
Section~\ref{subsub:Ts-vs-thetaB}), which was ignored in the
above-mentioned spectral models.

\begin{figure}[!]
  \centering
  \includegraphics[width=.4\textwidth]{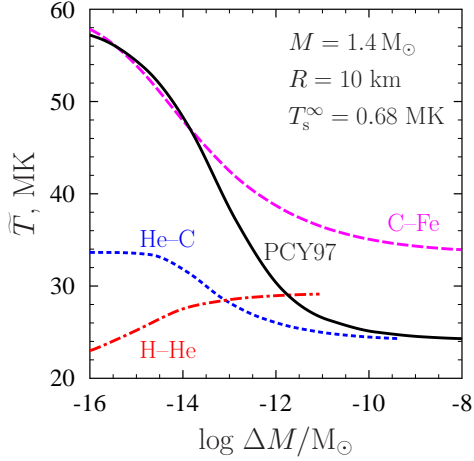}
  \caption{Redshifted internal temperature $\widetilde{T}$ of the canonical
    neutron star with $\Ts^\infty=6.8\times10^5$~K as a function
    of accumulated mass $\Delta M$ of light elements in the
    blanketing envelope. The solid line corresponds to the PCY97 model,
    while other lines refer to the models of envelopes composed of binary
    ion mixtures. After \citet{Beznogov_etal16}. See the text for    
    details.
}
  \label{fig:Vela_Tb-DM}
\end{figure}

Fig.~\ref{fig:Vela_Tb-DM} shows the dependence of the internal stellar
temperature $\widetilde{T}$ on the accumulated mass $\Delta M$ of
lighter elements in the blanketing envelope for different envelope
models. The short-dashed curve corresponds to the envelope composed of 
He\,--\,C mixture with $\rhob=10^{10}$ \gcc. It demonstrates the
dependence of $\widetilde{T}$ on the total mass of helium, $\Delta M =
\Delta M_\mathrm{He}$. The long-dashed line shows the same but for
C\,--\,Fe envelopes, with $\Delta M = \Delta M_\mathrm{C}$ being the total
mass of carbon. The dash-dot line refers to H\,--\,He envelopes, $\Delta
M = \Delta M_\mathrm{H}$ being the total mas of hydrogen; in this case 
$\rhob=10^{8}$ \gcc, $\Delta M_\mathrm{H} \lesssim 10^{-11}\, M_\odot$,
because hydrogen cannot survive at higher densities (cf.{}
Section~\ref{subsub:chemicalcompos}). The solid line is calculated for
the PCY97 model with $\rhob=10^{10}$~\gcc, and $\Delta M$ is the total
mass of H and He.

This figure is analogous to Fig.~\ref{fig:Tb-Rho*} in
Section~\ref{ssec:4:DiffEquilResults}. It also shows the ``anomalous''
behavior for the H\,--\,He mixture. For other envelope models, the
thermal conductivity and internal temperature increase with the growth
of $\Delta M$ at a fixed surface temperature (see
Section~\ref{ssec:4:DiffEquilResults} for details). According to Fig.\
\ref{fig:Vela_Tb-DM}, if the chemical composition of the envelope is
unknown, theoretical uncertainties of $\widetilde{T}$  due to unknown
$\Delta M$ are really large and hamper an accurate determination of
$\widetilde{T}$. The largest variation by a factor of $\sim 2.5$ is
achieved for the PCY97 model. This has been anticipated, because the
PCY97 model assumes the presence of larger spectrum of chemical
elements.  For binary mixtures, especially, for the H\,--\,He and
He\,--\,C envelopes, the variations  of $\widetilde{T}$ are smaller.
This is also natural, because the difference of charge numbers of ions
in the binary mixtures is smaller, hence variations of heat conduction
are weaker.

Since different curves in Fig.~\ref{fig:Vela_Tb-DM} 
are plotted for $\Delta M$ of different nature, a 
plain comparison of the curves may be misleading. However, in some
cases such a comparison is possible. For instance, the PCY97 and
He--C curves at $\Delta M \gtrsim 10^{-10}$ $M_\odot$ correspond to
the blankets which are mostly composed of He; these curves are 
in very good agreement with each other. Equally, the 
PCY and C--Fe curves at low $\Delta M$ correspond to the blankets
almost fully composed of Fe; they are also in good agreement.

Fig.~\ref{fig:Vela_Tb-Ts} shows thermal states of the same  star as in
Fig.\ref{fig:Vela_Tb-DM} but at different $\Ts$ (when the star is warmer
or  colder)   for different models of the heat blanketing envelopes. The
thermal states are characterized by the dependences  of  $\widetilde{T}$
on $\Ts^\infty$ which, in their essence, are analogous to the
dependences of  $\Tb$ on $\Ts$ (Fig.~\ref{fig:Tb-Ts}). The vertical
dotted line marks  $\Ts^\infty = 6.8\times10^5$~K (as in
Fig.~\ref{fig:Vela_Tb-DM}). The curves can be viewed as ``evolutionary
tracks'' of the star. The left panel is devoted to He--Fe and He--C
envelopes; the thick curves refer to one-component envelopes (the thick
dashed curve is for pure Fe, the solid curve for pure C, and the
dot-dashed curve for pure He). Thin curves of different styles
correspond to binary mixtures with different masses of lighter ions
($\Delta M/M_\odot = 10^{-16}$, $10^{-14}$, $10^{-12}$, $10^{-10}$ and
$10^{-8}$). The lowest value of $\Delta M$ corresponds to a thin surface
layer of a lighter element, while the highest value to a thin layer of a
heavier element at the bottom of the heat blanket. The
$\widetilde{T}(\Ts^\infty)$ relations vary in response to the variations
of the envelope's composition.

\begin{figure}[!]
  \centering
  \includegraphics[width=.4\textwidth]{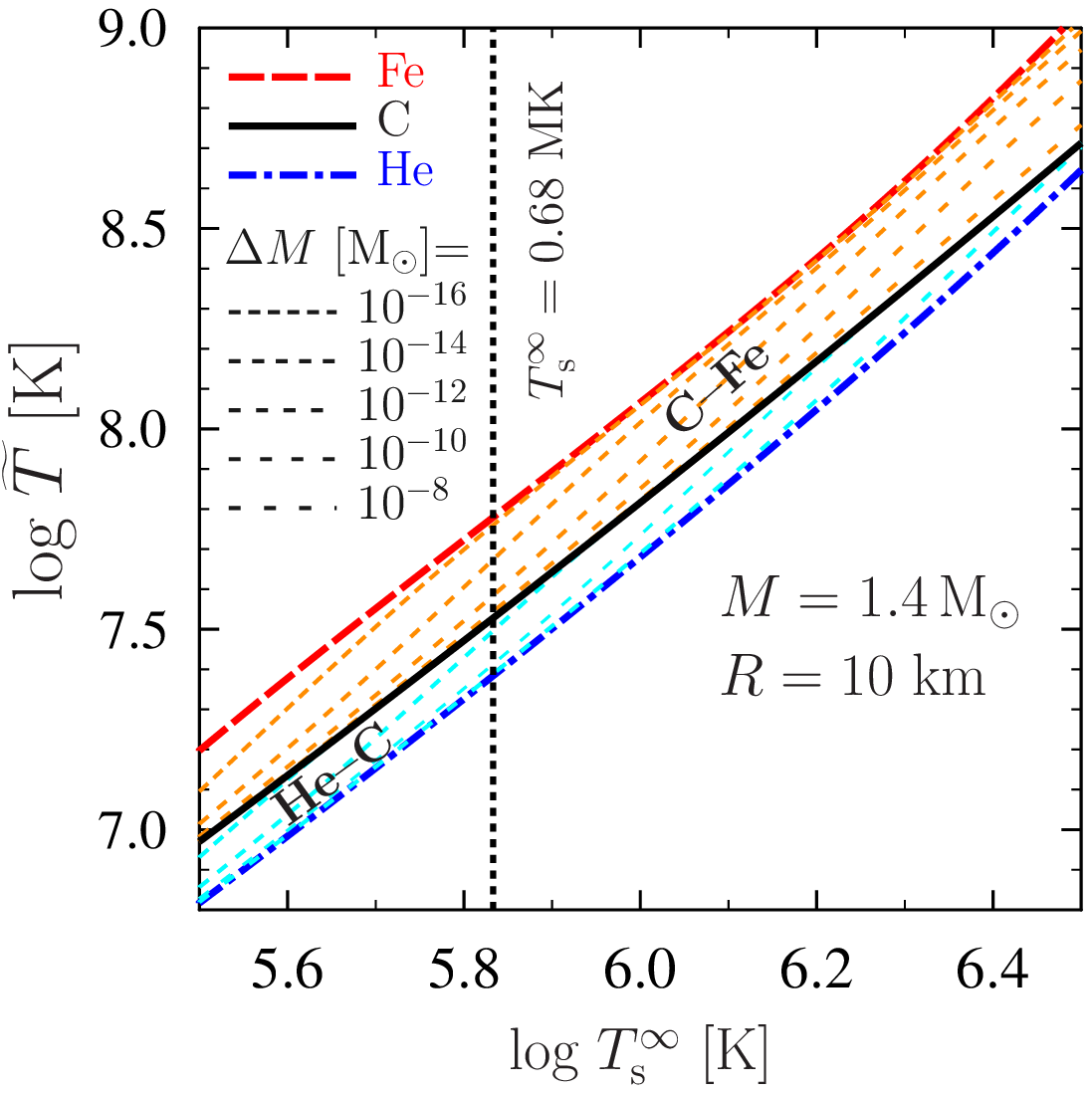}
  \includegraphics[width=.4\textwidth]{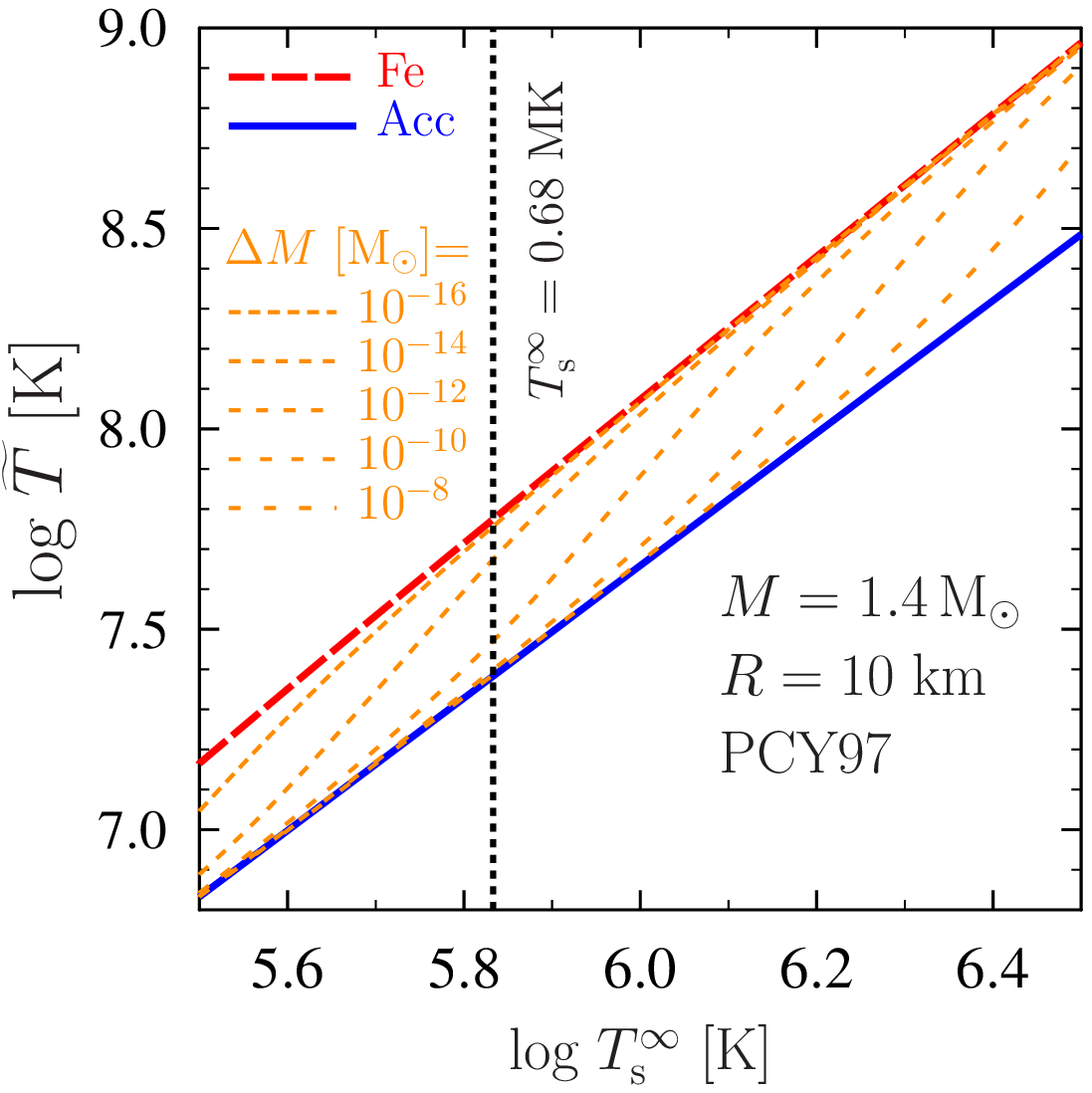}
  \caption{Thermal states ($\widetilde{T}-\Ts^\infty$ relations) 
    of the same star as in Fig.~\ref{fig:Vela_Tb-DM} (with 
    various blanketing models) but
    at different $\Ts^\infty$ (at different thermal surface states). The 
    basic surface state of Fig.~\ref{fig:Vela_Tb-DM} is
    plotted by the vertical dotted line. The curves to the 
    right of that line refer to
    hotter (younger) star, while the curves to the left
    refer to colder (older) star. 
     The left panel is for the C\,--\,Fe and He\,--\,C mixtures; the right panel is for the PCY97 model. See the
    text for details.}
  \label{fig:Vela_Tb-Ts}
\end{figure}

The right panel of Fig.~\ref{fig:Vela_Tb-Ts} demonstrates the same 
as the left panel but
for the PCY97 model. As before, the thick dashed line corresponds to the
envelope of pure iron. The thick solid line (denoted as ``Acc'') is for
the envelope with the maximum amount of H+He.  Thin dashed lines of
different styles refer to different masses of H+He. As expected, this
model gives the widest variations of $\widetilde{T}(\Ts^\infty)$, which
is clear from comparison of the right and left panels of
Fig.~\ref{fig:Vela_Tb-Ts}. Notice that difference in the values of
$\widetilde{T}$ for a partially accreted and non-accreted envelopes is
smaller in warmer stars and larger in colder stars, which is explained
by the shift of the sensitivity strip to lower densities as the star
cools down.

The most important result is that the $\Tb - \Ts$ relations for the
diffusive envelopes are nearly independent  of the structure of the
transition layer (on its width, distributions of ion fractions, presence
or absence of diffusive equilibrium). These relations depend only of
$\Delta M$ (or $\eta$). In particular, these results confirm the
validity of the PCY97 models    (Section~\ref{therm-env}), where the
envelopes were approximated  by the sequence of shells of ions of one
species (that is of H, He, C or Fe), with sharp boundaries between the
shells (the ``onion-like'' structure).

To summarize our comparison of the diffusive models of heat blankets with
the PCY97 models we would like to stress the
following:
\begin{itemize}

\item{}
Binary-mixture diffusive and PCY97 models
are based on almost the same microphysics. 
They are not diverse but complementary. 

\item{}
One can use the PCY97 model if the composition  of a given blanketing
envelope is formed via the quasi-stationary evolution of accreted
hydrogen-helium matter (with the layers of H, He, C, and Fe from top to
bottom). At $\Delta M \gtrsim 10^{-7}$ $M_\odot$ the  heat blanket will be
fully accreted (H, He, C). The position of the upper boundary of the Fe
layer can be shifted upwards by decreasing $\Delta M$. While using the
PCY97 fits presented in Section~\ref{sect-Tb-Ts}, one should bear in mind that
the positions of the interfaces between different elements are fixed. If
these positions are different, a heat blanket model should be
recalculated. 

\item{} PCY97  stated that replacing hydrogen with helium and carbon
with iron would have  almost no effect on $\Ts - \Tb$ relations. Here
we have
paid more attention to the effect of such replacement and confirmed that
the effect is small, compared with a replacement of light ions (H or He)
 by heavier ones (C or Fe). 

\end{itemize}

Nuclear reactions in the blanketing envelope can noticeably change
$\Delta M$ and the $\Tb-\Ts$ relation in the course of a neutron star
evolution. All the calculations in this section have neglected the
possibility of convection in the envelope. As discussed in
Section~\ref{sub:iron-blanket} (see Fig.~\ref{fig-therm-pr}), the
convection can occur in some parts of the envelopes, but it has almost
no effect on the $\Tb-\Ts$ relations.

The calculated
$\Tb-\Ts$ relations for diffusively equilibrated envelopes have been 
approximated by analytic expressions  (\ref{app:TsTb}),
which are convenient for simulating thermal evolution of neutron
stars and related phenomena.

The presented models of heat blankets are greatly simplified; real
envelopes  may contain ions of many species.  For example,
\citet{Fantina_20} studied the cooling and the equilibrium composition
of the outer layers of a non-accreting neutron star down to
crystallization and showed that  the sharp changes in composition
obtained in the one-component plasma approximation are smoothed out when
a full nuclear distribution is allowed. In the liquid part of the
envelope, however, stratification of ions will prevent the appearance of
regions containing many ion species at once. Realistic envelopes have
most probably shell structures with one type of ions in each shell and 
narrow diffusive transition layers of binary mixtures between the
shells. For calculating $\Ts - \Tb$ relations, it would be sufficient to
neglect diffusive broadening of the shell boundaries. However, the
``onion-like'' approximation could be insufficient for tracing the
evolution of nuclear composition  within the blanketing envelopes, for
example, with allowance for the diffusive nuclear burning
\citep{CB03,CB04,CB10,Wijngaarden_ea19}.

\section{Magnetic blanketing envelopes}
\setcounter{equation}{0}
\label{therm-magn-env}

\subsection{Statement of the problem}
\label{therm-magn-env-intro}

\subsubsection{Microphysics of matter in magnetic envelopes}
\label{subsub:microphysics-magnetic}

Microphysical properties of the matter in magnetic heat-blanketing
envelopes have been described in many publications (see, e.g.,
\citealt{YakovlevKaminker94,PC13,PPP2015}, and references
therein). In this section we briefly outline some important results.

Magnetic fields $\bm{B}$ in the heat blankets affect the properties of
electrons and  ions. As a rule, the effects on the properties of
electrons are most pronounced. These effects can be roughly separated in
two types. First, there are \emph{classical} effects associated with
electron rotation about $\vB$-lines. Secondly, there are
\emph{quantum-mechanical} effects produced by quantization of electron
motion across $\bm{B}$ and resulted in the appearance of the electron
Landau (or Rabi-Landau\footnote{The magnetic quantization was first
studied by \citet{Rabi28}.}) energy levels. The quantum effects are
usually pronounced at much higher magnetic fields than the classical
ones. The classical effects change mainly the electron transport
properties but leave the thermodynamic properties (e.g., the electron
pressure) unchanged. The quantum effects can modify the transport and 
thermodynamic properties. The effects of both types can be different in
non-degenerate and degenerate electron plasmas.

\paragraph{Classical effects}

 The most important classical effect is
that electron conduction  becomes anisotropic. The effect occurs at any
electron degeneracy and is controlled by the electron magnetization
parameter 
\begin{equation}
 \xi=\omega^* \tau_\mathrm{e},
\label{e:magnetiz} 
\end{equation} 
where $\tau_\mathrm{e}$ is the effective electron thermal-conduction
relaxation time, $\omega^* = \omc/\gammar$ is the characteristic
gyrofrequency of rotation of a conduction electron about the magnetic
field lines, $\omc=eB/(\mel c)$ is the electron cyclotron frequency,
and  $\gammar$ is the characteristic Lorentz factor of the conduction
electrons. In the degenerate matter, $\gammar=\sqrt{1+\xr^2}$, with
$\xr$ being determined by \req{eq:xr}. In this case, 
$
   \xi \approx 1760\,(B_{12}/\gammar)\tau_\mathrm{e}/(10^{-16}\mbox{~s}),
$
where $B_{12}\equiv B/10^{12}$~G. 

The electron heat conduction in a
magnetic field is determined by the three thermal conductivity
coefficients, specifically, by the thermal conductivities
$\kappa_\parallel$ and $\kappa_\perp$ along and  across $\bm{B}$, and by
the Hall thermal conductivity $\kappa_\mathrm{H}$ which describes the
heat flux component   perpendicular to $\bm{B}$ and to the temperature
gradient $\nabla T$.  If the quantum effects are small, the conductivity
$\kappa_\parallel$ appears to be almost independent of $B$. In the
regime of weak electron magnetization ($\xi \ll 1$, many collisions
during one gyro-rotation) the electron conduction is only slightly
anisotropic, with $\kappa_\parallel \approx \kappa_\perp$ and
$\kappa_\mathrm{H} \sim \kappa_\parallel \xi \ll \kappa_\parallel$. In
the opposite case of strongly magnetized electrons ($\xi \gg 1$, many
rotations between successive collisions),  $\kappa_\mathrm{H} \sim
\kappa_\parallel/\xi \ll \kappa_\parallel$ and  $\kappa_{\perp} \sim
\kappa_\parallel/\xi^2 \ll \kappa_\parallel$, so that thermal conduction
across $\bm{B}$-lines becomes greatly suppressed.  Therefore, the
magnetic field can  significantly affect the electron heat transport at
$\xi \gtrsim 1$. 

\paragraph{Quantum effects}

 In non-degenerate layers of neutron-star
envelopes, the electrons are usually non-relativistic; electron thermal
conduction is relatively unimportant because  radiative thermal
conduction is sufficiently high (resembling the non-magnetic case; e.g., 
Section~\ref{therm-analyt}). The magnetic field is called \emph{strongly
quantizing} for the electrons, if it forces the majority of
the electrons to occupy the ground Landau
level. In a non-degenerate matter, this occurs at $T \ll T_\mathrm{cycl}$, where
\begin{equation}
     T_\mathrm{cycl}= \hbar \omc / \kB \approx 1.343 \times 10^8 \, 
     B_{12}\mbox{K}.
    \label{eq:Tcycl}
\end{equation}
However, in the non-degenerate case the electron 
pressure still remains unaffected by $B$, being equal to
$P_\mathrm{e}=\nne \kB T$, although some other thermodynamic
functions are affected (for instance, the electron heat capacity is
reduced by a factor of 3). As for the radiative thermal conductivity, it
becomes  anisotropic. It is described by the two radiative thermal
conductivity coefficients, along and across $\bm{B}$, which are
enhanced, as compared to the non-magnetic case, proportionally to
$(T_\mathrm{cycl}/T)^2 \propto B^2$.

In the deeper layers of the heat blanket the
electrons become strongly degenerate (and possibly relativistic); 
the anisotropy of electron
thermal conduction is most important. It operates in the classical
and quantum regimes. In the quantum regime, the magnetic field modifies also
the electron gas thermodynamics, particularly, $P_\mathrm{e}$. 

The quantum effects of magnetic field are different in the two domains
of $\rho$ and $T$ (for details see, e.g., \citealt{HPY07}, Chapter~4).
The first is the domain of \emph{strongly quantizing} magnetic field,
which forces almost all the electrons to occupy the ground Landau
level.  It occurs at relatively low $\rho<\rho_B$ and low $T\lesssim
T_B$, where
\begin{equation} 
       \rho_B \approx 7 \times 10^3 \,(A/Z) \, B_{12}^{3/2}~~\gcc;
\qquad
     T_B=T_\mathrm{cycl}~\mbox{at}~~\rho< \rho_B ; \quad 
    T_B=T_\mathrm{cycl}/\gammar
~\mbox{~at}~~\rho> \rho_B.
       \label{e:rhoB}
\end{equation}
In this case all thermodynamic and kinetic properties of the electron
plasma can be strongly affected by the magnetic field.
The second is the domain of \emph{weakly quantizing} field, where $\rho
\gtrsim \rho_B$ and $T \lesssim T_B$. In this domain the electrons can populate many Landau
levels but the thermal energy  $\kB T$ is smaller than the distance
between neighboring Landau levels. Then the presence of the Landau
levels can still affect thermodynamic and kinetic properties of the
electron plasma. The bulk properties like the electron pressure,
internal energy, or chemical potential are affected only slightly. If
very high accuracy is not required, they can be replaced by
corresponding non-magnetic quantities. However, those quantities that
are determined by the electrons with energies near the Fermi energy (for
instance, electron specific heat or transport coefficients) can be
affected much stronger. Such quantities oscillate with increasing
density due to population of new  Landau levels by strongly degenerate
electrons.

At $\rho < \rho_B$ the quantizing magnetic field strongly reduces
the degeneracy temperature,
\begin{equation}
    T_\mathrm{F} \approx \frac{\sqrt{1+x_B^2}-1}{\sqrt{1+\xr^2}-1} \, T_\mathrm{F0}, \quad
    x_B= \left( \frac{4 \rho^2}{3 \rho_B^2} \right)^{1/3}\xr,
\label{e:TFB-xB}    
\end{equation}
where $T_\mathrm{F0}=T_\mathrm{F}(B=0)$ is given by \req{degenerateT},
and $x_B$ is the
Fermi momentum of degenerate electrons in the strongly quantizing
limit in units of $\mel c$.

In summary, the strongest effects  of a magnetic field on the
microphysics of plasma in heat blanketing envelopes  are expected at
sufficiently low densities and temperatures, $\rho \lesssim \rho_B$ and
$T \lesssim T_B$. For instance, at $B=10^{12}$ G, one has $T_B \sim
10^8$ K and $\rho_B \sim 10^4$ \gcc. If $B$ is more or less the same
within the envelope and we increase $\rho$ or $T$ beyond $\rho_B$ or
$T_B$, the magnetic effects will weaken.  In a weakly quantizing field, 
the quantum effects  make the thermodynamic functions to oscillate
around their classical values with changing $\rho$ or $B$
\citep[see][]{PC13}.

\subsubsection{Magnetic heat blankets for
1D and 2D thermal evolution codes}
\label{subsub:B-blanket-model}

High magnetic fields make thermal conduction in the outer layers of
neutron stars  strongly anisotropic. This affects thermal flows near the
stellar surface and the surface map of the effective temperature $\Ts$.
In addition,  the magnetic forces and quantum effects of the  $B$-field
on the pressure can change the hydrostatic structure of neutron star
layers. The problem of neutron star thermal evolution ceases to be
one-dimensional (1D) and becomes more complicated (2D or even 3D). 

It seems reasonably to assume that the blanketing envelope
remains thin and can be artificially divided into small 
domains in such a way that $\bm{B}$ is nearly constant in 
each domain (varying parametrically from one domain to another). 
Let us assume further that typical length-scales of these domains along
the surface are much larger than the heat blanket width. 
Then one can approximate any domain by a piece of plane-parallel layer
(like in Section~\ref{subsub:basic_eqns}), solve the corresponding 
heat transport problem and find a local $\Ts - \Tb$ relation. 
Here, $\Ts$ and $\Tb$ are, respectively, the local 
surface and internal temperatures for a given domain.
Generally, both temperatures  depend on the magnetic field $\bm{B}$
in the given domain.

The solution of the above problem for all domains can be used 
as a boundary condition to model (Section~\ref{therm-blanket})
the thermal and magnetic evolution 
in neutron star interiors ($r<R_\mathrm{b}$).

Naturally, the magnetic field penetrates into the entire star 
or into its essential part, the crust and the core. Then the 
magnetic effects, particularly, anisotropic heat conduction, 
have to be included into the equations which describe the 
thermal and magnetic evolution of the 
interiors. In this connection, one can employ the two 
heat-blanket descriptions which are good
either for 2(3)D or for 1D thermal
evolution codes. 

\paragraph{Heat blankets for 2D or 3D codes} 

These 1D heat-blanket
 models
based on the heat transport solutions for separate domains can serve
 as boundary conditions at appropriate 
internal domains in 2D or 3D codes to follow the thermal evolution
of magnetic  neutron stars.

\paragraph{Heat blankets for 1D codes}

 The 1D heat-blanket models provide natural boundary conditions for the
1D evolutionary codes. Their use can be based on the assumption that
neutron star interiors are isothermal; see \req{eq:isothermal} and the
discussion in Section~\ref{subsub:basic_eqns}. The isothermality can be
provided by high conductivity of the stellar interior, if the thermal
evolution time-scale is sufficiently long. The weaker assumption that
$\widetilde{T}$ depends only on $r$, but not on the angles in the
spherical coordinate system, is also sufficient. This weaker assumption
can be fulfilled, if the heat conduction at $\rho>\rhob$ is isotropic
and the thermal relaxation time of the non-isothermal zone is small
compared to the thermal evolution time-scale. There may be several
causes for this isotropy: the decrease of the Hall parameter $\xi$ with
increasing density, a small-scale (compared to $R$) configuration of the
magnetic field, or predominant conduction by neutrons if $\rhob$ reaches
the neutron-drip density (as in \citealt{PYCG03}). 

Then  $\Tb=T(R_\mathrm{b})$ is the same in all the domains of the
envelope at a given moment of time.  Therefore, one can integrate over
all the  domains and obtain  the total photon surface thermal luminosity
$L_\gamma$ as a function of $\Tb$. The $L_\gamma - \Tb$ relation,
derived in this way, plays the same role as    the $\Ts - \Tb$ relation
for non-magnetic stars.  In this case, it is convenient to define the
mean effective surface temperature $\bar{T}_\mathrm{s}$ by 
\beq
 L_\gamma = 4 \pi R^2 \sigma_\mathrm{SB}
\bar{T}_\mathrm{s}^4, 
\label{meanTs}
\eeq
 and  use the
$\bar{T}_\mathrm{s}-T_\mathrm{b}$ relation to study the thermal
evolution of the star with a  1D computer code. The heat blanket remains
essentially 2D, but the anisotropic temperature distribution in the heat
blanket is totally included in the appropriate 
$\bar{T}_\mathrm{s}-T_\mathrm{b}$ relation, making the internal thermal
evolution problem one-dimensional.

 Now the surface distribution of the
effective temperature $\Ts$ can be noticeably non-uniform and
the observable radiation flux can depend on observation direction.
However, we will see (Section~\ref{subsub:GravLens}) that 
the effect is almost smoothed out due to light bending  
in General Relativity, at least for a dipole surface magnetic field and
isotropic local surface emission model. A distant observer will detect nearly
the same bolometric thermal flux observing the star at any angle.  

\paragraph{1D versus multi-D}

 Since 2D or 3D codes are   more
complicated, they are used less often.  A review of such computations
is given by \citet{PonsVigano19}. For example, a 2D code has
been realized by \citet{2008Aguilera} and elaborated by
\citet{Vigano_13}.

1D codes are simpler; they have been employed in the majority  of
studies of cooling magnetized neutron stars. Their validity is
restricted by the requirement that $T$ is independent of angles at
$\rho>\rhob$, as discussed above. This requirement does not necessarily
imply that the effects of magnetic field  in the interiors are washed
out (for instance, the heat transport is isotropic  or the generation 
of Joule heat due to the electric current dissipation  is spherically
symmetric). The isothermality can be provided by the high thermal
conductivity even for anisotropic heat transport, sources, or sinks.
Then the main places of anisotropic  temperature distribution are the
heat insulating  envelopes. In these envelopes, the magnetic effects can
be especially strong and the heat conduction is not too fast to smear
out the anisotropy.

It would be difficult to formulate strict conditions for 
the validity of 1D codes. These conditions depend 
on specific problem, particularly, on magnetic field strength and 
geometry. We expect that 1D codes 
are especially accurate if the magnetic 
field in the neutron star crust is  $B \lesssim 10^{14}$~G.

In what follows we mainly discuss the heat blankets for 1D codes. 

\paragraph{Constructing magnetic heat blankets}

At the first step one needs to solve a heat blanket problem in a small
local part of the  insulating envelope.  If we assume a locally constant
$\bm{B}$ within this part, we will have no magnetic force there and the
hydrostatic equilibrium will be described by our familiar Eq.\
(\ref{dP/dz}), although the pressure $P$ can  depend on $B$ due to the
quantum effects outlined in Section~\ref{subsub:microphysics-magnetic}. 

As we discussed above, the heat transport in the magnetic heat blanket
is generally described by the three thermal conductivity coefficients:
$\kappa_\parallel$,  $\kappa_\perp$, and $\kappa_\mathrm{H}$.
Nevertheless, one can show that, in our  approximation of a local thin
plane-parallel blanket with a locally fixed $\bm{B}$, \req{therm-Tcrust}
remains valid, if  $\kappa$ means the effective radial thermal
conductivity, given by
\beq
\kappa=\kappa_\|\cos^2\theta_B+\kappa_\perp\sin^2\theta_B,
\label{kappa-sum}
\end{equation}
where $\theta_B$ is the angle between $\vB$ and the normal to the
surface. Therefore, 
the $\Ts - \Tb$ problem for a local domain reduces to
solving the same two equations (\ref{dP/dz}) and (\ref{therm-Tcrust}),
which have been used for non-magnetic heat blankets, but with
more complicated physics involved.
  
The magnetic field affects the thermal structure of the blanketing
envelope in several ways (Section~\ref{subsub:microphysics-magnetic}). 
First, it makes the thermal conductivity anisotropic. Here, the effects
are twofold.

 \textit{(i)}~Classical effects of electron rotation about
field lines can strongly reduce $\kappa_\perp$ but they do not affect
$\kappa_\parallel$. They are especially important near the magnetic
equator [$\theta_B \approx 90^\circ$, $\kappa \approx \kappa_\perp$ in
Eq.\ (\ref{kappa-sum})].  Such equatorial regions  become poor heat
conductors, which lowers the local effective temperature $\Ts$ for a
given $\Tb$. 

\textit{(ii)}~Quantization of electron motion into Landau levels can
strongly modify both $\kappa_\parallel$ and $\kappa_\perp$. If the heat
is mostly transported by degenerate electrons and the magnetic field is
strongly quantizing,  then the quantum effects enhance
$\kappa_\parallel$.  These effects are most pronounced  near the
magnetic poles, where $\theta_B \approx 0$ and $\kappa \approx
\kappa_\parallel$. Then the quantum effects increase the local $\Ts$ for
a given $\Tb$.

In the domains of strongly quantizing magnetic field, the
classical  and quantum effects on thermal conduction act in opposite
directions. They are mainly important in different parts of the
neutron  star surface. In addition, one needs much stronger $B$-fields
to make the quantum effects pronounced. Note that the quantum effects
modify also the plasma pressure, and hence the $\rho(z)$ profiles.

\paragraph{The approximations of the model}

Evidently, the formulated heat blanket model is not perfect. First of
all, the approximation of locally constant (force-free) magnetic fields
can be too crude. Indeed, the $\bm{B}$ field in the outer neutron star
layer  is likely to be nearly force-free, but even relatively small
corrections may produce magnetic forces which could affect the structure
of the outer layers. 

Second, the formulated model deals actually with radial heat fluxes.
Within the same model, there are also tangential fluxes. Such fluxes may
be insignificant at one domain of the surface, but combining the domains
we will obtain heat fluxes circulating under the surface. They can be
locally constant along their circulation lines. However, globally,
tangential heat circulations may redistribute some amount of heat from
one domain of the heat blanket to others and affect thus $\Ts$. For
example, in the dipole field model, the heat is transported along the
field lines from hotter polar regions to cooler equatorial domains, so
that the equator temperature becomes higher than predicted by the
plane-parallel approximation. This becomes important in superstrong
magnetic fields $B\gtrsim10^{14}$~G (e.g., \citealt{PPP2015}). However,
this can hardly affect the observed luminosity, because, in strong
magnetic fields, the equatorial region is cold and gives negligible
contribution to the total flux anyway. One can also
anticipate that the temperature gradients along the surface may render
some parts of the envelope baroclinically unstable, although a strong
magnetic field may partly stabilize it.

Such effects have been almost not considered in the literature and will
be ignored below. They could be good subjects for future projects.

\subsection{Analytic model}
\label{therm-magn-analyt}

The main effects of strong magnetic fields on the heat blankets can be
understood using a simplified fully analytic model,  analogous to that
considered in Section~\ref{therm-analyt} for the non-magnetic case.  For
a strongly quantizing magnetic field, such a model was constructed by
\citet{VP}; it is outlined below. Many results are useful for
understanding the main features of the  heat blankets for 2D codes. More
elaborated numerical models are discussed in Section~\ref{sect-therm-numeric-magnetic}, mostly for the heat blankets designed
for 1D codes.

\subsubsection{Equation of state}
\label{subsub:EOS-B}

As in the non-magnetic case, we consider the envelope  composed of a
fully ionized electron-ion plasma. The equation of  state of such a
plasma has been discussed  in \citet{PC13}.\footnote{See footnote
\ref{EIP} on page \pageref{EIP}.} Under
typical conditions in the heat blanketing envelopes, the
Landau quantization of ion motion can be neglected, whereas the
electrons can be quantized.

First consider the case of low temperatures, where the electrons are
degenerate. As mentioned in Section~\ref{subsub:microphysics-magnetic},
in the regime of weak quantization, the electron pressure
oscillates around its non-magnetic values with increasing density.
Replacing the accurate EoS by its non-magnetic counterpart will not
noticeably affect the structure of the envelope. 

In the strongly quantizing limit ($T\ll
T_B$ and $\rho<\rho_B$) the pressure of degenerate electrons becomes
much lower than at $B=0$. As a result, the electrons remain non-degenerate
along the radiation-dominated part of the envelope. Since the
pressure of the non-degenerate gas is independent of the magnetic
field, we can use the classical non-degenerate EoS in this part of
the envelope.

\subsubsection{Radiative opacities}
\label{subsub:RadOpacB}

The radiative thermal conductivity in a magnetized plasma was described,
e.g., by \citet{PC18}. The radiative conductivity becomes anisotropic 
(Section~\ref{subsub:microphysics-magnetic}), but the difference between
its longitudinal ($\kappa_\mathrm{r,\|}$) and transverse
($\kappa_\mathrm{r,\perp}$) components is not too large, so that we can
neglect the difference for a qualitative analysis.  If the radiative
opacities are mediated by free-free transitions,  then in a strongly
quantizing magnetic field they tend to
\beq
  K_\mathrm{r}(B)\approx \left(\frac{23.2\,T }{ T_B} \right)^2
  K_\mathrm{r}(0)
  \simeq 2.2\,\frac{ \bar{g}_\mathrm{eff}\,
  Z^3  \rho }{ A^2 \,T_6^{1.5}\,
  B_{12}^{2}}
  \mbox{~~cm$^2$~g$^{-1}$}, 
\label{K-mag}
\end{equation}
where $K_\mathrm{r}(0)$ is given by \req{K_0},
 $x_B$ is
given by Eq.\ (\ref{e:TFB-xB}), 
and $\rho$ is expressed in \gcc. This estimate may be
used if only $T_6\lesssim B_{12}$ and $x_B \ll 1$.
We will use it in Section~\ref{subsub:T(rho,B)}.
According to Eqs.~(\ref{therm-Eddington}) and (\ref{K-mag}), the
strong magnetic fields, $B_{12}\gtrsim T_6$, push  the 
radiative surface to higher densities, $\rho_\mathrm{s} \propto
B$.

\begin{table}
\caption{Coefficients $a_n$, $b_n$ and $c_n$
in Eq.~(\protect\ref{KRmag}).
}
\label{tab-KRmag}
\begin{center}
\begin{tabular}{lllllll}
\toprule
$n$ & 1 & 2 & 3 & 4 & 5 & 6\\
\midrule
$a_n$ & 0.0949 & 0.1619 & 0.2587 & 0.3418 & 0.4760 & 0.2533\\
$b_n$ & 0.0610 & 0.1400 & 0.1941 & 0.0415 & 0.3115 & 0.1547\\
$c_n$ & 0.090 & 0.0993 & 0.0533 & 2.15 & 0.2377 & 0.231 \\
\bottomrule
\end{tabular}
\end{center}
\end{table}

Another approximation, which
takes both free-free transitions and
Thomson scattering into account, has been developed by \citet{PY01}
following numerical calculations of \citet{SY80}.
At fixed $\rho$ and $T$,
it reads
\beq
  \frac{\kappa_\mathrm{r,\|}(B)}{\kappa_\mathrm{r}(0)} = 
   \left(\frac{K_\mathrm{r,\|}(B)}{K_\mathrm{r}(0)}\right)^{-1} = 
     1+ \frac{A_1\,u+(A_2\,u)^2 }{ 1+A_3\,u^2}\,u^2 ,
\qquad
  \frac{\kappa_\mathrm{r,\perp}(B)}{\kappa_\mathrm{r}(0)} = 
   \left(\frac{K_\mathrm{r,\perp}(B)}{K_\mathrm{r}(0)}\right)^{-1} = 
   \frac{1+ (A_4\,u)^{3.5}+(A_5\,u)^4 }{ 1+A_6\,u^2},
\label{KRmag}
\eeq
where
\bea&&\hspace*{-1em}
   u \equiv \frac{T_\mathrm{cycl}}{2T},
\qquad
  A_n = a_n - b_n\,f^{\,c_n},
\qquad
   f\equiv \frac{K_\mathrm{ff}}{K_\mathrm{ff}+K_\mathrm{T}},
\qquad
   K_\mathrm{T} = 
      \sigma_\mathrm{T}\,
      \frac{\nne}{\rho}
      = 0.4\,\frac{Z}{A}\,\frac{\mbox{cm}^2}{g},
\nonumber\\&&\hspace*{-1em}
 K_\mathrm{ff} = 2\times10^4
\frac{1+0.502\,T_\mathrm{Ry}^{0.355}+0.245\,T_\mathrm{Ry}^{0.834}}{
      108.8+77.6\,T_\mathrm{Ry}^{0.834}}
\,\frac{Z^2 }{ A}\,
          \frac{\rho}{ T_6^{7/2}} K_\mathrm{T},
\qquad
   K_\mathrm{r}(0) =
   (K_\mathrm{ff}+K_\mathrm{T})\,A(f,T)\,A_\mathrm{pl}(\rho,T),
\nonumber\\&&\hspace*{-1em}
   A(f,T) = 1+\frac{1.097+0.777\,T_\mathrm{Ry} }{
    1+0.536\,T_\mathrm{Ry}}
           \,f^{0.617}\,(1-f)^{0.77},
\qquad
   A_\mathrm{pl} = \exp\left\{ 0.005
     \left[
     \ln\left(1+ \frac{0.5}{T_6}\,\sqrt{\frac{Z}{A}\,\rho}
     \right) \right]^6 \right\}.
\nonumber
\eea
Here, mass density $\rho$ is measured in \gcc,  $K_\mathrm{r,\|}(B)$ and
$K_\mathrm{r,\perp}(B)$ are the Rosseland opacities in a strongly
quantizing magnetic field $B$ for propagation of photons along and
across $\bm{B}$, respectively, which are related by \req{K-kappa} to
$\kappa_\|$ and $\kappa_\perp$ in \req{kappa-sum}; $K_\mathrm{T}$ is the
Thomson scattering opacity at $B=0$ 
(determined by the Thomson scattering cross section 
$\sigma_\mathrm{T}$), $T_\mathrm{Ry}=\kB T/Z^2\,\mbox{Ry}
= 6.33\,T_6/Z^2$, where  $\mbox{Ry}=13.605$ eV is the Rydberg energy;
$K_\mathrm{ff}$ is the free-free opacity at $B=0$. 
The fit parameters $a_n$, $b_n$ and $c_n$ given
in Table~\ref{tab-KRmag} ensure an average fit error of 5.5\% with the
maximum error of 11\% to the numerical results of \citet{SY80}. The
factor $A_\mathrm{pl}(\rho,T)$, which effectively eliminates the
radiative transport at large densities, has been introduced by \citet{PYCG03}; it
mimics the suppression of radiative transport at photon frequencies
below the electron plasma frequency. 

The scattering opacities are modified by the electron degeneracy at
high $\rho$ and by the Compton effect at $T\gtrsim10^8$~K. An accurate
analytic description of both these effects is given by
\citet{Poutanen17}.
The free-free opacities are suppressed by electron degeneracy. In the
absence of the Landau quantization, the free-free opacities at arbitrary
degeneracy have been fitted by \citet{Schatz_ea99}, based on numerical
calculations of \citet{Itoh_ea91}.
The fit of \citet{Schatz_ea99} is inapplicable in the case of quantizing
magnetic fields. On the other hand, the fit (\ref{KRmag}) is only
applicable for non-degenerate non-relativistic plasmas. 
A smooth interpolation between the different regimes has been suggested
by \citet{PC18}.

Caution is necessary however while using these fit expressions. 
A strong
magnetic field shifts the ionization equilibrium toward a lower
ionization degree by increasing the electron binding energies. Therefore, even if
the plasma is fully ionized at some $\rho$ and $T$ in the absence of
the magnetic field, it can be only 
partially ionized at the same $\rho$
and $T$ for high $B$. This increases the contribution of 
bound-bound and bound-free transitions and can increase the
radiative opacity well above the values given by \req{K-mag}
\citep[see][]{PLP2015}.

\subsubsection{Electron thermal conductivities}
\label{subsub:econdB}

\begin{figure*}[t]
\centering
\includegraphics[width=.7\textwidth]{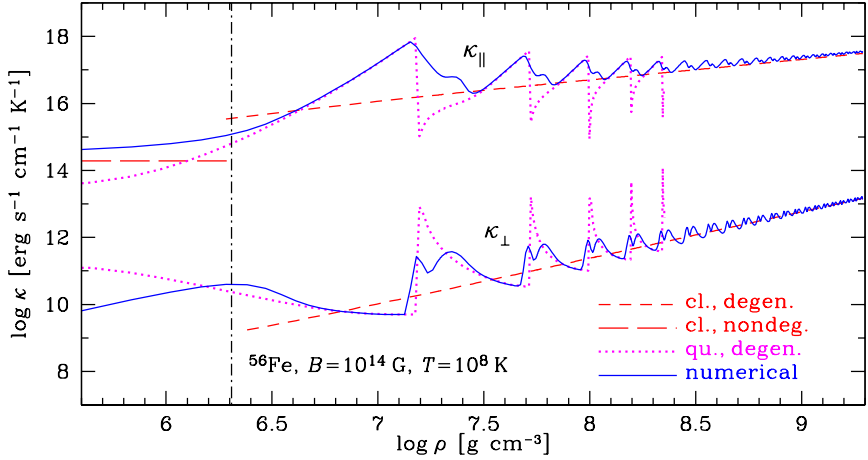}
\caption{Longitudinal ($\kappa_\|$) and transverse ($\kappa_\perp$)
electron thermal conductivities in the outer neutron-star
 envelope composed of iron at 
$T=10^8$~K and $B=10^{14}$~G: 
comparison of accurate numerical results (solid lines)
with different approximations. 
The electrons are degenerate to the right of
the vertical dot-dashed line, which marks $\rho_\mathrm{F}$.
In this domain, the classical approximations for 
$\kappa_\|$ [\req{therm-kappa_ei}] and $\kappa_\perp\sim\kappa_\|/\xi^2$
are shown by
short-dashed lines. 
The long-dashed horizontal line in the non-degenerate region
shows the conductivity $\kappa_\mathrm{e}^\mathrm{nd}$
given by \req{kappa-ND}
with $F_Z/\Lambda=1$.
The dotted lines show the approximations
including the Landau quantization effects but
without thermal averaging
of the effective electron relaxation time (i.e., with the step-like
approximation of the Fermi distribution function; cf.~\citealt{PY96}).
After \citet{VP}.}
\label{fig:cond}
\end{figure*}

Unified expressions for the electron thermal conductivities in a fully
ionized degenerate plasma with arbitrary magnetic field are discussed,
e.g., in \citet{PPP2015}. As we already mentioned in  Section~\ref{subsub:microphysics-magnetic}, 
these conductivities undergo quantum
oscillations  at $\rho\gtrsim\rho_B$. At $B\gg10^{10}$~G, the electron
transport across the field is typically suppressed by orders of
magnitude. This allows one to neglect $\kappa_\perp$, which is a good
approximation everywhere except in the domains
where $\theta_B\approx\pi/2$. In this approximation,
\req{kappa-sum} reduces to 
\beq 
\kappa\approx\kappa_\|\cos^2\theta_B. 
\label{e:kappa_eff-approx}
\end{equation} 
This formula holds not only in the degenerate but also in the
non-degenerate electron gas. In the simplest
approximation, in which the
effective electron relaxation time
is calculated for fully degenerate matter, 
the conductivity $\kappa_\|$ decreases and $\kappa_\perp$
increases towards lower densities at $\rho < \rho_B$,
as is shown in Fig.~\ref{fig:cond}. However,
averaging over the finite thermal width of the Fermi level terminates
the growth of $\kappa_\perp$ and moderates the decrease of $\kappa_\|$,
before they become comparable.

In order to construct a temperature profile, we can calculate
$\kappa_\|$ in the classical (non-magnetic) approximation
(\ref{therm-kappa_ei}) at high densities, where the magnetic field is
weakly quantizing. At lower densities, the quantizing nature of the
field must be taken into account. At $\rho<\rho_B$, as long as the
electrons are strongly degenerate and the ions form a strongly coupled
Coulomb liquid, 
one obtains the order-of-magnitude estimate \citep[see][]{VP}
\beq
\kappa_\|\simeq \frac{ 4 }{ 3b} \,
\kappa_\mathrm{nm}\,\Lambda\gammar^2\,x_B^2
\simeq 5\times10^{15}\,\frac{T_6\, x_B^3 }{ Z}
  \mbox{~erg~cm$^{-1}$~s$^{-1}$~K$^{-1}$},
\label{kappa-mag}
\end{equation}
where  $b=\hbar\omega_\mathrm{c}/\mel c^2 = B_{12}/44.14$ is magnetic
field strength in relativistic units, $\kappa_\mathrm{nm}$ is the
non-magnetic conductivity given by \req{therm-kappa_ei}, and $\Lambda$
is the non-magnetic Coulomb logarithm. This estimate gives the values of
$\kappa_\|$ not much different from numerical results, provided that
$\rho < \rho_B$ and $T \ll T_\mathrm{F}$.

As noted above, very strong fields push the onset of electron
degeneracy to higher $\rho$. Therefore, the turnover from radiative
to electron thermal conductivity may occur in the non-degenerate
regime. In that case, $\kappa_\|$ can be evaluated from
\req{kappa-ND}.

\subsubsection{Temperature profile}
\label{subsub:T(rho,B)}

In Section~\ref{subsub:microphysics-magnetic}, we have defined several
regimes regulating the EoS and opacities in strong magnetic fields. To
construct an approximate analytic temperature profile, it is sufficient
to use the non-magnetic radiative and longitudinal electron thermal
conductivity $\kappa_\|$ unless the field is strongly quantizing.
Magnetic oscillations around the classical thermal conductivities will
be smoothed out by integration while obtaining the temperature profile
from \req{therm-Tcrust}.

In the domain of strongly quantizing magnetic field, the opacities are
appreciably modified. However, in the liquid degenerate part of the heat
blanket, which is of our primary interest here, the
analytic expressions for $\kappa_\|$ can be again approximated by a
power law, Eqs.~(\ref{kappa-mag}) and (\ref{kappa-ND}). As follows from
\req{K-mag}, the same is true for the radiative opacity in the extreme
quantizing limit (provided that the free-free opacity 
dominates). In the non-degenerate regime, the magnetic field does not
affect the EoS. In this case, we recover the solution given by
\req{therm-T-rho} with the new values of $\beta=4.5$ and $\kappa_0$, 
\beq
 T_6\approx 0.95\, (\bar{g}_\mathrm{eff}\,q)^{2/9} \,(\rho
Z/A)^{4/9}\,B_{12}^{-4/9}, 
\label{nd-mag}
\eeq
 where $q$ is given by \req{nd-solution-q}. Thus the temperature is
reduced (its profile becomes less steep) with increasing $B$ as long as
the field is strongly quantizing ($\rho<\rho_B$ and $T\ll T_B$).

Interestingly, the value of the constant
conductivity along the thermal track, \req{therm-kappa=const}, is
independent of the magnetic field, while its numerical value is only
slightly lowered as a result of changing the parameter $\beta$.

\subsubsection{Sensitivity strip}
\label{subsub:sensitivityStrip}

As in the non-magnetic case, the sensitivity strip is placed near the
point, where $\kappa_\mathrm{r}=\kappa_\mathrm{e}$, and the radiative
conduction is overpowered by the electron one. In a strongly quantizing
field, using Eqs.~(\ref{K-mag}) and (\ref{kappa-mag}), we have
\beq
\rho\approx 250\,(A/Z)\,(\bar{g}_\mathrm{eff})^{-0.2}\, T_6^{0.7}\,
|\cos\theta_B|^{-0.4} B_{12}~~\gcc,
\label{turn-mag}
\end{equation}
instead of \req{turn1}. With the temperature profile (\ref{nd-mag}),
we now obtain 
\beq
T_\mathrm{t}\approx \frac{3.5\times10^7}{|\cos\theta_B|^{8/31}}
\,\bar{g}_\mathrm{eff}^{6/31}\,
q^{10/31}\,\mbox{~K},
\quad
\rho_\mathrm{t}
\approx
\frac{3\times10^3 \, q^{7/31}\,B_{12}A
}{
|\cos\theta_B|^{18/31}\, \bar{g}_\mathrm{eff}^{2/31} Z}~~\gcc.
\label{e:TtrhotB}
\end{equation}

If, however, the electrons are non-degenerate along the turning
line, then $\kappa_\|$ is given by \req{kappa-ND} instead of
\req{kappa-mag}, and we obtain the turning point at
\beq
\rho_\mathrm{t} \approx 52\,
\sqrt{\Lambda/F_Z}\,(A/Z)\,\bar{g}_\mathrm{eff}^{-1/2}\, T_6\,
|\cos\theta_B|^{-1}\,B_{12}~~\gcc.
\label{turn-nd}
\end{equation}
Combining with \req{nd-mag}, we get
\begin{equation}
T_\mathrm{t}\approx \frac{2.2\times10^7}{|\cos\theta_B|^{0.8}}
\left( \! \frac{\Lambda q}{F_Z} \!  \right)^{0.4}\mbox{K},
\quad
\rho_\mathrm{t}
\approx
\frac{1.1\times10^3 \, q^{0.4}\,B_{12} A}{
|\cos\theta_B|^{1.8}\,\bar{g}_\mathrm{eff}^{0.5}Z}
\left(\! \frac{\Lambda}{ F_Z} \! \right)^{0.9}~~\gcc.
\label{e:Tt:rhot1}
\end{equation}

Thus, in the cases of degenerate and non-degenerate electrons we
obtain quite similar expressions for $\rho_\mathrm{t}$ and $T_\mathrm{t}$.
Comparing them with \req{turn} we see that in the non-magnetic case
$T_\mathrm{t}$ has the same order of magnitude as in the magnetic field
at $\theta_B=0$. However $T_\mathrm{t}$ increases with increasing
$\theta_B$ in the magnetized envelope. Notice that in a strongly
quantized magnetic field $T_\mathrm{t}$ is independent of the field
strength, while $\rho_\mathrm{t}$ grows linearly with $B$. 
One can see that $\rho_\mathrm{t}$ lies
in the region of strong magnetic quantization. Assuming that
$\theta_B$ is not close to $\pi/2$ and neglecting the factors about
unity, we see that $\rho_\mathrm{t}<\rho_B$ for $B_{12}\gtrsim
(Z\,T_\mathrm{s6}^4/g_\mathrm{s14})^{14/31}$, which corresponds to the
high-field pulsars and magnetars (Section~\ref{sub:2Dmagnet}).

Let us also estimate the point at which the electrons become
degenerate. For simplicity, we assume that the electrons are
non-relativistic. Note that the
condition $T=T_\mathrm{F}$ in the strongly quantizing magnetic field is
equivalent to $\rho\approx608\,(A/Z)\,\sqrt{T_6}\,B_{12}$ g
cm$^{-3}$. Then from \req{nd-mag} we obtain
\beq
\rho_\mathrm{F}\simeq3700\,(\bar{g}_\mathrm{eff}\,q)^{1/7} (A/Z)
B_{12}~~\mbox{g~cm}^{-3}.
\label{e:rhoF}
\end{equation}
Thus, in analogy to the non-magnetic case, turning from 
radiative to electron thermal conduction occurs not far from the
degeneracy onset, $\rho_\mathrm{t}\sim\rho_\mathrm{F}$. Depending on
$\theta_B$, it occurs either in the non-degenerate (at
$\theta_B\approx0$) or in the degenerate (at $\theta_B\gtrsim
60^\circ$) electron gas.

The integration of the temperature profile beyond the turning point
(for obtaining $T_\mathrm{b}$) can be done in the same way as in the
non-magnetic case. However, the integration path should be divided
in two parts: \textit{(i)} $\rho<\rho_B$, where \req{kappa-mag} for the
thermal conductivity can be used, and \textit{(ii)} $\rho>\rho_B$, where
\req{dT/dx} can be used with the right-hand side divided by
$\cos^2\theta$. The result is similar to \req{therm-T1(z)}, but
contains a profound dependence on the inclination angle: the thermal
gradient grows rapidly as $\theta$ approaches $\pi/2$.

\subsection{Numerical results}
\label{sect-therm-numeric-magnetic}

We will mainly outline the models for heat blankets made of iron
\citep{PY01} and of partly accreted (PCY97-like) matter \citep{PYCG03}.
Since the isothermality in high-$B$ fields may be reached at larger
$\rho$, the bottom density $\rhob$ in these models is shifted to $4
\times 10^{11}$ \gcc. The analytic $\bar{T}_\mathrm{s}-\Tb$ fits for
magnetic models are constructed in such a way to reproduce the $\Ts-\Tb$
fits derived (Section~\ref{sect-Tb-Ts}) for $B \to 0$ at $\rhob=10^{10}$
\gcc. The models are designed for 1D codes. The  results for $B \gtrsim
10^{14}$~G are illustrative (may be improved with 2D or 3D
codes, as discussed above; it was demonstrated, for instance, by
\citealt{PPP2015}).

\subsubsection{Equation of state and opacities}
\label{subsub:EOSandOPACmag}

In the deep layers of the blanketing envelope, where the plasma is
fully ionized, the pressure is mostly determined by free electrons
 with small
corrections due to ions (Chapter 4 of \citealt{HPY07}).
The transport properties of such a plasma have been reviewed
by \citet{PPP2015}.

A considerable complication at lower densities is introduced by
bound species. As discussed in Chapter 4 by \citet{HPY07},
different approaches to the EoS yield appreciably
different models of atmospheric layers. We will mainly
use the models of iron (non-accreted) blanketing envelopes
based on the Thomas-Fermi EoS derived by \citet{Thorolfsson}.

\begin{figure*}[t]
\centering 
\includegraphics[width=\textwidth]{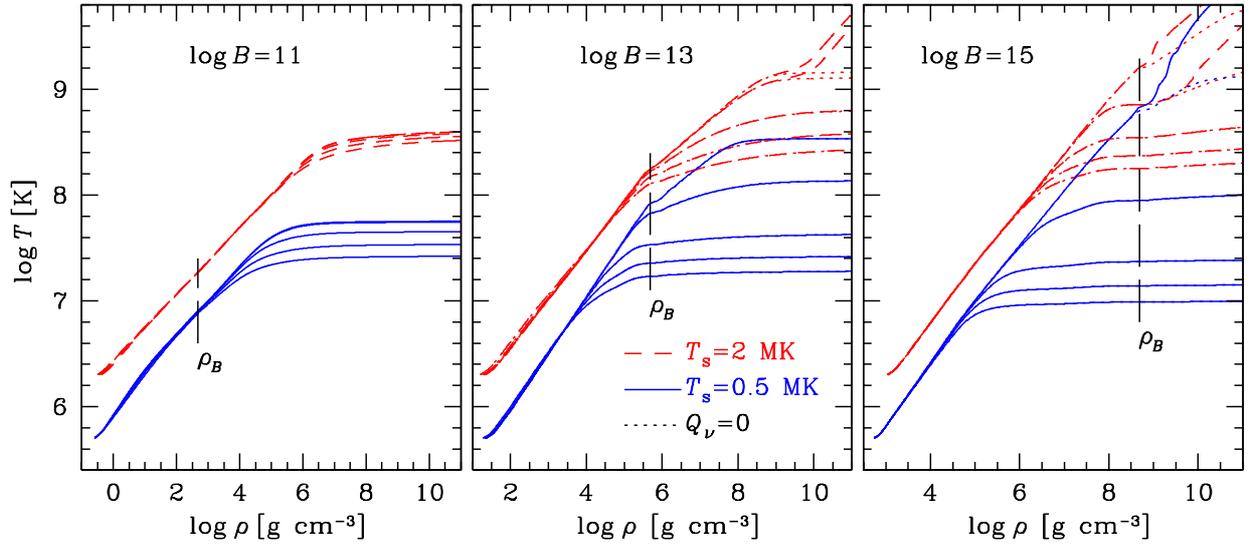} 
\caption{Temperature  profiles through an iron envelope of   a neutron
star with the surface gravity $g_\mathrm{s}=10^{14}\mbox{~cm\,s}^{-2}$,
effective surface temperature $T_\mathrm{s}=5\times10^5$~K (solid lines)
and $2\times10^6$~K (dashed lines), and different magnetic field
strengths (marked in the panels). Different lines in each bunch show
different magnetic field inclinations: $\cos\theta_B=0$ (the top line in
each bunch), 0.1, 0.4, 0.7,  and 1 (the bottom line). 
Dotted lines show the profiles calculated without
neutrino emission. The long vertical dashes mark the density $\rho_B$ at
which the first Landau level starts to be populated.
}
\label{fig-profmag}
\end{figure*}

\subsubsection{Temperature profiles}
\label{therm-temperature-profiles-B}

Fig.~\ref{fig-profmag} shows the calculated temperature profiles in
the envelopes of neutron stars at three surface magnetic fields,
$B=10^{11}$~G, $10^{13}$~G, and $10^{15}$~G, two effective surface
temperatures, $\Ts=0.5$~MK and 2~MK, and five angles $\theta_B$ between the field
and the
normal to the surface, from 0 to 90$^\circ$. The curves
start at the radiative surface, where $T=\Ts$. These results confirm the
qualitative conclusions of Section~\ref{therm-magn-analyt}. Strong
dependence on the magnetic field inclination,  $\theta_B$,  starts to be
pronounced near turning points. In accordance with our estimates
(Section~\ref{therm-magn-analyt}), they are shifted to higher
densities with increasing $B$. The linear dependence of the
radiative-surface density, $\rho_\mathrm{s}\propto B$, obtained
analytically, is seen to be realized for $B\gtrsim10^{11}$~G.

The higher the temperature, the wider is the density region, where
$\kappa_\|$ and $\kappa_\perp$ do not differ strongly from the
scalar thermal conductivity $\kappa$ at $B=0$. Therefore the
dependence of the profiles on the magnetic field $\bm{B}$ is less
pronounced at higher $T_\mathrm{b}$ showing  convergence to the
$B=0$ case.

The temperature profiles, which are calculated with allowance for
neutrino emission (solid and dashed lines), are compared with the
results of calculations assuming $Q_\nu=0$ (dotted lines). In the
$\rho-T$ domains where the difference between these results is
noticeable, the thermal flux is not constant through the envelope, so
that \req{LrTs} is not applicable.

It should be noted that the profiles in Fig.~\ref{fig-profmag} have been
calculated assuming a neutron-star photosphere without phase
transitions. However, some theoretical results hint that the strong
magnetic field may cause the so-called magnetic condensation, which implies
formation of a physical solid or liquid surface instead of imaginary
radiative surface inside an extended atmosphere (see
\citealt{MedinLai07} and references therein). \citet{PCY07}
compared thermal profiles with and without 
the magnetic condensation
and demonstrated that the effect of the condensation
on the $\Ts-\Tb$ relation is small (see
their Fig.~7). Nevertheless, this effect can be visible on the neutron
star cooling curves \citep{PC18}.

\subsubsection{Surface temperature at the magnetic pole and equator}
\label{subsub:Ts-pole-equator}

The relation between the internal and surface temperatures,
considered for non-magnetic envelopes in Section\ \ref{sect-Tb-Ts}, is
strongly affected by the magnetic fields (Fig.~\ref{fig-profmag}).
To study these effects we introduce the  ratios 
$\mathcal{R}=\Ts(\bm{B})/T_\mathrm{s0}$ of the surface temperature $T_\mathrm{s}$ at a
given field $\bm{B}$ to the value $T_\mathrm{s}=T_\mathrm{s0}$ at $B=0$
for the same $T_\mathrm{b}$.

\begin{table}[t]
\centering
  \caption{Parameters of Eqs.~(\ref{Xpar}), (\ref{Xperp})
 \label{tab:PYCG}}
  \begin{tabular}{c l c c c l l}
    \toprule
 &\multicolumn{2}{c}{Iron envelope} & \multicolumn{2}{c}{Accreted envelope} \\
 $n$   &   $a_n$      &  $b_n$      &           $a_n$     &  $b_n$          \\
\midrule
  {1}  & $1.76\times10^{-4}$ &  159 & $4.50\times10^{-3}$ &  172            \\
  {2}  &   0.038      &  270        &           0.055     &  155            \\
  {3}  &   1.5        &  172        &           2.0       &  383            \\
  {4}  &   0.0132     &  110        &           0.0595    &  94             \\
  {5}  &   0.620      &  0.363      &           0.328     &  0.383          \\
  {6}  &   0.318      &  0.181      &           0.237     &  0.367          \\
  {7}  & $2.3\times10^{-9}$ &  0.50 &  $6.8\times10^{-7}$ &  2.28           \\
  {8}  &   3          &  0.619      &           2         &  1.690          \\
  {9}  &   0.160      &             &           0.113     &                 \\
  {10} &   21         &             &           163       &                 \\   
  {11} & $4.7\times10^{5}$ &        &   $3.4\times10^{5}$ &                 \\
\bottomrule
\end{tabular}
\end{table}
  
Let us start with the two most important cases of the magnetic
field which is either normal or tangential to the surface (i.e., at the
magnetic pole or equator, respectively). In the first case the heat
is transported through the heat-blanketing surface by the
longitudinal thermal conductivity along the magnetic field lines,
$\kappa=\kappa_\|$; we will call this case the parallel
conduction case. In the second case the heat is transported by the
transverse thermal conductivity across the field lines,
$\kappa=\kappa_\perp$, which will be referred to as the 
transverse conduction case. Extensive
calculations
of the temperature profiles for both cases have been performed by
\citet{PY01} in the case of iron envelope and by \citet{PYCG03} for  partially and
fully accreted envelopes. These authors produced 
the following analytic fits:
\bea
   \mathcal{R}_\| &=& \left( 1 + \frac{a_1+a_2\,T_\mathrm{b9}^{a_3}
                }{
                T_\mathrm{b9}^2 + a_4 T_\mathrm{b9}^{a_5} }\,
                \frac{B_{12}^{a_6}
                   }{
                   (1+a_7 B_{12} 
                   / T_\mathrm{b9}^{a_8})_{\phantom{b9}}^{a_9} } \right)
       \left( 1+\frac{1}{3.7+ (a_{10}+a_{11}\,B_{12}^{-3/2})\,
          T_\mathrm{b9}^2  }  \right)^{-1} \!\!\!\!\!,
\hspace*{2em}
\label{Xpar}\\
   \mathcal{R}_\perp &=& 
    \frac{\left[ 1 + b_1\,B_{12}/(1+b_2\,T_\mathrm{b9}^{b_7}) \right]^{1/2}
       }{
 \left[ 1 + b_3\,B_{12}/(1+b_4\,T_\mathrm{b9}^{b_8}) \right]^{\beta\phantom{/}}
       } ,
\quad
   \beta=\left( 1+b_5\,T_\mathrm{b9}^{b_6} \right)^{-1},
\label{Xperp}
\eea
with the parameters $a_i$ and $b_i$ given in Table~\ref{tab:PYCG}. Here,
as before, $T_\mathrm{b9}=T_\mathrm{b}/10^9\mbox{~K}$.  These fits were
checked against calculations for input parameters restricted by the
conditions $6.5 < \log T_\mathrm{b}<9.5$, $\log T_\mathrm{s}>5.3$, and
$10<\log B<16$, where $\Tb$ and $\Ts$ are expressed in K and $B$ is G. 
The numerical values of $T_\mathrm{s}$ are reproduced with residuals up
to 5--10\%. The authors emphasized that these  results are uncertain at
superstrong fields ($B \gtrsim 10^{14}$~G).

\begin{figure}
\centering 
\includegraphics[width=.47\textwidth]{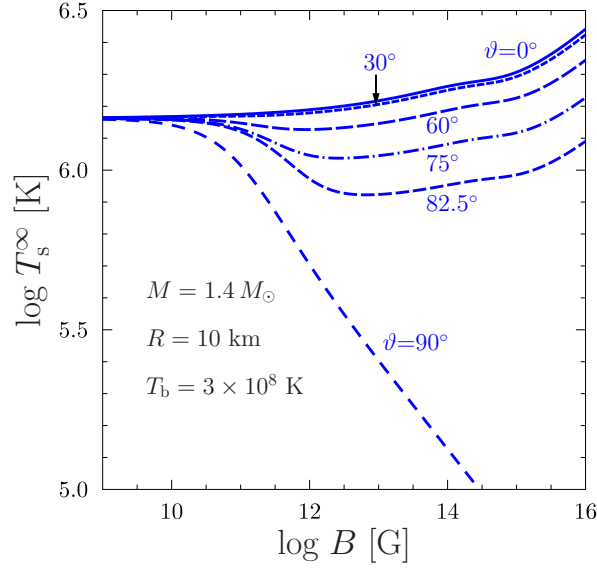} 
\caption{Redshifted local surface temperature of the canonical neutron
star with a dipole magnetic field versus the field strength
$B=B_\mathrm{pole}$ at the magnetic pole for the internal temperature
$T_\mathrm{b}= 3 \times 10^8$ K and iron heat blanketing envelope at six
values of magnetic colatitude $\vartheta$ in the 1D approximation.
} 
\label{fig-therm-angle}
\end{figure}

The effect of the magnetic field on the local effective temperature
$T_\mathrm{s}$ is illustrated in Fig.~\ref{fig-therm-angle}. The figure
shows the distribution of the redshifted temperature
$T_\mathrm{s}^\infty$ over the surface of the canonical neutron star
with the dipole magnetic field versus the magnetic field strength at the
pole.  The upper and lower curves of the same style present
$T_\mathrm{s}$ at the magnetic pole and equator, respectively. The
electron-quantization effects amplify the longitudinal thermal
conductivity, which is fully responsible for the heat transport near the
pole. These effects make the polar regions of the heat-blanketing
envelope more heat-transparent, increasing $\Ts$ for a given
$T_\mathrm{b}$. On the contrary, the classical Larmor-rotation effects
strongly reduce the transverse conductivity which is most important near
the equator. The equatorial regions become less heat transparent, which
lowers the local effective temperature. These results are in qualitative
agreement with the earlier results of \citet{KVR88}, \citet{Schaaf90a},
and \citet{hh01-multi}, although there are quantitative differences
discussed by \citet{PY01}.

\subsubsection{Variation of temperature over the stellar surface}
\label{subsub:Ts-vs-thetaB}

The dependence of $T_\mathrm{s}$ on the angle $\theta_B$ is most easily
described by the model of \citet{Greenstein} which implies a
superposition of ``longitudinal'' and ``transverse'' heat fluxes:
\beq
T_\mathrm{s}^4(B,\theta_B)=T_\mathrm{s \|}^4(B)\cos^2\theta_B
+T_\mathrm{s \perp}^4(B)\sin^2\theta_B.
\eeq
 This approximation has been
used, e.g., by \citet{page95}, \citet{ShibYak}, and \citet{HeylHern}.
Numerical calculations of \citet{PY01} confirmed that it accurately
(within $\approx30$\%) reproduces the dependence of $T_\mathrm{s}$ on
$\theta_B$. However, a replacement of the power-law index 4 with
$\alpha$ according to
\beq
   \mathcal{R} = \left(\mathcal{R}_\|^\alpha \cos^2\theta_B
   + \mathcal{R}_\perp^\alpha \sin^2\theta_B\right)^{1/\alpha},
\quad
   \alpha = \left\{  \begin{array}{ll}
   4+\sqrt{\mathcal{R}_\perp/\mathcal{R}_\|}
    & \mbox{(for iron)}, 
                           \cr\noalign{\smallskip}
         ( 2 +  \mathcal{R}_\perp/\mathcal{R}_\|)^2
                 & \mbox{(fully accreted)}.
\end{array} \right.
\label{therm-fit-angle}
\eeq
 yields better accuracy (see \citealt{PYCG03}).

According to \req{therm-fit-angle}, the flux density for any angle
$\theta_B$ is expressed through the solutions for the cases of parallel
and transverse conduction. Using Eqs.~(\ref{Xpar}) and (\ref{Xperp}),
one can thus find the flux densities at arbitrary $B$ and $\theta_B$ for
the iron ($F_r^\mathrm{(Fe)}$) and fully accreted envelope
($F_r^\mathrm{(a)}$). For a partially accreted envelope, one can use the
interpolation (\ref{F20a}), which remains reasonably accurate  for
strong magnetic fields. With this solution, one can easily calculate the
distribution of the flux density and the effective temperature $T_\mathrm{s}$
[\req{TsF20}] over the neutron star surface for  any strength and
geometry of the surface magnetic field.

Consider, for instance, a dipole surface magnetic field
\citep{1964GO},
\beq
B(\vartheta) = B_\mathrm{pole}\,\sqrt{ \cos^2\vartheta
+ a_\mathrm{g}^2\,\sin^2\vartheta},\qquad
\mathrm{tan} \, \theta_B = a_\mathrm{g}\,\mathrm{tan} \,\vartheta,
\label{B(theta)}
\end{equation}
where $\vartheta$ is the polar angle of a local element on the
surface measured from the magnetic axis, $B_\mathrm{pole}$ is the field
strength at the pole, and $a_\mathrm{g}$ is a factor due to 
General Relativity,
\beq
a_\mathrm{g} = \frac{\psi(r_\mathrm{g}/R)}{ 2 f(r_\mathrm{g}/R)},
\label{e:a_g}
\end{equation}
where
\begin{equation}
f(x) = -\frac{3}{ x^3}\left[\ln(1-x)+x+\frac{x^2 }{ 2}\right],
\quad
\psi(x) = \sqrt{1-x} \left[\frac{3}{ 1-x}-2f(x)\right].
\label{equat:f(x)}
\end{equation}
One has $a_\mathrm{g} \to 1$ in the flat-space geometry 
($r_\mathrm{g}/R \to 0$).

The distribution of $T_\mathrm{s}$ over the surface of the star with a
dipole magnetic field can be deduced from Fig.~\ref{fig-therm-angle} 
(for $T_\mathrm{b}= 3 \times 10^8$ K). This
distribution drastically depends on the magnetic field strength. The
fields $B \lesssim 3 \times 10^{10}$ G weakly affect the thermal
conductivity and, hence, the surface temperature distribution. The
fields $3 \times 10^{10}\mbox{G} \lesssim B \lesssim 3 \times
10^{13}$ G influence the transverse thermal conductivity much
stronger than the longitudinal one due to the classical
effects of electron Larmor rotation 
(Section~\ref{subsub:microphysics-magnetic}). A wide equatorial region of the
star becomes much colder than at $B=0$ while a smaller polar region
becomes slightly warmer. For the higher fields, $ B \gtrsim
10^{14}$ G, the situation is inverted: the increase of the
longitudinal thermal conductivity by 
the quantizing magnetic field becomes more
important. A large region of the surface with the center at the
magnetic pole becomes hotter than at $B=0$, but a narrow strip near
the equator stays much colder than at $B=0$. In this strip, the
approximation of plane-parallel layer used for calculating the
temperature profiles may become inaccurate. Large tangential heat
flows may smear the temperature gradients in this equatorial
``valley of the cold'' which hopefully does not affect an overall
surface temperature distribution.

Note that the anisotropy of the temperature distribution depends on
the internal temperature $T_\mathrm{b}$. With increasing $T_\mathrm{b}$
the anisotropy becomes smaller, as a result of the 
convergence to the $B=0$ case (Section~\ref{therm-temperature-profiles-B}).

\subsubsection{Total photon luminosity}
\label{subsub:TotalLs}

A $\Ts - \Tb$ relation is basic and extremely useful for non-magnetic
neutron stars; $\Ts$ is a reliable observable, and the relation allows
one to infer $\Tb$. In a magnetic star, $\Ts$ varies over the surface
and is not a robust observable  any more. Instead,  it is more
instructive to use the total surface luminosity of the star, $L_\gamma$,
which seems to be really robust (see Section~\ref{subsub:GravLens}
below).  The mean surface temperature  $\bar{T}_\mathrm{s}$ can be
conveniently  defined by \req{meanTs}.  In this way the $\Ts(\Tb)$
relation in non-magnetic stars is replaced by the $L_\gamma(\Tb)$ (or,
equivalently, $\bar{T}_\mathrm{s}(\Tb)$) relation for magnetic stars.

The total photon luminosity $L_\gamma$ is obtained by integrating the local
radiated flux, $\sigma_\mathrm{SB} T_\mathrm{s}^4$, over the entire
stellar surface.  \citet{PY01}
numerically calculated $L_\gamma$ for iron blanketing envelopes with
the dipole magnetic field and isotropic surface emission model
and fitted the result as
\begin{equation}
\frac{L_\gamma(B) }{ L_\gamma(0)} = 
\frac{1 + a_1 \beta^2 + a_2 \beta^3
+ 0.007 \, a_3 \beta^4
}{
1 + a_3 \beta^2},
\label{fit-F}
\end{equation}
where
\begin{eqnarray}
\beta &=&  0.074 \,\sqrt{B_{12}} \,T_\mathrm{b9}^{-0.45},
\nonumber\\
a_1 &=& \frac{5059 \,T_\mathrm{b9}^{3/4}
}{
(1 + 20.4 \,T_\mathrm{b9}^{1/2}
+ 138 \,T_\mathrm{b9}^{3/2}
+ 1102 \,T_\mathrm{b9}^2)^{1/2} },
\nonumber\\
a_2 &=& \frac{1484 \, T_\mathrm{b9}^{3/4}
}{
(1 + 90\,T_\mathrm{b9}^{3/2}
+ 125 \,T_\mathrm{b9}^2 )^{1/2} },
\nonumber\\
a_3 &=& \frac{5530 \,(1 - 0.4 \,r_\mathrm{g}/R)^{-1/2}
\,T_\mathrm{b9}^{3/4}
}{
(1 + 8.16 \,T_\mathrm{b9}^{1/2}
+ 107.8 \, T_\mathrm{b9}^{3/2}
+ 560 \,T_\mathrm{b9}^2)^{1/2} },
\nonumber
\end{eqnarray}
and $B_{12}$ relates to the magnetic pole. The maximum fit error  is
6.1\%\ (i.e., 1.5\%\ for the mean effective temperature
$\bar{T}_\mathrm{s}$). Note that the fit expressions used by the
authors for calculating $L_\gamma(B)$ were less accurate by themselves.
This lowered the real accuracy of the presented fits. Nevertheless this
accuracy seems sufficient for cooling simulations.

\begin{figure}
\centering 
\includegraphics[width=.6\textwidth]{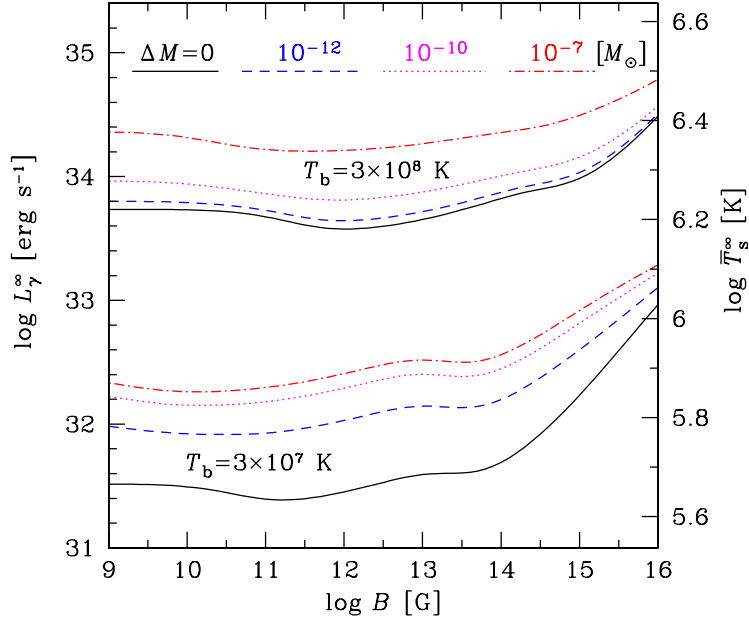} 
\caption{Redshifted photon surface luminosity (left
vertical axis) and mean effective temperature (right vertical axis)
of the canonical neutron star  with 
a dipole magnetic field, for two
values of $T_\mathrm{b}$ and four models of the heat-blanketing
envelope (accreted mass $\Delta M=0$, $10^{-12}$, $10^{-10}$,
and $10^{-7} \, {M}_\odot$) versus magnetic field strength
at the pole $B=B_\mathrm{pole}$.
(After \citealt{PYCG03}.)
}
\label{fig-teb}
\end{figure}

Fig.~\ref{fig-teb} displays the photon luminosity versus
$B=B_\mathrm{pole}$ for two values of $T_\mathrm{b}$ and four values of
$\Delta M$, mass of accreted matter in the blanketing envelope. The
magnetic field affects the photon luminosity at $B \gtrsim 3 \times
10^{10}$ G. In the range of $B \lesssim 3 \times 10^{13}$ G, the
equatorial decrease of the heat transport dominates, and the luminosity
is lower than at $B=0$. For $B \gtrsim 10^{14}$~G, the polar increase of
the heat transport becomes more important, and the magnetic field
enhances the photon luminosity.

The joint effect of the accreted envelope and the magnetic field is
demonstrated by the dot-dashed, dotted, and dashed lines. As in the
non-magnetic case, the accreted material makes the envelope more
heat-transparent, increasing the luminosity at given $\Tb$. Therefore,
at $B\sim10^{10}$--$10^{13}$ G, the magnetic field and the accreted
envelope affect the thermal insulation in the opposite directions. At
higher $B$, both effects increase the luminosity. However, as evident
from Fig.~\ref{fig-teb}, the dependence of this increase on $B$ and
$\Delta M$ is complicated. In particular, at $B \gtrsim 10^{14}$ G, the
effect of the accreted envelope is weaker than in the
non-magnetic case.

Note one important feature: the effect of the magnetic field on
$L_\gamma$ becomes weaker with growing $T_\mathrm{b}$. It is explained
by the convergence to the $B=0$ solution discussed in Section~\ref{therm-temperature-profiles-B}. Accordingly, the luminosity of a
hot neutron star cannot be strongly affected even by very high
magnetic fields.

\subsubsection{Gravitational lensing and observed flux}
\label{subsub:GravLens}

Since the temperature distribution over the surface of a magnetized
neutron star is non-uniform, the flux of radiation $F$ detected from
the star depends on observation direction. Calculating the
flux, it is important to take into account gravitational bending of
light rays propagating from the stellar surface to a distant
observer. We will illustrate this effect of General Relativity for a
spherically symmetric neutron star (with the Schwarzschild
space-time geometry outside the star) using the results of
\citet{pfc83}
and \citet{zsp95},
and employing the heat-blanket model designed for 1D cooling code.
The most important consequence of light bending is that an observer
will collect radiation from a larger part of the stellar surface. The
maximum colatitude of the surface element (with respect to line
of sight) visible at infinity, $\vartheta_\mathrm{max}^\mathrm{viz}$, is
determined by the compactness parameter $r_\mathrm{g}/R$,
\beq
\vartheta_\mathrm{max}^\mathrm{viz}= \int_0^{r_\mathrm{g}/2R} \frac{ \dd u }{
\sqrt{ (1-r_\mathrm{g}/R)(r_\mathrm{g}/2R)^2 - (1-2u) u^2}}.
\label{therm-light-bending}
\end{equation}
One has evidently $\vartheta_\mathrm{max}^\mathrm{viz}=90^\circ$ and the
visible fraction of the stellar surface $s=0.5$ for a flat space,
$r_\mathrm{g}/R \to 0$. One has $\vartheta_\mathrm{max}^\mathrm{viz}=112^\circ$
and $s=0.686$ for a neutron star with mass $M=1.4 \, {M}_\odot$
and radius $R=15$ km ($r_\mathrm{g}/R=0.275$); and $\vartheta_\mathrm{max}^\mathrm{viz}=132^\circ$ and $s=0.833$ for the 
canonical neutron star ($r_\mathrm{g}/R=0.413$). The
observer would see the entire surface ($\vartheta_\mathrm{max}^\mathrm{viz}=180^\circ$) at $r_\mathrm{g}/R=0.568$. With increasing $r_\mathrm{g}/R$, the photons emitted from certain places of
the surface nearly tangentially to it should move along more tightly
curved spirals to reach the observer. At $r_\mathrm{g}/R=2/3$, the
tangentially emitted photons would travel along closed circular
orbits ($\vartheta_\mathrm{max}^\mathrm{viz}=\infty$). For a 
canonical neutron star, the cases of $\vartheta_{\max}^{
viz}=180^\circ$ and $\infty$ would realize at $R=7.27$ and $6.19$
km, respectively. The propagation of light rays emitted from rapidly
rotating neutron stars was considered in many publications
(e.g.,
\citealt{brr00,Cadeau_07,Baublock_15,NattilaPihajoki18,Poutanen20,SuleimanovPW20},
and references therein).

\begin{figure}
\centering \includegraphics[width=.47\textwidth]{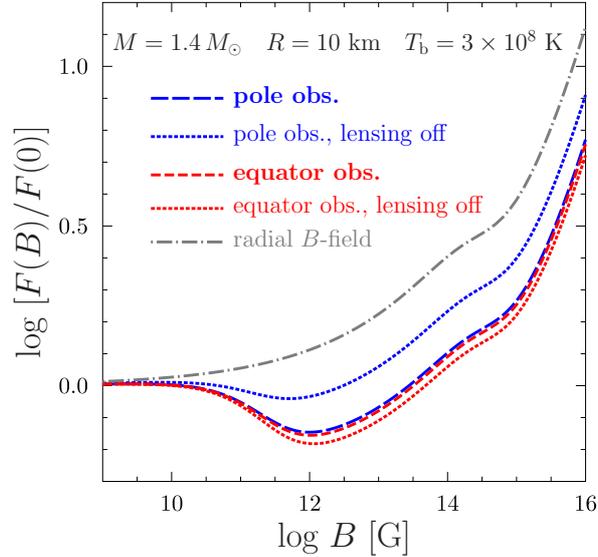}
\caption{Effect of dipole or radial surface magnetic fields on the flux
of electromagnetic radiation $F(B)$ detected from a  canonical neutron
star for the internal temperature $T_\mathrm{b}=3 \times 10^8$ K at the
bottom of the iron heat blanket. Long-dashed line presents the flux
observed in the direction towards the magnetic pole and short-dashed
line towards the magnetic equator as functions of the field strength $B$
at the pole.  The upper and lower dotted lines show the same fluxes
calculated neglecting gravitational bending of light rays outside the
neutron star.  The dot-dashed line is for the radial field of the same
strength as the field at the magnetic pole.
}
\label{fig-gravlens}
\end{figure}

The gravitational bending effect is illustrated in
Fig.~\ref{fig-gravlens} for the canonical neutron star. For comparison,
by the dot-dashed line we show the flux from the star where the magnetic
field is radial everywhere on the surface; the surface temperature
distribution is then isotropic (although $\Ts$ depends on $B$), the
detected flux $F$ (radiated from a visible part of the surface)  is
evidently independent of observation direction. All other lines in the
figure show the fluxes $F(B)$ for the dipole magnetic field. The
dash-and-dot lines refer to the fluxes detected either along the
magnetic pole or along the equator (assuming the magnetic axis coincides
with the rotational one). These lines demonstrate the largest difference
of the fluxes detected under different angles. They are obtained taking
proper account of bending of light rays. For comparison, we present also
the fluxes calculated neglecting the gravitational ray-bending effect
(as if space-time were flat outside the star). In the absence of light
bending, the difference of the fluxes observed under different angles
would be quite noticeable. For instance, at $B=4 \times 10^{12}$ G the
difference would be about 57\%. The light bending reduces it to 4\%,
making it almost negligible. The observer detects the flux from a large
fraction of the surface, that is close to the flux averaged over
observation directions.

Therefore, the light bending
for neutron stars is usually strong and crucial. It allows one to
neglect weak dependence of the observed fluxes on the detection
direction and use the average fluxes and the mean effective
temperatures $\bar{T}_\mathrm{s}$ in the theories of neutron-star
thermal evolution. 
\emph{Although the local surface temperature of a strongly
magnetized neutron star largely varies over the surface, 
General Relativity
disguises these variations: the
flux of thermal radiation, as detected by a distant observer, 
is mostly determined by the total luminosity of 
the star.}
Therefore, if phase-resolved observations of some neutron star
demonstrate noticeable variations of the bolometric flux, they cannot
be attributed to the blackbody (isotropic) thermal radiation emergent
from the star with a dipole magnetic field and isothermal interior. 

Nevertheless, the  bolometric flux showing large variations with
rotation phase still can be attributed to the thermal radiation emergent
from the star with a dipole magnetic field and isothermal interiors,
provided that radiative transfer in the strongly magnetized neutron star
photosphere is treated accurately. In contrast with the isotropic
blackbody radiation, discussed above, radiation of a magnetic
photosphere consists of a narrow ($<5^\circ$) pencil beam along the
magnetic field and a broad fan beam with typical angles 
$\sim20^\circ-60^\circ$ \citep{Zavlin_95}.  For example,
\citet{Storch_14} have demonstrated that the large X-ray pulse fraction
of PSR B0943+10 can be explained by including the beaming effect of a
magnetic atmosphere, while remaining consistent with the dipole field
geometry constrained by radio observations. 

\section{Heat blanketing envelopes and cooling of isolated 
neutron stars}
\setcounter{equation}{0}
\label{sec:5}

In the final sections we outline the applications of heat blanket models
for numerical simulations of observational manifestations of neutron
stars. In this section, we consider the most familiar  application to
cooling isolated middle-aged neutron stars  neglecting the effects of
magnetic fields. We do not pretend to  give a detailed description of
the neutron star cooling  theory and observations. The neutron-star
cooling theory was described in a number of detailed reviews
\citep{YP04,Page09,Tsuruta09,PPP2015,Geppert17}, and an
up-to-date survey of the observations of thermally emitting cooling
neutron stars with references to original works can be found in
\citet{COOLDAT}.\footnote{An updated list of the basic properties of the
thermally emitting neutron stars, extracted from observations, is
available at
\url{http://www.ioffe.ru/astro/NSG/thermal/}.\label{COOLDAT}}

\subsection{Cooling simulations}
\label{ssec:5:Intro}

Evidently, the interpretation of
observations of neutron stars is greatly complicated by
uncertainties in the chemical composition of heat blankets. 
Here, following \citet{Beznogov_etal16}, 
we illustrate the effects of
these uncertainties on thermal evolution of isolated neutron stars.

The main objects of study will be not very young 
(age $t \gtrsim 100$ yr) cooling isolated neutron stars. These stars
have already passed the early stage of internal relaxation
(e.g., \citealt{GYP01,YP04} and references therein). Their internal regions are
already isothermal, with  large temperature gradients 
remaining only in the heat-insulating 
blankets. 

Aside of the envelopes consisting of binary ion 
mixtures (Section~\ref{sec:4}), it is
important to study the PCY97 envelope model 
(Section~\ref{therm-env}) containing sequences of
spherical shells of pure H, He, C, and Fe. 
The corresponding $\Ts-\Tb$ relation is governed by $\Delta M$,
the accumulated mass of H and He. 

The PCY97 model was elaborated further by \citet{PYCG03} by including
the effects of the magnetic fields (Section~\ref{therm-magn-env}) and
the effect of temperature growth to $\rhob>10^{10}$ \gcc{}
(Section~\ref{sect-Tb-Ts}).  \citet{Beznogov_etal16} compared the
diffusive-equilibrium models with the widely used PCY97 model and
employed the neutron star models with  the BSk21 EoS \citep{GCP10,
PCGD12, Potekhin_etal13}  in their interiors (at $\rho>\rhob$).  In this
case, the maximum mass for stable neutron star models is
$M_\mathrm{max}=2.27\,M_\odot$ (with $R=11.04$ km and the central
density $\rho_\mathrm{c}=2.29 \times 10^{15}$ \gcc). The most powerful
direct Urca process of neutrino emission \citep{LPPH91,Haensel95} in
the  cores of such stars is allowed at $M > M_\mathrm{Durca}=1.59
\,M_\odot$ ($\rho_\mathrm{c}> 8.21 \times 10^{14}$ \gcc).  Furthermore,
we fixed $M=1.4\,M_\odot$ star (with $R=12.60$ km) at which the direct
Urca process is forbidden, and we  neglect the effects of superfluidity
of nucleons in the core and the crust, focusing on the effects of heat
blankets. The main neutrino cooling process for such a star would be 
the modified Urca process from the core, which is treated following
\citet{YKGH01}. Note that the efficiency of this process can be enhanced
by in-medium effects (e.g., \citealt{SHTERNIN2018}). This enhancement is
not included into our illustrative calculations here.

Fig.~\ref{fig:CoolCompar} presents some  computed cooling curves. Each
panel of this figure refers to one model of the blanketing envelope. The
upper left panel shows the envelopes made of  H\,--\,He mixtures, the
upper right panel is for He\,--\,C mixtures, the bottom left panel for 
C\,--\,Fe mixtures and the bottom right panel is for the PCY97 model. A
shaded strip in each panel is composed of possible cooling curves
for a given heat blanket (only $\Delta M$ varies whereas other
parameters are fixed). These strips fill the areas between thick lines
showing the cooling of neutron stars with
nearly pure hydrogen and helium, helium and carbon, carbon
and iron, iron and the PCY97 envelopes with the highest amount of light
elements (labeled as ``Acc''). The dashed lines are calculated for some
intermediate values of $\Delta M$. They demonstrate that the strips are
really filled with the cooling curves. Because of the reasons discussed
in Section~\ref{ssec:4:HeatBlanketsModel}, different envelope models are
taken for different values of $\rhob$. That is why the cooling curve
for pure He in the upper left panel is slightly different from the
analogous curve in the upper right panel.

\begin{figure}[!]
  \centering
  \includegraphics[width=.48\textwidth]{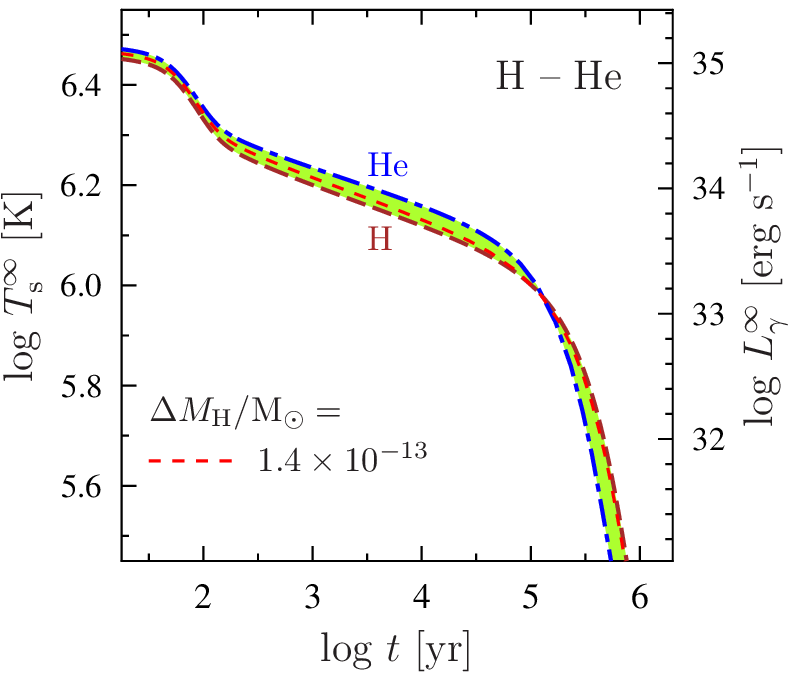}~
  \includegraphics[width=.48\textwidth]{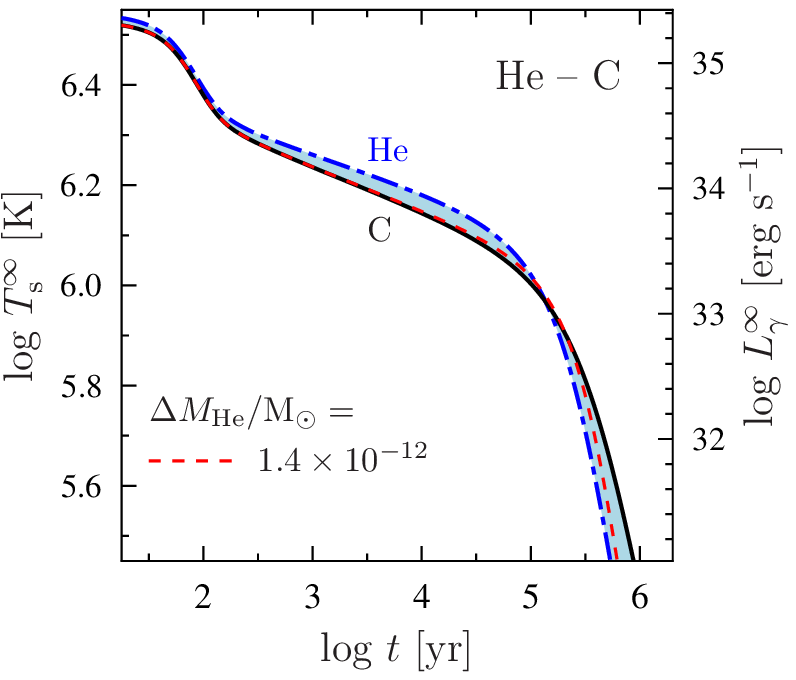}
\\ \vspace{1ex}%
  \includegraphics[width=.48\textwidth]{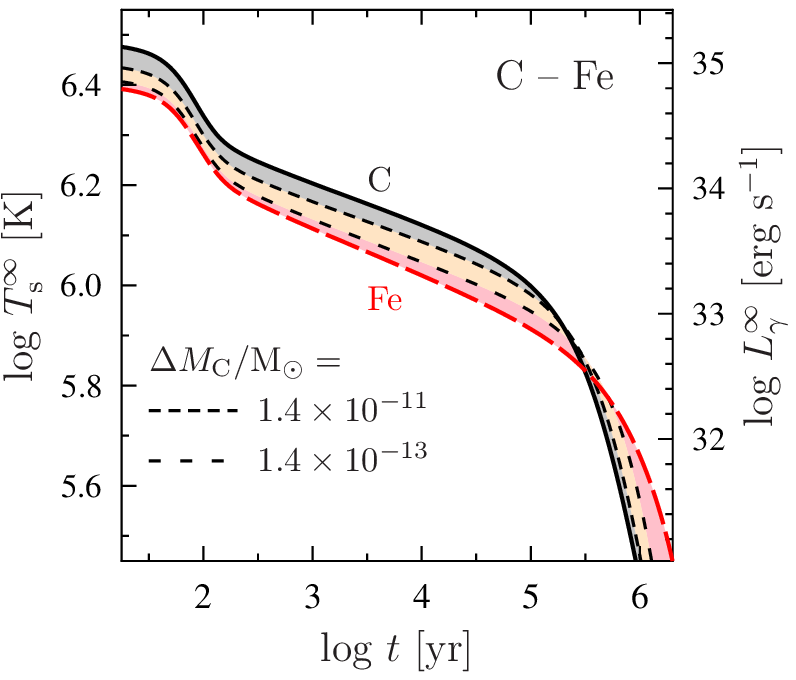}~
  \includegraphics[width=.48\textwidth]{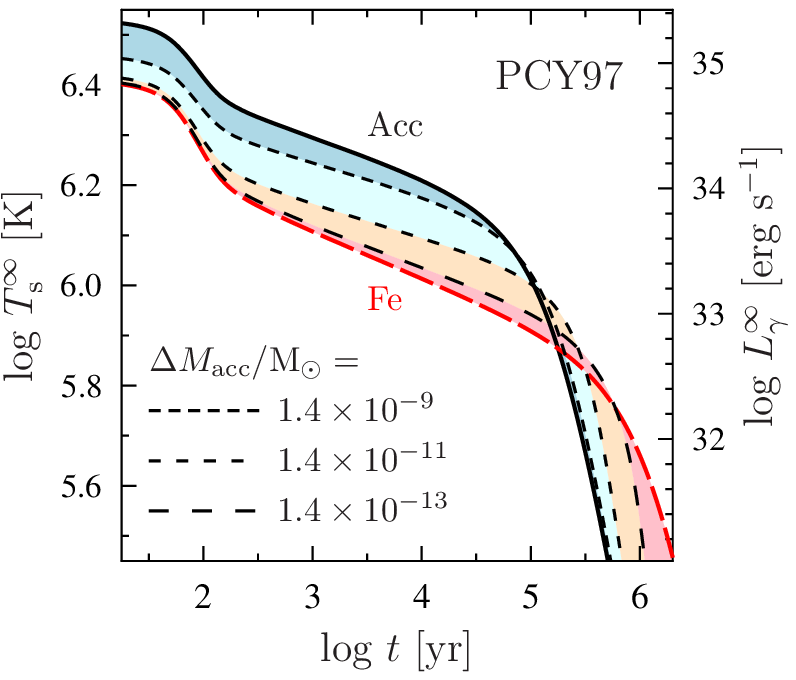}
  \caption{Cooling curves $\Ts^\infty(t)$ (left vertical axes)
   and $L_\gamma^\infty(t)$ (right vertical axes)
   for the 1.4\,$M_\odot$ star
    with the BSk21 EoS and with different models of
    heat blanketing envelopes. The star is 
    non-superfluid and the direct Urca process
    of neutrino emission is forbidden. See text for details.
    (After \citealt{Beznogov_etal16}.)
}
  \label{fig:CoolCompar}
\end{figure}

The strips shown in Fig.~\ref{fig:CoolCompar} can be viewed  as some
cooling curves, ``broadened'' due to unknown $\Delta M$. Clearly, the
strip widths depend on the envelope type. For the envelopes made of
H\,--\,He and He\,--\,C mixtures, the broadening is rather weak; for the
C\,--\,Fe envelopes it is wider; the largest broadening is naturally
provided by the PCY97 envelope. 

\subsection{Blanket composition and internal structure of neutron stars}
\label{ssec:5:errors}

The larger the broadening, the higher the uncertainty of the internal
temperature of a neutron star
$\widetilde{T}$ inferred from observations using
heat-blanket models. This 
uncertainty leads even to higher uncertainties in the
neutrino cooling function,
\begin{equation}
   \ell(\widetilde{T})= \frac{L_\nu^\infty(\widetilde{T})}{C(\widetilde{T})},
\label{e:ell}  
\end{equation}
that is the ratio of the total neutrino luminosity to the
total heat capacity of the star. Both, 
$L_\nu^\infty(\widetilde{T})$ and $C(\widetilde{T})$, are
mostly determined by neutron star cores. From  
observations of cooling middle-aged neutron stars one 
can estimate $\tilde{T}$. Then, using the cooling theory,
one can estimate (constrain) the cooling functions
$\ell(\widetilde{T})$ which contain the most important
information on microphysical properties of superdense
matter in neutron star cores. 

The attempts to evaluate $\ell(\widetilde{T})$ for some  cooling neutron
stars have been done in numerous publications, particularly by
\citet{Beznogov_etal16}.  The uncertainties of  inferring
$\widetilde{T}$ from observable $\Ts^\infty$ due to unknown chemical
composition of heat blankets can reach a factor of $\sim$2.5 (see
Section~\ref{ssec:4:Concl}). This leads to uncertainties of inferring
$\ell(\widetilde{T})$ by a factor of $\sim 10^2$ because of a strong
temperature dependence of $\ell(\widetilde{T})$. Since, theoretically,
$\ell(\widetilde{T})$ can vary within $\sim$10 orders of magnitude 
(e.g., \citealt{Beznogov_etal16}), depending on composition and
superfluid properties of neutron stars cores, the uncertainties by a
factor of $\sim 10^2$ should not be treated as ``too enormous to be
meaningless'', but the problem to reduce them via understanding the
composition of heat blankets looks very important.

\subsection{Cooling neutron stars with different envelopes}
\label{ssec:5:PhotonCooling}

Another characteristic feature of cooling curves seen from
Fig.~\ref{fig:CoolCompar} is their inversion: at a certain age $t$ 
a strip of any curve becomes narrow but then 
widens again; the cooling curves intersect and
interchange. For instance, for the C\,--\,Fe mixture before the
inversion the curve for pure C heat blanketing envelope goes higher
than for pure Fe envelope, while after the inversion it becomes
lower. The inversions occur at $t \approx 10^5$ yr
for the envelopes containing H\,--\,He and He\,--\,C mixtures, and at $t
\approx (2-3) \times 10^5$ yr for the C\,--\,Fe and PCY97 envelopes.
Notice that for the envelopes composed of binary ionic mixtures, all
cooling curves intersect at almost one and the same $t$, while for
the  PCY97 envelopes they intersect in a small area in the
$\Ts^\infty - t$ plane.

Such inversions are well known in the literature (e.g.,
\citealt{YP04}). They manifest the transition of the star from the
neutrino cooling stage to the photon cooling stage. The transition
time is relatively short.  At the
neutrino cooling stage, the star is mainly cooling via neutrino
emission from the entire stellar interior; it has lower surface
temperature if its heat blanket consists of heavy elements with low
thermal conduction. At the photon cooling stage, the star is mostly
cooling via thermal photon emission from the surface; the neutrino
emission stops to affect the cooling which is now regulated by the heat
capacity of the core and the thermal conductivity of the heat blanket. The lower the thermal conductivity, the higher $\Ts$.

\begin{figure}[!]
  \centering
  \includegraphics[width=.48\textwidth]{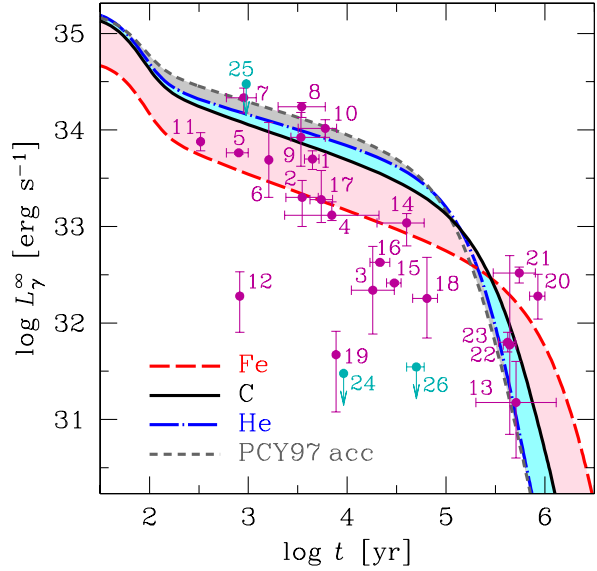}
  \caption{Some selected cooling curves of neutron stars
    from Fig.~\ref{fig:CoolCompar} compared with observations.
    The observational data are plotted
for the neutron stars with known kinematic ages and estimated
thermal luminosities from the surface.
    See the text for details. }
  \label{fig:CoolCompar-2}
\end{figure}

 \begin{table*}[!t]
    \centering
    \caption{Cooling neutron stars with known ages and thermal luminosities.}
    \label{tab:AgeLshort}
    \begin{tabular}{r @{\hspace{1ex}} l @{\hspace{1ex}} c @{\hspace{1ex}} c @{\hspace{1ex}} c @{\hspace{1ex}} c}
      \toprule
 no.  & Object               &      Age             & $L_\gamma^\infty$           & ${\kB}T^\infty$        & Reference numbers\\
      &                      &      (kyr)           & ($10^{32}$ erg s$^{-1}$)    &    (eV)                &                  \\
      \midrule
      \multicolumn{6}{c}{Weakly magnetized thermally emitting neutron stars}\\
 1 & RX J0822.0$-$4300 (in Puppis A) & $4.45\pm0.75$        & $50\pm11$                   & $265\pm15$    & 1(a), 2(s)       \\[.2ex]
 2 & CXOU J085201.4$-$461753 (``Vela Jr.'') & 2.1\,--\,5.4         & $20\pm10$                   & $90\pm10$              & 3(a), 4(s)       \\[.2ex]
 3 & 2XMM J104608.7$-$594306 & 11\,--\,30           & 0.77\,--\,6.2               & 40\,--\,70             & 5(a,s)           \\[.2ex]
 4 & 1E 1207.4$-$5209 (in G296.5+10.0) & $7^{+14}_{-5}$       & $13.1^{+4.9}_{-1.6}$        & 90\,--\,250            & 6(a), 7(s)       \\[.2ex]
 5 & CXOU J160103.1$-$513353 & $0.8\pm0.2$          & $58\pm2$                    & $118\pm1$              & 8(a), 9(s)       \\[.2ex]
 6 & 1WGA J1713.4$-$3949 (in G347.3$-$0.5) &  1.608               & $\sim20-120$                & $138\pm1$              & 10(a), 11(s)     \\[.2ex]
 7 & XMMU J172054.5$-$372652 & $0.9\pm0.3$          & $216^{+55}_{-66}$           & $161\pm9$              & 12(a), 11(s)  \\[.2ex]
 8 & XMMU J173203.3$-$344518 &  2\,--\,6            & $174^{+19}_{-39}$           & $153^{+4}_{-2}$        & 13(a), 14(s)     \\[.2ex]
 9 & CXOU J181852.0$-$150213 & $3.4^{+2.6}_{-0.7}$  & $84^{+68}_{-42}$            & $130\pm20$             & 15(a), 16(s)     \\[.2ex]
10 & PSR J1852+0040 (in Kes 79) & $6.0^{+1.8}_{-2.8}$  & $104^{+24}_{-20}$           & $133\pm1$              & 17(a), 18(s)     \\[.2ex]
11 & CXOU J232327.8+584842 (in Cas A)~& 0.320\,--\,0.338     &  61\,--\,94                 & 123\,--\,185           &19(a), 20\,--\,20(s)\\[.2ex]
      \multicolumn{6}{c}{Middle-aged pulsars}\\
12 & PSR J0205+6449 (in 3C 58)&    0.819             & $1.9^{+1.5}_{-1.1}$         & $49^{+5}_{-6}$         & 23,24(a), 11(s)  \\[.2ex]
13 & PSR J0357+3205 (``Morla'')& 200\,--\,1300        & $0.15^{+0.25}_{-0.11}$      & $36^{+9}_{-6}$         & 25(a,s)          \\[.2ex]
14 & PSR J0538+2817 (in Sim 147) & $40\pm20$            & $10.9^{+2.7}_{-4.6}$        & $91\pm5$               & 26(a), 27(a,s)   \\[.2ex]
15 & CXOU J061705.3+222127 (in IC 443) &  $\sim30$            & $2.6\pm0.1$                 & $58.4^{+0.6}_{-0.4}$   & 28(a), 29(s)     \\[.2ex] 
16 & PSR B0833$-$45 (Vela pulsar) & 17\,--\,23           & $4.24\pm0.12$               & $57^{+3}_{-1}$         & 30(a), 31(s)     \\[.2ex]
17 & PSR J1119$-$6127 (in G292.2$-$0.5)& 4.2\,--\,7.1         & $19^{+19}_{-8}$             & $\sim80$\,--\,210      & 32(a), 33(s)     \\[.2ex]
18 & PSR B1951+32 (in CTB 80)&   $64\pm18$          & $1.8^{+3.0}_{-1.1}$         & $130\pm20$             & 34(a), 35(s)     \\[.2ex]
19 & PSR B2334+61 (in G114.3+0.3)&   $\sim7.7$          & $0.47\pm0.35$               & $38^{+6}_{-9}$         & 36(a), 37(s)     \\[.2ex]
      \multicolumn{6}{c}{Strongly magnetized X-ray emitting isolated neutron stars (XINSs)}\\
20 & RX J0720.4$-$3125       & $850\pm150$          & $1.9^{+1.3}_{-0.8}$         & 90\,--\,100            & 38(a), 39,40(s)  \\[.2ex]
21 & RX J1308.6+2127         & $550\pm250$          & $3.3^{+0.5}_{-0.7}$         & $\sim50$\,--\,90       & 41(a), 42(a,s)   \\[.2ex]
22 & RX J1605.3+3249         & $440^{+70}_{-60}$    & 0.07\,--\,5                 & 35\,--\,120            & 43(a), 44,45(s)  \\[.2ex]
23 & RX J1856.5$-$3754       & $420\pm80$           & 0.5\,--\,0.8                & 36\,--\,63             &46(a), 47\,--49(s)\\[.2ex]
      \multicolumn{6}{c}{Useful upper limits}\\
24 & PSR J0007+7303 (in CTA 1) &    $\approx9.2$      & $<0.3$                      &   $<200$               & 50(a), 51(s)     \\[.2ex]
25 & PSR B0531+21 (Crab pulsar) &    0.954             & $<300$                      &   $<180$               & 52(a), 53(s)     \\[.2ex]
26 & PSR B1727$-$47 (in RCW 114) &  $50\pm10$           & $<0.35$                     &   $<33$                & 54(a), 11(s)     \\[.2ex]
\hline
\multicolumn{6}{@{}p{\linewidth}}{
\rule{0pt}{3ex}
\textbf{References:} 
 1.~\citet{Becker_ea12};
 2.~\citet{DeLuca_ea12};
 3.~\citet{Allen_15};
 4.~\citet{Danilenko_15};
 5.~\citet{Pires_ea15};
 6.~\citet{Roger_ea88};
 7.~\citet{Mereghetti_ea02};
 8.~\citet{Borkowski2018-age-cco-snrg330};
 9.~\citet{Doroshenko2018-cco-snrg330};
10.~\citet{CassamChenai_04};
11.~\citet{COOLDAT};
12.~\citet{Lovchinsky_11};
13.~\citet{CuiPS16};
14.~\citet{Klochkov_etal15};
15.~\citet{Sasaki_ea18};
16.~\citet{Klochkov_16};
17.~\citet{Sun_ea04};
18.~\citet{Bogdanov14};
19.~\citet{Ashworth80};
20.~\citet{HH10};
21.~\citet{PosseltPavlov18};
22.~\citet{Wijngaarden_ea19};
23.~\citet{Stephenson71};
24.~\citet{Kothes13};
25.~\citet{Kirichenko_etal14};
26.~\citet{Kramer_03};
27.~\citet{Ng_ea07};
28.~\citet{Chevalier99};
29.~\citet{Swartz_15};
30.~\citet{Aschenbach02};
31.~\citet{2018VELA};
32.~\citet{KumarSHG12};
33.~\citet{Ng_ea12};
34.~\citet{Migliazzo_02};
35.~\citet{LiLuLi05};
36.~\citet{Yar-Uyaniker_UK04};
37.~\citet{McGowan_ea06};
38.~\citet{Tetzlaff_ea11};
39.~\citet{Hohle_ea12};
40.~\citet{Hambaryan_17};
41.~\citet{Motch_ea09};
42.~\citet{Hambaryan_ea11};
43.~\citet{Tetzlaff_ea12};
44.~\citet{Pires_ea19};
45.~\citet{Malacaria_19};
46.~\citet{Mignani_ea13};
47.~\citet{Ho_etal07};
48.~\citet{Sartore_12};
49.~\citet{Yoneyama_ea17};
50.~\citet{Martin_16};
51.~\citet{Caraveo_10};
52.~\citet{StephensonGreen03};
53.~\citet{Weisskopf_ea11};
54.~\citet{Shternin_19}.
}
\\
      \bottomrule
    \end{tabular}
\end{table*}
\vfill

Fig.~\ref{fig:CoolCompar-2} presents some selected cooling curves from
Fig.~\ref{fig:CoolCompar}. The long-dashed curve refers to the
blanketing envelope of pure iron; the solid curve is for the pure
carbon; the gray short-dashed line (``PCY97 acc'') is for the PCY97
envelope with the maximum amount of light elements. The shaded region
between the Fe and C curves is filled by the cooling curves calculated
assuming the C\,--\,Fe heat blanket with different $\Delta M$. The
region between the C and He cooling curves is filled by cooling curves
for the He\,--\,C envelope. The region between the He and ``PCY97 acc''
lines is filled by those cooling curves for the PCY97 envelope, which
avoided other filled regions. To simplify the figure we do not show the
cooling curves for the blanketing envelopes composed of H\,--\,He
mixtures. In addition, in Fig.~\ref{fig:CoolCompar-2} we plot some
observational data on isolated neutron stars whose thermal emission has
been detected or upper limits have been obtained (see footnote
\ref{COOLDAT} on page \pageref{COOLDAT}). We have chosen to plot the
estimated luminosities,  rather than temperatures, for the reasons
discussed in \citet{COOLDAT} (also see \citealt{Vigano_13}). Here, we
show only the objects with available ``kinematic ages'', estimated
independently of pulsar timing, and we adopt them as the observational
estimates of the true ages. Such estimate can be based on proper motion
of the star, on physical properties of the associated SNR or surrounding
nebula, or, in a few cases, on historical supernova dates. For
convenience of the reader, we list the estimated ages, luminosities, and
temperatures of these neutron stars in Table~\ref{tab:AgeLshort}
(details and discussion are given in \citealt{COOLDAT}). All the
quantities in Table~\ref{tab:AgeLshort} and Fig.~\ref{fig:CoolCompar-2}
are shown as measured by a distant observer (i.e., ``redshifted'').
Errorbars in the figure correspond to the measured values with
uncertainties at the $1\sigma$ confidence level), and downward arrows
mark upper limits (at the $3\sigma$ confidence level). The last column
of the table indicates the references for the given estimates, the
reference numbers being supplemented by a letter ``a'' for the kinematic
age and ``s'' for the results of spectral analysis.

Let us start with the stars which are currently at the neutrino cooling
stage. The long-dashed cooling curve  is for the iron blanketing
envelope. Calculations show (e.g., \citealt{YKGH01,PC18}) that this
curve is almost independent of neutron star mass as long as the direct
Urca process in the stellar core is forbidden. As seen from the
Fig.~\ref{fig:CoolCompar-2}, variations of chemical composition in the
blanketing envelope allow one to explain much more objects, but not all
of them. In order to explain other objects, the effects of nucleon
superfluidity in the stellar core are  required (see \citealt{Page_14}
for a review of these effects). For instance, the neutron stars XMMU
J172054.5$-$372652, XMMU J173203.3$-$344518, and CXOU J185238.6+004020
(objects 7, 8, and 10) are significantly hotter than they should be at
their ages according to the model of a cooling neutron star with iron
heat blanket, but the allowance for accreted blankets, composed of light
elements, brings the theory to agreement with the observations. On the
other hand, the coldest stars at the neutrino cooling stage, such as the
Vela pulsar (object 16 in the figure), PSR J0205+6449 (object 12), or
PSR B1727$-$47 (object 26) are much colder than they should be according
to the iron heat blanket model. Their low thermal luminosities can
actually be explained by the direct Urca reactions, if these neutron
stars are sufficiently massive (\citealt{LPPH91,Haensel95}; see
Section~\ref{ssec:5:Intro}; the possibility of this explanation for the
listed objects is demonstrated in \citealt{COOLDAT}).

Now let us focus on the objects which are at the photon cooling stage.
According to Fig.~\ref{fig:CoolCompar-2}, all of them are more or less
compatible with the  modified Urca neutrino cooling in  non-superfluid
stars. Generally, they can be explained by variations of chemical
composition in heat blanketing envelopes. For instance, warmer objects
at this cooling stage, such as RX J0720.4--3125 (object 20), should have
predominantly iron blanketing envelopes, while colder objects may have
heat blankets made of light elements.

The evolution of neutron stars at the photon cooling stage does not
depend directly on their neutrino emission.  However, one can state that
the objects observed at the photon cooling stage could not have neutrino
emission   strongly enhanced  at the previous neutrino cooling stage
(with respect to the modified Urca emission of non-superfluid stars). 
Otherwise, at the neutrino cooling stage they should have been cooling
fast and they would transit to the photon cooling stage earlier. By $t
\sim 10^5$ yr their surface luminosity would be too week to be observed.
Therefore, all the objects observed at the photon cooling stage may
have  strongly suppressed neutrino emission; the hottest of them should
have blanketing envelopes made of iron.

\section{Other models of heat blankets and their applications}
\setcounter{equation}{0}
\label{sec:OtherBlankets}

Here we outline some other applications of heat-blanket
models in neutron star physics.

\subsection{Ordinary magnetic neutron stars with isothermal interiors}
\label{sub:1Dmagnet}

Let us consider a passively  cooling  ordinary isolated neutron star
with not too strong  magnetic field ($B \lesssim 10^{14}$
G)  provided its interiors are isothermal after the internal thermal
relaxation.  As discussed in Section~\ref{therm-magn-env}, the internal
thermal evolution of  such stars can be simulated with 1D cooling codes;
the main effects  of the magnetic field there consist in redistributing
the heat flow emerging  through the heat blanket to the surface. These
effects are incorporated into the models of heat blankets. 

The effects can be formally quite substantial making the magnetic poles much warmer
than the equatorial surface belts. However, they do 
not greatly affect the total thermal surface
luminosity $L_\mathrm{s}^\infty$ for a given temperature $\Tb$ 
at the heat blanket bottom (Section~\ref{subsub:GravLens}),
at least for a dipolar surface magnetic field configuration.

Cooling of neutron stars in this approximation has been  studied in many
publications (relevant reviews have been listed above). It is the
simplest way to include the effects  of magnetic fields into the cooling
theory of neutron stars but one  should bear in mind that the approach
is restricted by not too strong $B$-fields.  Cooling of these stars is
mostly regulated by the nuclear  composition of  their cores (which
opens or forbids enhanced neutrino emission like direct Urca process)
and by  baryon superfluidity of the core that greatly affects the
neutrino luminosity and heat capacity of the stars. 

\subsection{Neutron stars with very strong magnetic fields; magnetars}
\label{sub:2Dmagnet}

Cooling of neutron stars with superstrong magnetic fields is different. As
argued in Section~\ref{therm-magn-env}, their internal regions ($\rho
\gtrsim \rhob$) can be essentially non-isothermal mainly because of
strong anisotropic thermal conduction there. Anisotropic internal
temperature distributions, $\widetilde{T}(\bm{r},t)$, have to be
determined by the  internal thermal evolution equations with 2D (or 3D)
cooling codes using  local differential heat-blanket models as boundary
conditions.  These calculations are complicated. The results appreciably
depend on magnetic field strength and geometry  throughout the star. The
thermal and magnetic field evolution may become strongly coupled so that
one should study the  united thermomagnetic evolution of strongly
magnetized neutron stars (see, e.g., \citealt{PonsVigano19}, for
review). This science may be needed to explore the evolution  of the
high-$B$ pulsars and X-ray (``dim'') isolated neutron stars (XINSs) and
especially the evolution of soft gamma repeaters (SGRs) and anomalous
X-ray pulsars (AXPs; see, e.g., \citealt{2013Mereghetti,2017KasB} and
references therein). 

There is no clear difference between SGRs and AXPs \citep{GavriilKW02}.
Their most popular models assume $B \sim(10^{14}-10^{16})$~G (see
\citealt{2017KasB}, for review and references).  The stars with such
strong magnetic fields are called \emph{magnetars}
\citep{DuncanThompson92}. Although estimates of dipole magnetic fields
for a few of them are not so high, they may possess superstrong
small-scale fields near the surface \citep[e.g.,][]{2015Mereghetti}. An
alternative interpretation of the properties of the SGRs/AXPs (e.g.,
\citealt{ZezasTK15,BisnoIkhsanov15}, and references therein) is based on
the assumption that they are neutron stars with ``normal'' magnetic
fields $B\sim10^{11}-10^{12}$~G, slowly accreting matter from a
residual disk (left after a supernova  explosion or after a high-mass
X-ray binary evolutionary stage). However, an observational argument
against the latter scenario has recently been given by
\citet{Doroshenko_20}, based on the absence of aperiodic spectral
variability in the SGRs/AXPs, unlike in known accreting sources.

The SGRs/AXPs appear overall much hotter than ordinary cooling neutron
stars and spontaneously show violent bursting activity. In frames of the
magnetar paradigm, these features are usually associated with the
persistent and  explosive processes of internal energy release powered
somehow by magnetic fields. In this scenario, the SGRs/AXPs are usually
thought to be relatively young, $t \sim 10^3-10^5$ yr
\citep[see][]{2017KasB}. Observations show that some SGRs/AXPs
intermittently behave as high-$B$ pulsars and vice versa. One also
observes bursting activity in some ``ordinary'' neutron stars. 

The history of the development of 2D codes to follow the evolution of
strongly magnetized neutron stars is nicely reviewed by
\citet{PonsVigano19}, who also describe the most relevant numerical
methods and main results of such simulations. It seems that the use of
2D codes becomes especially important for surface magnetic fields $B \gtrsim
10^{14}$~G (as we have already mentioned above). 

Note that in some cases one needs to consider magnetized neutron stars 
which are so hot that the neutrino emission in their
heat blankets becomes important along with the
heat conduction. If so, the assumption of thermal
flux conservation, Eq.\ (\ref{therm-Tcrust}), used
in the standard heat-blanketing models, is violated and one
should account for the neutrino energy losses within the blanket
(as already mentioned in Section~\ref{therm-analyt}).

For instance, according to  \citet{2012KKPY}, this happens  when the
temperature $\Tb$ at the bottom of the blanket exceeds about $10^9$ K, which
may occur in magnetars. The appropriate heat-blanket models have been
constructed and used in a number of publications as reviewed by
\cite{PPP2015}. 

The heat transport through such envelopes can 
be essentially reduced owing to the neutrino emission.
This may affect numerical solution of the thermal evolution
problem within the neutron star interior ($\rho>\rhob$). 
In addition to the $\Ts -\Tb$ relations, which are
sufficient for ordinary cooling codes, it may be profitable  
to have the tables of the heat flux density at $\rho=\rhob$
and impose the condition of flux continuity at the bottom
of the heat blanket. 

\subsection{Hot neutron stars: newly born and old merging}
\label{sub:youngAndmergingNSs}

Strong neutrino emission may affect the structure of the
heat blankets not only in magnetars but also in other 
neutron stars with hot surface layers. 

First of all, they are \emph{neo-neutron stars} -- newly born neutron
stars which descended from the protoneutron-star stage (either after a
core-collapse supernova or accretion-induced collapse of a white dwarf)
but have not yet reached the stage of internal thermal relaxation (e.g.,
\citealt{LRPP94,GYP01}).  This evolutionary stage is called the
neo-neutron star phase and it lasts for $\sim 10^4$~s until the outer
crust ``forgets'' its initial conditions. Thermal evolution at this
stage has to be followed using the elaborated heat blanket models, which
take into account neutrino emission, contributions of photons and
electron-positron pairs into the pressure as well as the contraction of
the outer layers (see \citealt{BeznogovPR20} for details). Such stars
have not been observed so far, although they could be observed in the
future in a lucky chance of nearby core-collapsed supernova explosion.

Another example of neo-neutron stars can be provided by merging neutron 
stars in compact double neutron star binaries. One such merging 
event GW170817 has been detected by the LIGO/Virgo 
collaboration of gravitational  observatories \citep{2017GW} and in 
electromagnetic waves \citep{2017ApJGW170817}. One expects to 
observe many such events. Observational prospects of 
detecting neo-neutron star in mergers are discussed in 
\cite{BeznogovPR20}. Before two neutron stars merge, they can be 
heated up by strong tidal forces and become efficient neutrino 
emitters (e.g., \citealt{2002RD,2003RL,2018Alford}).

\subsection{Accreting neutron stars}  
\label{sub:accrete}

Constructing heat blanketing envelopes of accreting neutron
stars is a complicated task. A neutron star may accrete interstellar
matter or a matter from a companion star in a binary system. 
The infalling matter, composed usually of light
elements (hydrogen and helium), affects chemical composition of surface 
layers and produces energy release there due to the transformation
of the infall energy into the heat. If accretion is a
(quasi) persistent process, the freshly accreted 
material becomes eventually 
buried under the weight of newly accreted matter. This can be viewed  as
if the accreted matter sinks into the deeper layers of the star. 
The sinking can be accompanied by gravitational separation,
diffusion of ions and nuclear burning. Depending on the parameters,
the burning can be stable or explosive. If explosive, it creates
bursting activity of neutron stars by triggering X-ray bursts and
superbursts.  

As a rule, all these processes occur in heat blanketing
envelopes of accreting neutron stars. They have been studied 
for many years in numerous publications. Usually such
studies have not been focused on the heat insulating problem 
but have been mostly devoted to bursting activity of neutron 
stars (see, e.g., \citealt{2017Galloway}, \citealt{2017Zand}, 
and references therein).

Heat blanketing envelopes ($\Ts - \Tb$ relations)
of accreting neutron stars can be essentially time-dependent;
their chemical composition, and hence heat-insulating 
properties, can be variable. The assumption of stationary 
thermal flux conservation, Eq.~(\ref{therm-Tcrust}), may break down
because of the thermonuclear energy generation within the
heat blankets and strong associated neutrino emission. For instance, according to \citet{BBC02},
thermal luminosities of accreting neutron stars 
in different quiescent epochs (between accretion episodes)
can vary by a factor of $2-3$ because of variable chemical
composition in the blanketing layer for a constant internal
temperature of the star. While considering thermonuclear
burning of hydrogen and helium in the envelope, it may be 
important to include diffusive nuclear burning of
protons and He nuclei. They diffuse into the deeper 
and hotter layers, where their burning is essentially intensified
\citep{CB03,CB04,CB10,Wijngaarden_ea19}.

The models for heat blankets containing 
light (accreted) elements described in Sections~\ref{therm-env} and~\ref{sec:4}
can also be applied to some scenarios of evolution of accreting neutron stars. 

The models of heat blankets for neutron stars
which accrete either from interstellar matter
or from companion stars and
change chemical composition due to
diffusive nuclear burning of H, He or C
have been developed by \citet{Wijngaarden_ea19}.

A special case of the accreting neutron stars, for which the
heat blanketing envelopes can be most useful,
are soft X-ray transients
(SXTs) with intermittent active and quiescent periods 
of accretion. During high-state accretion episodes, compression
of the crust under the weight of newly accreted matter
results in deep crustal heating, driven by exothermic
nuclear transformations \citep{HZ90,HZ08,Fantina_18,GusakovChugunov21}.
There is a close correspondence
between the theory of thermal states of transiently
accreting neutron stars and the theory of neutron star
cooling \citep{YLH03}. Comparing the heating curves with a
measured equilibrium thermal luminosity, one can constrain
parameters of dense matter \citep[e.g.][]{YLPGH,Ho11,PCC19}; see
\citet{WDP13}, who discussed prospects of application of
such an analysis to various classes of X-ray transients.
A survey of neutron stars in SXTs with evaluated
average accretion rates and thermal luminosities in quiescence
has been presented by \citet{PCC19}.
For such SXTs, the models of heat blanketing envelopes
can be most useful.

\subsection{Rotating neutron stars}
\label{sub:rotate}

All neutron stars rotate, and their rotation may affect 
the structure of their heat blankets. A rotating star becomes
oblate with respect to the spin axis, which produces non-uniform
effective temperature distribution over the surface.
To be specific, we consider rigid rotation.

The general technique for solving the problem of rotating envelopes is
somewhat in line with that for magnetic envelopes 
(Section~\ref{therm-magn-env-intro}). One can divide the heat blanket
into small domains and apply the  approach of a locally flat,
plane-parallel layer  with some properly determined effective surface
gravity $g_\mathrm{eff}$ in each domain. The total luminosity of the
star is the sum of the surface emissivities of all the domains. Global
space-time metric is no longer spherically symmetric, being deviated
from the spherically symmetric metric (\ref{str-metrics}) due to
rotation.  The common and most reasonable approximation is to assume
that a $\Ts - \Tb$ solution in each domain is  the same as in a
virtual non-rotating star which has the same effective surface
gravity $g_\mathrm{eff}$ as in the local domain. Then the solution can
be found using self-similarity  relations
(Section~\ref{subsub:basic_eqns}) discussed throughout this paper many
times. While calculating the stellar luminosity for a distant observer, one
should take into account gravitational light bending outside the star
(like in Section~\ref{subsub:GravLens}).    

The first, simplified models of heat blankets for spinning
neutron stars were constructed by 
\citet{1985Geppert} and \citet{1986Geppert,1988Geppert} 
who were inspired by the discovery of millisecond
pulsars. The general solution was constructed by \citet{1993Miralles}
in the Hartle approximation, that is by treating the
rotation as sufficiently slow. This approximation seems
sufficient for all observable neutron stars. The 
local effective surface temperature $\Ts$ at the equator
of a spinning neutron star with isothermal interiors
appears slightly lower than at the pole. The effect
seems so weak that it modifies neutron star cooling
only slightly and is commonly ignored in cooling simulations.

Note that the above procedure to construct the heat blankets  neglects
meridional circulation of heat flows in stars deformed by rotation
(e.g., \citealt{Schwarzschild58,Kippenhahn}). Such circulations in
neutron stars are typically weak \citep{1978PavYak}.

\subsection{Old and cold neutron stars}
\label{sub:old}

At late cooling ages $t\gtrsim10$~Myr, isolated neutron stars
become really cold if they are not
strongly reheated, for instance, by accretion or by internal exothermal
non-equilibrium processes. According to estimates by 
\citet{YP04}, a non-superfluid 
passively cooling neutron star of age $t\sim 20$ Myr
would have the surface temperature $\Ts \sim 10^3$ K.
The presence of superfluidity in the core would lower $\Ts$ 
even more, by a factor of several. This seems to be the
lowest temperature limit for an old neutron star,
which is certainly purely academic. 

Observations of thermal emission of old neutron stars  (with
characteristic ages $\sim$ 1 Gyr) with the \textit{Hubble Space
Telescope (HST)} and large ground-based telescopes in the ultraviolet
and optical bands substantially supplement the results obtained with the
\textit{XMM-Newton} and \textit{Chandra} orbital observatories in the
X-rays. Most of the results are rather uncertain. The thermal radiation
of the nearest millisecond  pulsar J0437$-$4715 has been observed and
analyzed by many authors, in particular by
\citet{2004Kargaltsev,2012Durant,GonzalezCaniulef_GR19}. The
characteristic (spindown) age of this pulsar is 6.64 Gyr. Modeling the
cool thermal component of its spectrum yields $R=13^{+0.9}_{-0.8}$ km
and $T^\infty=(2.3\pm0.1)\times10^5$~K \citep{GonzalezCaniulef_GR19},
which indicates that the pulsar has been reheated during its evolution. 
The effective surface temperature $\Ts\sim(1-3)\times10^5$~K and
bolometric thermal luminosity $L^\infty=8^{+7}_{-4}\times10^{29}$ erg
s$^{-1}$,  obtained by \citet{Pavlov_ea17} for the ordinary radio pulsar
B0950+08 with characteristic age $17.5$ Myr, should also be caused by
reheating. On the other hand,  \citet{2019ColdNS} derived the estimate
$T < 42000$~K from the non-detection of PSR J2144$-$3933 with
characteristic age 0.3 Gyr in deep \textit{HST} observations. This
result may be related to slow rotation of the PSR J2144$-$3933, whose
spin period of 8.5~s is the longest among known radio pulsars.

Possible reheating mechanisms have been discussed in a number of
publications and summarized by \citet{2010Gonzalez}, who showed that the
rotochemical heating and superfluid vortex creep are preferable. 
Comparison of the reheating theory with observations was performed,
e.g., by \citet{GJPR15,YanagiNH20}. More exotic hypothetical heating
mechanisms include, for example, annihilation of dark matter particles
inside a neutron star \citep[e.g.,][]{Hamaguchi_NY19}.

Even these, really cold neutron stars cannot be absolutely isothermal, from the
surface to the center. Nevertheless, it is 
quite reasonable to assume that their internal regions
are highly isothermal and heat blankets are very thin. 
The ions in these blankets may be in the gaseous, liquid, 
solid or amorphous state. Their effects of partial ionization
and strong Coulomb coupling can be dominant. In contrast to
many cases considered above, the electrons may not constitute
an almost incompressible (uniform) background. The electron
system can be rather compressible, because the densities of these
cold heat blankets are relatively low. Both, bound and unbound,
electrons may coexist like in a terrestrial condensed
matter but the presence of huge, non-terrestrial magnetic
fields can make the properties of such a matter very peculiar.

\section{Conclusions and outlook}
\setcounter{equation}{0}
\label{sec:Conclude}

The models for heat blanketing  envelopes are most important for
numerical simulations of thermal evolution of neutron stars of many
types. These are ordinary cooling isolated middle-aged neutron  stars,
accreting neutron stars, magnetic and non-magnetic neutron stars
(particularly, magnetars and high-$B$ pulsars), very young neutron stars
which are recently born in supernova explosions, and old merging 
neutron stars in compact binaries. The heat blanket models are also
required to interpret observations of old and cold currently isolated
neutron stars as well as merging neutron stars which are expected to
become hot due to intensified tidal  interactions. 

These blanket models can be of different types and
flexible, for instance, for solving a specific
problem. For example, one can distinguish the
models for spherically symmetric neutron stars with
isotropic temperature distribution from the differential
models for strongly magnetized stars (that provide 
boundary conditions for computing the thermal evolution
problem in a star with very strong 
magnetic fields and highly anisotropic thermal conduction
in its interiors). 

We have presented heat-blanketing envelope models
composed of binary ion mixtures in diffusive equilibrium
and compared them with the traditional models
where the ion species are assumed to be strictly separated.
The diffusive equilibrium for non-isothermal
envelopes has been treated in thermodynamically consistent manner,
which is an advance compared with early works where the diffusive
equilibrium was evaluated neglecting temperature gradients.
This is a step towards the physically consistent treatment
of thermal states of neutron stars.

Nevertheless, current models for heat
blanketing envelopes are not perfect and
can be improved. For instance, 
one can include thermal diffusion, which is currently neglected.
Besides, it may be
important to include the magnetic force
into the hydrostatic balance in magnetic envelopes.
It would also be interesting to consider nucleosynthesis
and associated energy generation in the blankets
for accreting stars. More refined heat-blanket
models are desirable, containing 
ion mixtures of more than two species
 in and out of diffusive equilibrium. 

Even a perfect knowledge of  the heat-blanket model for an assumed
chemical composition of neutron star surface layers would not allow one
to unambiguously interpret the observations. If the effective surface
temperature is accurately measured, the uncertainties due to the unknown
chemical composition within the envelope can translate into a factor of
2--3 uncertainties in our theoretical predictions of the internal
temperature of the star $\Tb$, and in a factor of $\sim 100$
uncertainties in our predictions of the neutrino luminosity of the star.
Besides, the neutron stars with strong magnetic fields have strongly
non-uniform surface temperature distribution. The theoretical $\Ts-\Tb$
relations cannot be directly used to interpret observations of such
stars. Instead, they should be only used to calculate the total photon
luminosity, with the result depending on a largely unknown temperature
pattern on the surface.

To move further, it would be perfect to combine the modeling of the
neutron star thermal structure and evolution with  the modeling of
chemical evolution of the heat blanket. This program is  difficult to
realize, especially if the network of nuclear reactions within the
envelope has to be solved. Its realization would be equivalent to
solving all the necessary equations together, which will render the
artificial separation of  heat-blanketing envelopes unnecessary. Some
steps towards realization of this program have been made
\citep[e.g.,][]{Wijngaarden_ea19}, but its fulfillment will
obviously take a long time, during which the heat blanket models will
stay as an indispensable ingredient in the theory of neutron-star
thermal evolution.

\section*{Acknowledgements}

The authors are grateful to Pawel Haensel who participated at the very
initial state of this project.  M.V.B.{} acknowledges financial support
by the Mexican Consejo Nacional de Ciencia y Tecnolog\'{\i}a with a
CB-2014-1 grant \#240512. M.V.B.{} also acknowledges support from a
postdoctoral fellowship from UNAM-DGAPA. The work of A.Y.P.{} was
partially supported by the Ministry of Science and Higher Education of
the Russian Federation (Agreement with Joint Institute for High
Temperatures RAS No.\,075-15-2020-785) and by the Russian Foundation for
Basic Research (RFBR) jointly with Deutsche Forschungsgemeinschaft (DFG)
according to the research project 19-52-12013. The work of D.G.Y.{} was
partially supported by the grant 14.W03.31.0021 of the Ministry of
Science and Higher Education of the Russian Federation.

\newcommand{\NewAppendix}{
\setcounter{equation}{0}
\setcounter{figure}{0}
\setcounter{table}{0}
}

\begin{appendix}
\makeatletter
\renewcommand{\@seccntformat}[1]{\csname the#1\endcsname.~~}
\makeatother
\renewcommand{\thesection}{Appendix \Alph{section}}
\renewcommand{\thesubsection}{\Alph{section}.\arabic{subsection}}
\renewcommand{\thefigure}{\Alph{section}.\arabic{figure}}
\renewcommand{\theequation}{\Alph{section}.\arabic{equation}}
\renewcommand{\thetable}{\Alph{section}.\arabic{table}}

\NewAppendix
\section[\qquad\qquad Analytic approximations for Coulomb
logarithm]{Analytic approximations for Coulomb logarithm}
\label{app:Lambda_Eff}

In order to approximate the Coulomb logarithm $\Lambda_{\mathrm{eff}}$,
which determines the diffusion coefficient $D_{12}^{*}$
[\req{e:Lambda-eff-def}], we use the expression
\begin{equation}
\Lambda_{\mathrm{eff}}\left( \Gamma_0, x_1 \right) =
\ln{\left(1+\frac{p_1 x_1^2 + p_2 x_2^2 + p_3}{\Gamma_0^{p_4 x_1+p_5}} \right)}\,,
\label{e:Lambda-fit}
\end{equation}
containing five parameters  $p_1,\ldots p_5$. These parameters are
listed in Table \ref{tab:DiffCoeff-FitParam}.
Table \ref{tab:DiffCoeff-FitParam1}.
gives the root
mean squared (rms) relative deviation $\delta_\mathrm{rms}$ and the
maximum relative error $\delta_\mathrm{max}$.

For each mixture, all fit parameters have been determined from
the values of $\Lambda_{\mathrm{eff}}$ computed on
some grid points $(\Gamma_0, x_1)$. The target
function to minimize has been $\delta_\mathrm{rms}$. The grid of
$x_1$ points has been chosen as $x_1 = 0.01,0.1,0.2,0.3,\dots,0.9,0.99$; the
grid of $\Gamma_0$ points has been different for each binary
mixture. It is presented in Table  \ref{tab:FitDataRange}. For each
binary mixture, these grid points have been divided into three
intervals of $\Gamma_0$, denoted in Table \ref{tab:FitDataRange} as
I, II, and III. These ranges refer to the weak, intermediate, and
strong Coulomb coupling of ions, respectively. Note that the real
measure of the Coulomb coupling is  $\bar{\Gamma}$ (not
$\Gamma_0$). In interval I, the grid points are distributed
uniformly (each next point is larger than the previous one by
$\Delta^+$). In intervals II and III, logarithmic scale has been
used (each next point is larger than the previous one by a factor of
$\Delta^\times$).

\renewcommand{\arraystretch}{1.05}
\setlength{\tabcolsep}{6pt}
  \begin{table}[!t]
    \centering \caption{Fit parameters in Equation \eqref{e:Lambda-fit}.}
    \begin{tabular}{l c c c c c}
      \toprule
      \multicolumn{1}{ c }{Mixture} &  $p_1$  &  $p_2$  &  $p_3$  &  $p_4$  &  $p_5$  \\
      \midrule
      $^{1}$H -- $^{4}$He & $7.43\!\times\! 10^{-2}$ & $-1.13\!\times\! 10^{-2}$ & $1.72\!\times\! 10^{-1}$ & $8.57\!\times\! 10^{-2}$ & $1.45$ \\
      $^{1}$H -- $^{12}$C & $3.80\!\times\! 10^{-2}$ & $6.57\!\times\! 10^{-3}$ & $2.52\!\times\! 10^{-2}$ & $1.39\!\times\! 10^{-1}$ & $1.34$ \\
      $^{4}$He -- $^{12}$C & $7.01\!\times\! 10^{-3}$ & $9.08\!\times\! 10^{-4}$ & $1.09\!\times\! 10^{-2}$ & $1.17\!\times\! 10^{-1}$ & $1.41$ \\
      $^{12}$C -- $^{16}$O & $9.95\!\times\! 10^{-5}$ & $-6.35\!\times\! 10^{-6}$ & $1.61\!\times\! 10^{-3}$ & $3.96\!\times\! 10^{-2}$ & $1.48$ \\
      $^{16}$O -- $^{79}$Se & $7.22\!\times\! 10^{-5}$ & $5.00\!\times\! 10^{-5}$ & $1.14\!\times\! 10^{-4}$ & $1.33\!\times\! 10^{-1}$ & $1.38$ \\
      \bottomrule
    \end{tabular}
    \label{tab:DiffCoeff-FitParam}
  \end{table}
  \setlength{\tabcolsep}{6pt}
  \renewcommand{\arraystretch}{1.0}

\renewcommand{\arraystretch}{1.05}
\setlength{\tabcolsep}{10pt}
\begin{table}[!t]
  \centering \caption{Fit errors:
    the rms deviation $\delta_\mathrm{rms}$ and the maximum relative fit error $\delta_\mathrm{max}$.
    The last column gives the point, where the error reaches maximum.}
  \begin{tabular}{l c c l}
    \toprule
    \multicolumn{1}{ c }{Mixture} &   $\delta_\mathrm{rms}$  & $\delta_\mathrm{max}$  &  \multicolumn{1}{ c }{$(x_1, \Gamma_0)_{\mathrm{max}}$}\\
    \midrule
    $^{1}$H -- $^{4}$He & $0.031$ & $0.10$ & $(0.7,0.4)$\\
    $^{1}$H -- $^{12}$C &  $0.056$ & $0.18$ & $(0.99,0.729)$\\
    $^{4}$He -- $^{12}$C &  $0.040$ & $0.13$ & $(0.9,5.785)$\\
    $^{12}$C -- $^{16}$O &  $0.026$ & $0.10$ & $(0.9,0.015)$\\
    $^{16}$O -- $^{79}$Se &  $0.041$ & $0.16$ & $(0.9,0.187)$\\
    \bottomrule
  \end{tabular}
  \label{tab:DiffCoeff-FitParam1}
\end{table}
\setlength{\tabcolsep}{6pt}
\renewcommand{\arraystretch}{1.0}

  \renewcommand{\arraystretch}{1.05}
  \setlength{\tabcolsep}{6pt}
  \begin{table*}[!t]
    \centering
    \caption{Grid points $\Gamma_0$ used for calculating and fitting $\Lambda_\mathrm{eff}$.
      For each binary mixture the grid points are divided into three groups
      I, II and III; $\Delta^{+}$ and $\Delta^\times$ determine distance between two neighboring
      points in every group (see the text for details). The lower boundary of each group is
      given exactly, while the upper boundary is given approximately.}
    \begin{tabular}{l l l l}
      \toprule
      \multicolumn{1}{c}{Group}  &  $\Gamma_0$ in group I   &  $\Gamma_0$ in group II  &  $\Gamma_0$ in group III  \\
      \midrule
      $^{1}$H -- $^{4}$He    & $[10^{-4},0.05],\Delta^{+}\! =\! 0.002$ & $[0.4,1.6],\Delta^{\times}\! =\!1.25$ & $[1.7,52],\Delta^{\times}\! =\! 1.3$\\
      $^{1}$H -- $^{12}$C    & $[10^{-4},0.01],\Delta^{+}\! =\!0.001$ & $[0.15,0.4],\Delta^{\times}\! =\!1.2$ & $[0.4,6],\Delta^{\times}\! =\! 1.35$\\
      $^{4}$He -- $^{12}$C   & $[10^{-4},0.005],\Delta^{+}\! =\!3.5\!\times\!10^{-4}$ & $[0.06,0.2],\Delta^{\times}\! =\!1.25$ & $[0.2,5.8],\Delta^{\times}\! =\! 1.4$\\
      $^{12}$C -- $^{16}$O   & $[10^{-4},0.003],\Delta^{+}\! =\!10^{-4}$ & $[0.015,0.05],\Delta^{\times}\! =\!1.35$ & $[0.055,3.2],\Delta^{\times}\! =\! 1.4$\\
      $^{16}$O -- $^{79}$Se  & $[10^{-5},2.5\!\times\!10^{-4}],\Delta^{+}\! =\!10^{-5}$ & $[0.003,0.01],\Delta^{\times}\! =\!1.22$ & $[0.01,0.2],\Delta^{\times}\! =\! 1.34$\\
      \bottomrule
    \end{tabular}
    \label{tab:FitDataRange}
  \end{table*}

\NewAppendix
\section[\qquad\qquad Analytic approximations of $\Tb-\Ts$ relations]{Analytic approximations of $\Tb-\Ts$ relations}
\label{app:TsTb}

For convenience of using the $\Tb(\Ts, \rho^*)$ relations in
applications, all these
relations (Section~\ref{sec:4}) have been approximated by analytic
expressions of the form
\begin{align}
\Tb \left(Y, \rho^* \right) =10^7~\mathrm{K} \times
\left(f_4(Y) + \left(f_1 (Y)  - f_4(Y) \right)
\left[ 1 + \left( \frac{\rho^*}{f_2(Y)}\right)^{f_3(Y)} \right]^{f_5(Y)}\right),
\label{e:GenFit}
\end{align}
where the functions $f_1,\ldots, f_5$ are different for any binary
mixture, and $Y = (\Ts / 1~\mathrm{MK}) \allowbreak \times
\left({g_{s0}}/{g_{s}}\right)^{\slfrac{1}{4}}$. The expression for $Y$
allows one to rescale the values of  $\Tb$ for any value of the surface
gravity $g_{s}$ (see, e.g., \citealt{GPE83});
$g_{s0}=2.4271\times10^{14}$ cm~s$^{-2}$ is the surface gravity for the
canonical neutron star model; $Y$ has meaning of the
surface temperature (expressed in MK) for a star with
$g_{s}=g_{s0}$.

For H--He envelopes,
\begin{align}
\begin{split}
&f_1(Y) = p_1 Y^{p_2} \sqrt{1+p_3 Y^{p_4}}, \quad
f_4(Y) = p_5 Y^{p_6} \sqrt{1+p_7 Y^{p_8}}, \\
&f_2(Y) = \frac{p_9 Y^{p_{10}}}{\left(1-p_{11} Y + p_{12}Y^2\right)^2}, \quad
f_3(Y) = p_{13} Y^{-p_{14}} , \quad
f_5(Y) = -0.3.
\end{split}
\label{e:H-He-Fit}
\end{align}

For He\,--\,C envelopes,
\begin{align}
\begin{split}
&f_1(Y) = p_1 Y^{p_2 \lg Y + p_3}, \quad
f_4(Y) =  p_4 Y^{p_5 \lg Y + p_6}, \\
&f_2(Y) = p_7 Y^{p_8 \left(\lg Y \right)^2 + p_9}, \quad
f_3(Y) = p_{10} \sqrt{\frac{Y}{Y^2+p_{11}^2}} ,\quad
f_5(Y) = -0.2.
\end{split}
\label{e:He-C-Fit}
\end{align}

For C\,--\,Fe envelopes,
\begin{align}
&f_1(Y) = p_1 Y^{-p_2}\left(p_3 Y^2 + p_4 Y^4 - 1 \right), \quad
f_4(Y) = p_5 Y^{p_6}\left(1+p_7 Y^2 -p_8 Y^4 \right), \allowdisplaybreaks[4] 
\nonumber \\
&f_2(Y) = p_9 Y^{p_{11}-p_{10} \left( \lg Y \right)^2}, \allowdisplaybreaks[4] 
\nonumber \\
&f_3(Y) = p_{12} \sqrt{\frac{1}{Y^2+p_{13}^2}}\left(1-p_{14}Y^2\right) , \quad
f_5(Y) = -0.4.
\label{e:C-Fe-Fit}
\end{align}

All fit parameters and fit errors are listed in Tables
\ref{tab:FitParam} and \ref{tab:FitErr}, respectively; $\mathrm{lg}Y=\log_{10}Y$.

For each binary mixture the fit parameters have been calculated  from
the values of $\Tb$ computed on a grid of $(Y, \rho^*)$ points. The
target function to minimize has been the rms relative deviation. The
range of the data has been as follows. For all binary mixtures,  the
parameter $Y$ has been varied from 0.32 to $\approx 2.865$ with a
uniform step in logarithmic scale, with 24 grid points. The range of
$\rho^*$ has been different for different mixtures. For the H -- He
mixture, $\rho^*$ has been varied from $\approx 19.42$ \gcc{} to
$\approx 3.737\times 10^6$ \gcc{} with non-uniform steps; 41 grid points
in total. For He\,--\,C mixtures, $\rho^*$ has been varied from $\approx
280.5$ \gcc{}  to $10^8$ \gcc{} (the largest span). Since helium cannot
exist at $\rho>10^9$ \gcc, all values $\rho^* > 10^8$ \gcc{} have been
excluded. Therefore, for different $Y$ one has different number of grid
points $\rho^*$. For C\,--\,Fe mixtures, $\rho^*$ has been varied from
$\approx 1459$ \gcc{} to $\approx 10^9$ \gcc; 40  grid points $\rho^*$
in total.

Note that for all binary mixtures the grid points do not fill a
rectangular region in the  $(Y, \rho^*)$ plane. This region has the
shape of a quadrilateral with two parallel sides (along $Y$ axis).
The ranges of $\rho^*$ mentioned above form boundaries of the
regions (for each specific $Y$ the range is smaller and depends on
$Y$). However, this does not restrict the applicability of the
approximations. Because of their form  \eqref{e:GenFit}, which
describes a smooth temperature transition from  $f_1$ to $f_4$, they
can be safely extrapolated along $\rho^*$ outside the initial
region. The extrapolation along $Y$ should be done with care.

\begin{table*}
  \setlength{\tabcolsep}{3pt}
\caption{Fit parameters for the $T_\mathrm{b}(T_\mathrm{s})$ relations 
for different mixture types and bottom densities ($\rho_\mathrm{b}$ in \gcc).}
\label{tab:FitParam}
\begin{tabular}{l | c | c c c | c c c}
\toprule
   &  H\,--\,He mixtures, & \multicolumn{3}{c|}{He\,--\,C mixtures,} & \multicolumn{3}{c}{C\,--\,Fe mixtures,} \\
   &  \req{e:H-He-Fit}   & \multicolumn{3}{c|}{\req{e:He-C-Fit}} & \multicolumn{3}{c}{\req{e:C-Fe-Fit}} \\
$\rho_\mathrm{b}$ &  $10^8$ &  $10^8$ &  $10^9$ &  $10^{10}$ &  $10^8$ &  $10^9$ &  $10^{10}$ \\
\hline
 $p_1$      &3.150\rule{0ex}{2.5ex}& 5.161                & 5.296                 & 5.386                  & 0.2420               & 0.1929                & 0.1686                \\
 $p_2$      & 1.546                & 0.03319              & 0.07402               & 0.1027                 & 0.4844               & 0.4239                & 0.3967                \\
 $p_3$      & 0.3225               & 1.654                & 1.691                 & 1.719                  & 38.35                & 48.72                 & 55.94                 \\
 $p_4$      & 1.132                & 3.614                & 3.774                 & 3.872                  & 0.8680               & 1.423                 & 1.992                 \\
 $p_5$      & 1.621                & 0.02933              & 0.08210               & 0.1344                 & 5.184                & 5.218                 & 5.208                 \\
 $p_6$      & 1.083                & 1.652                & 1.712                 & 1.759                  & 1.651                & 1.652                 & 1.651                 \\
 $p_7$      & 7.734                & $1.061\!\times\!10^5$&  $1.057\!\times\!10^5$&  $1.056\!\times\!10^5$ & $-0.04390$           & 0.001037              & 0.03235               \\
 $p_8$      & 1.894                & 1.646                & 1.915                 & 1.881                  & 0.001929             & 0.004236              & 0.005417              \\
 $p_9$      & $2.335\!\times\!10^5$& 3.707                & 3.679                 & 3.680                  & $3.462\!\times\!10^4$& $3.605\!\times\!10^4$ & $3.652\!\times\!10^4$ \\
 $p_{10}$   & 7.071                & 4.011                & 3.878                 & 3.857                  & 2.728                & 2.119                 & 1.691                 \\
 $p_{11}$   & 5.202                & 1.153                & 1.110                 & 1.102                  & 4.120                & 4.014                 & 3.930                 \\
 $p_{12}$   & 10.01                &     ---              &      ---              &         ---            & 2.161                & 1.943                 & 2.021                 \\
 $p_{13}$   & 2.007                &     ---              &      ---              &         ---            & 2.065                & 1.788                 & 1.848                 \\
 $p_{14}$   & 0.4703               &     ---              &      ---              &         ---            & 0.008442             & 0.01758               & 0.02567               \\
\bottomrule
\end{tabular}
\end{table*}

\setlength{\tabcolsep}{6.5pt}
\begin{table}
  \caption{Fit errors for H\,--\,He, He\,--\,C and C\,--\,Fe mixtures;
    $\delta_{\mathrm{rms}}$ is the rms relative deviation,
    $\delta_{\mathrm{max}}$ is the maximum error. The last column gives the points
    where the error is maximal.}
  \centering
  \begin{tabular}{l l c c c}
    \toprule
    Mixture & $\log \rhob$ & $\delta_{\mathrm{rms}}$   &  $\delta_{\mathrm{max}}$  &  $(Y,~\rho^*~[\gcc])$\\
    \midrule
    H\,--\,He &  8.0     &  0.0031  &  0.015  &  $(2.865, 3.345\!\times 10^5)$        \\
    \hline
    He\,--\,C & 8.0     &  0.0036  &  0.011  &  $\left(0.32, 1.245\!\times\!10^3\right)$        \\
    He\,--\,C & 9.0     &  0.0036  &  0.011  &  $\left(0.32, 1.657\!\times\!10^3\right)$        \\
    He\,--\,C & 10.0    &  0.0035  &  0.010  &  $\left(0.32, 1.245\!\times\!10^3\right)$      \\
    \hline
    C\,--\,Fe & 8.0     &  0.0051  &  0.017  &  $\left(2.865, 1.528\!\times\!10^4\right)$        \\
    C\,--\,Fe & 9.0     &  0.0048  &  0.015  &  $\left(0.4259, 1.772\!\times\!10^3\right)$        \\
    C\,--\,Fe & 10.0   &  0.0047  &  0.014  &  $\left(0.3872, 1.637\!\times\!10^3\right)$      \\
    \bottomrule
  \end{tabular}
  \label{tab:FitErr}
\end{table}
\setlength{\tabcolsep}{6pt}
\renewcommand{\arraystretch}{1.0}

\end{appendix}

\bibliographystyle{elsarticle-harv}

\newcommand{\aap}{Astron. Astrophys.}
\newcommand{\aj}{Astron. J.}
\newcommand{\apj}{Astrophys. J.}
\newcommand{\apjs}{Astrophys. J. Suppl. Ser.}
\newcommand{\apss}{Astrophys. Space Sci.}
\newcommand{\araa}{Annu. Rev. Astron. Astrophys.}
\newcommand{\mnras}{Mon. Not. R. Astron. Soc.}
\newcommand{\nat}{Nature}
\newcommand{\pasa}{Publ. Astron. Soc. Australia}
\newcommand{\pasj}{Publ. Astron. Soc. Japan}
\newcommand{\pasp}{Publ. Astron. Soc. Pacific}
\newcommand{\physrep}{Phys. Rep.}
\newcommand{\prc}{Phys. Rev. C}
\newcommand{\prd}{Phys. Rev. D}
\newcommand{\pre}{Phys. Rev. E}
\newcommand{\prl}{Phys. Rev. Lett.}
\newcommand{\qjras}{Quarterly J. R. Astron. Soc.}
\newcommand{\sovast}{Sov. Astron.}
\newcommand{\ssr}{Space Sci. Rev.}

\newcommand{\eprint}{}

\end{document}